\documentclass[superscriptaddress,amsmath,amssymb,aps,prb,twocolumn,10pt]{revtex4-2}

\usepackage{todonotes}

\usepackage{dcolumn}% Align table columns on decimal point
\usepackage{bm}% bold math
\usepackage{float}
\usepackage{xcolor}
\usepackage{multirow}
\usepackage{hyperref}
\usepackage{graphicx}% Include figure files

\renewcommand{\vec}[1]{\ensuremath{\mathbf{#1}}}
\newcommand{\tper}{t_{\perp}}
\newcommand{\tpar}{t_{\parallel}}
\newcommand{\phiper}{A_{\perp}}
\newcommand{\phipar}{A_{\parallel}}
\renewcommand{\i}{\text{i}}
\renewcommand{\l}{\ell}

\newcommand{\nn}{\nonumber \\}

\begin{document}

\title{Dynamical correlations and nonequilibrium sum rules in photodoped Hubbard ladders}%\\ after pump-probe pulses}% Force line breaks with \\

\author{E.~Merhej}
\affiliation{
 School of Physical Sciences, The Open University,\\ Milton Keynes, MK7 6AA, United Kingdom 
}%
\affiliation{ PASQAL SAS, 24 rue Emile Baudot, 91120 Palaiseau, Paris, France}

\author{J.~P.~Hague}
\affiliation{
 School of Physical Sciences, The Open University,\\ Milton Keynes, MK7 6AA, United Kingdom 
}%

\author{R.~M.~Konik}
\affiliation{
 Condensed matter physics and Material Science Division,\\Brookhaven National Laboratory, Upton, NY 11973-5000, USA
}%

\author{A.~J.~A.~James}
\email{andrew.james@open.ac.uk}
\affiliation{
 School of Physical Sciences, The Open University,\\ Milton Keynes, MK7 6AA, United Kingdom 
}%

\date{\today}
\begin{abstract}
Using matrix product state techniques we study the nonequilibrium dynamical response of the half-filled Hubbard ladder when subject to an optical pump.
Optical pumping offers a way of producing and manipulating new strongly correlated phenomena by suppressing existing magnetic correlations.
The ladder allows the effects of pump directionality to be investigated, and compared to a single chain it has strong spin-charge coupling and a fully gapped excitation spectrum, promising different nonequilibrium physics.
We compute time-dependent correlations, including the nonequilibrium dynamical structure factors for spin and charge.
By deriving a combined spin-charge sum rule that applies both in and out-of-equilibrium, we show that spectral weight is pumped directly from the antiferromagnetic spin response into a low energy $\omega\sim 0$ charge response below the Mott gap.
The transfer of weight is pump direction dependent: pumping directed along the legs disrupts magnetic correlations more than pumping in the rung{-}direction, even if the post pump energy density is similar.
The charge correlation length is dramatically enhanced by the pump, whilst the spin correlations are most strongly suppressed at nearest and next-nearest neighbour spacings.
After the pump the system is in a nonthermal correlated metallic state, with gapless charge excitations and approximately equal spin and charge correlation lengths, emphasising the importance of treating these degrees of freedom on an equal footing in nonequilibrium systems.
\end{abstract}
\maketitle
\section{Introduction}
The dynamical response of a many-body system is a sensitive probe of the energy and momentum dependence of its collective excitations.
For equilibrium systems, the evolution of the dynamical response on tuning a parameter (such as temperature, doping, or an applied field) can reveal subtle phase transitions and the behaviour of emergent collective phenomena, and act as a precise test of theoretical models~\cite{lake2013multispinon}.
In nonequilibrium condensed matter systems, the dynamical response can be interrogated in real time, thanks to the advent of ultrafast experimental probes, including time resolved resonant inelastic X-ray scattering (tr-RIXS)~\cite{dean2016ultrafast,cao2019ultrafast,mitrano2019ultrafast,mitrano2019evidence,mitrano2020probing,wang2021x,wang2021single,paris2021probing,mazzone2021laser,baykusheva2022ultrafast}. %{\color{red}more references are needed to tr-RIXS}
This allows for the realisation and characterisation of nonequilibrium states in systems tuned by optical { (including infrared)} pumping, also known as `photodoping'.
These techniques have been used to demonstrate the emergence of nonequilibrium collective phenomena when an equilibrium correlated state is suppressed by pumping~\cite{fausti2011light,mitrano2016possible}.

{ An area of significant interest is the application of ultrafast pump–probe techniques to Mott and charge-transfer insulators~\cite{Giannetti2016}.
Ultrafast pump–probe work on these materials has mapped a consistent experimental storyline:  femtosecond photodoping shifts spectral weight from interband Hubbard transitions into in-gap/intraband channels and can transiently yield metallic responses on sub-ps scales~\cite{Basov2017}.
Seminal optical reflectivity experiments on halogen-bridged Ni chains have shown photoinduced collapse of the gap and the appearance of Drude-like weight within 130 fs~\cite{PhysRevLett.91.057401}.
Organic one-dimensional Mott salts then provided clean model systems: in ET-F$_2$ TCNQ, pump–probe spectroscopy resolved a transient metallic state mediated by spin–charge separation, coherent holon–doublon dynamics, and pressure-tunable recombination, with later work identifying biexciton features~\cite{PhysRevLett.112.117801,Wall2010,Mitrano2014}. Strong-field experiments have added a complementary axis: mid-IR/THz driving modulates the effective Hubbard 
U, and intense single-cycle THz fields can trigger an impulsive dielectric breakdown of the Mott phase and even sub-ps insulator–metal switching~\cite{Yamakawa2017,PhysRevB.107.085147,Singla2015}.  In parent cuprates, femtosecond absorption/reflectivity measurements have established photoinduced metallization and ultrafast carrier relaxation in Nd$_2$CuO$_4$ and La$_2$CuO$_4$, with $7$-fs probes directly capturing spin-sector dynamics~\cite{Okamoto2010,PhysRevB.83.125102}.
Momentum-resolved x-ray probes are now able visualize how photocarriers reshape correlations: tr-RIXS on Sr$_2$IrO$_4$ revealed rapid suppression of long-range magnetism with persistent short-range correlations~\cite{dean2016ultrafast,mazzone2021laser}.
On the electronic structure side, tr-ARPES has followed gap collapse and in-gap band formation in spin-orbit Mott iridates from the first pump-probe spectra to more recent light-induced insulator–metal transitions~\cite{PhysRevB.93.241114,Choi2024}.
Complementary all-optical and x-ray studies of photodoped spin–orbit Mott magnets have quantified sub-ps demagnetization and recovery pathways, helping tie the charge and spin channels together~\cite{PhysRevX.9.021020}.

In this work we simulate pump-probe studies on ladder materials (those in which pairs of 1D chains are coupled to produce a ladder like geometry).
There now exists a wide body of work studying the out-of-equilibrium behavior in cuprate-based telephone number ladders, Sr$_{14-x}$Ca$_x$Cu$_{24}$ (SCCO).
In such ladder materials, ultrafast experiments have progressed from all-optical snapshots of carrier coherence to momentum-resolved views of light-stabilized phases.
Femtosecond reflectivity on SCCO first showed reversible photo-conversion between insulating and metallic responses and even a hidden insulating state controlled by trains of sub-ps pulses, establishing that ladder carrier coherence can be steered on ultrafast timescales~\cite{Fukaya2015}.
Building on this work, optical-pump/THz-probe measurements combined with femtosecond resonant soft-x-ray scattering then uncovered a symmetry-protected metastable electronic state: near-infrared pumping was shown to transfer holes from chain “reservoirs” to the ladders, partially melting equilibrium charge order, and yielding long-lived, nonthermal conductivity changes.
Most recently, tr-RIXS at the oxygen K edge directly resolved a gapless, dispersive charge-order mode emerging in this light-induced state, with dispersion up to 0.8 eV from the charge-order wavevector and a slope comparable to the quasiparticle velocity.
This constitutes momentum-resolved evidence that photodoping can endow ladder carriers with itinerant character while charge order becomes dynamically fluctuating~\cite{Padma}.

We theoretically determine the momentum and energy resolved nonequilibrium response of a half-filled Hubbard ladder (a system of two coupled 1D Hubbard chains) due to simulated pumps.
A key aspect of our work is an investigation of how this response is constrained by sum rules.  Sum rules} relate integrals of dynamic response functions - moments of spectral weight - to static quantities.
A well known case occurs in many-body spin systems, for which the integrated spin response over all energies and momenta can be related to the (constant) magnitude of the local moment.
Analytical methods can be verified by checking that their integrated dynamics saturate this spin sum rule~\cite{caux2005computation}.
Moreover, as the total spin spectral weight is conserved, if the spin response is reduced in one region of energy-momentum space by tuning, then there is a concomitant increase in spin response elsewhere.
For example, in gapped spin systems with hard core excitations, increasing temperature leads to suppression of interband spectral weight, and the emergence of a low energy intraband response due to scattering from a thermal population of spin excitations~\cite{james2008finite,goetze2010low,tennant2012anomalous}.

When there is coupling between spin and other degrees of freedom such as charge, the magnitude of the local moment may not be conserved.
In these cases the spin sum rule will be static in equilibrium but can be time-dependent for nonequilibrium systems.
This allows for the possibility of pumping spectral weight not just to a different region within the spin response channel, but to a different response channel altogether, creating a fundamentally different correlated state.

We analyse { the optical pumping induced} transfer of dynamical response from the spin channel to the charge channel for the half filled Hubbard ladder.
{ As t}he Hubbard model provides a minimal theoretical description of the interplay between spin and charge degrees of freedom in a many-body quantum system { it is of significant theoretical interest in addition to its relevance to experiment}.
At half filling (one electron per site) the Hubbard model is a Mott insulator (an insulator driven by interaction rather than band structure effects), with strong antiferromagnetic correlations characterising the equilibrium state.

Exact diagonalization (ED) and matrix product state (MPS) methods have been used to simulate pump-probe experiments on the 1D Hubbard chain, showing that optical pumping suppresses the strong antiferromagnetic spin correlations, and produces a nonequilibrium metallic state with a low energy charge response~\cite{wang2017producing,kaneko2019photoinduced,ejima2020photoinduced,ejima2023entanglement}.
A nonequilibrium phase diagram for the 1D Hubbard chain at half filling, featuring eta-pairing, charge and spin density wave phases, has been produced for photodoped Mott insulators at late times, by using a generalised Gibbs ensemble approach~\cite{murakami2022exploring}.
However, it should be noted that the Hubbard chain has a number of unusual properties.
In particular it is an integrable system for which excitations are fractionalized, with a gapless spin sector and a gapped charge sector.

In contrast to the chain, the 2D half filled Hubbard model has strongly coupled spin and charge degrees of freedom: the motion of a hole in an antiferromagnetic background leaves a trail of flipped spins which is energetically penalised.
Hence, the response of the 2D system to optical pumping is expected to reflect this coupling.  
The photodoped 2D Hubbard model has been investigated using the dynamical cluster approximation (DCA)~\cite{eckstein2016ultra}, and the spin-charge coupling was found to affect the relaxation rate of the system after pumping.
However this method, other cluster dynamical mean field theory (DMFT) extensions, and ED can only provide limited information on the spatial correlation effects that are accessible to tr-RIXS (for an excellent recent review covering a variety of methods and results see Ref.~\cite{murakami2023photo}).

The Hubbard ladder, like the chain, is amenable to matrix product state numerical methods so that both short and long ranged correlations can be accurately probed as a function of time.
Unlike the chain, but similar to the 2D model, the ladder is nonintegrable and its spin and charge sectors are coupled, with both being gapped.
A further distinction between the ladder and the chain is that the former allows the effects of anisotropy to be explored.
Therefore, the ladder is a promising system for obtaining new insights into the nonequilibrium behaviour of the photodoped Hubbard model, and understanding the differences between the 1D and 2D cases.

We use matrix product state techniques to study infinite and finite ladders, and examine the interplay of spin and charge degrees of freedom for different pump directions by extracting the appropriate equal and unequal time correlation functions.
In particular we calculate time-dependent dynamical structure factors, which are relevant to tr-RIXS type experiments.

In the limit of strong repulsion the low-energy properties of the Hubbard model at half filling are effectively given by the spin degrees of freedom alone, resulting in a Heisenberg model description.
A recent field theory analysis of the Heisenberg ladder in the weak rung coupling limit~\cite{ren2023ultrafast}, has examined the suppression and partial recovery of antiferromagnetic correlations for square form quenches of the rung exchange interaction, but this method by construction cannot reveal the effects of charge excitations.
Consequently, in the Heisenberg model spectral weight can only be redistributed within the spin sector, and the total integrated spin spectral weight is constant.

Here we { determine a nonequilibrium sum rule for single band Hubbard models, showing that the combined total spin and charge spectral weight is a conserved (static) quantity.} 
%even for the nonequilibrium Hubbard model.
{}
This combined sum rule can be used to understand that optical pumping of the ladder does not redistribute spectral weight within the spin and charge channels independently (by creating spin and charge excitations respectively).
Instead there is a net transfer of spectral weight to the charge channel, in the form of a strong response at low energies associated with intraband scattering processes in the upper Hubbard band.
{ This nonequilibrium reduction in the total spin response is in stark contrast to spin only models, where the suppression of antiferromagnetic correlations requires an enhancement of the spin response at other energies and momenta.}

We show that pump anisotropy has a strong effect on the suppression of spin correlations{: for a given fluence, when the pump is along the leg{-}direction the larger number of nearest neighbours leads to enhanced energy transfer and therefore more disruption of spin correlations relative to pumping along the rung{-}direction.}
We find that the suppression of the spin response is characterised by both a decrease in the antiferromagnetic correlation length and an additional large decrease in the spin correlations at short distances, relative to equilibrium.
Concomitantly, as optical pumping transfers spectral weight to the charge response, the charge correlation length grows dramatically until it is similar to the spin correlation length { and the charge response becomes gapless}, leaving the system in a correlated metallic state far from the Heisenberg model limit.

This paper is organised as follows.
In Section~\ref{sec:model} we describe the specifics of our model and pump protocol, and some aspects of the MPS methods used including the use of conserved quantities.
Section~\ref{sec:realtime} reports our results for real time dynamics during and after different simulated pumps, including the magnetisation, local doublon/holon density and the equal time spin and charge correlation functions.
In Section~\ref{sec:dsf} we present { the corresponding results} for the nonequilibrium spin and charge dynamical structure factors as a function of time.
We summarise and discuss our findings in Section~\ref{sec:discussion}, and provide some further technical details in the appendices.

\section{Model and pump protocol}
\label{sec:model}
The two-leg Hubbard ladder is described by the Hamiltonian
\begin{align}
\label{eq:hubbardladder}
    H_0 =& -t_{\parallel}\sum_{i,l,\sigma}[c^{\dagger}_{i,l,\sigma}c_{i+1,l,\sigma} + h.c] \nonumber \\
    & -t_{\perp}\sum_{i,\sigma}[c^{\dagger}_{i,1,\sigma}c_{i,2,\sigma} + h.c] +
    U \sum_{i,l}n_{i,l,\uparrow}n_{i,l,\downarrow},
\end{align}
for a system of $L$ rungs and therefore $N=2L$ sites.
Here $c^{\dagger}_{i,l,\sigma}$($c_{i,l,\sigma}$) creates (annihilates) an electron on rung $i$, leg $l=1,2$, with a spin $z$-component $\sigma \in \{\uparrow,\downarrow\}$.
The number operator $n_{i,l,\sigma} = c^{\dagger}_{i,l,\sigma}c_{i,l,\sigma}$.
The first term of the Hamiltonian describes the hopping of the electrons along the same leg with hopping parameter $t_{\parallel}$, while the second term describes the electron hopping along the rungs of the ladder with hopping parameter $t_{\perp}$.
The third term represents the repulsive on-site interaction, with strength $U>0$.

In equilibrium we assume the hopping is isotropic with $t_\parallel=t_\perp=t_h$ (we set the energy scale using $t_h=1$).
However, we will consider optical pumping that can affect the hopping parameters anisotropically so that $t_\parallel \ne t_\perp$ during the pump.

The equilibrium properties of the Hubbard ladder have been studied using analytical and numerical techniques, including bosonization (see e.g. Ref~\cite{giamarchi2003quantum} and references therein) and density matrix renormalization group algorithms~\cite{noack1996ground,yang2019spectral}.
At half filling (one electron per site) any finite $U$ results in an insulating ground state with  antiferromagnetic correlations.
Under these circumstances both charge and spin excitations are gapped. In a simplified description, photodoping the half-filled system produces vacancies on some sites and double occupancy of others.
The resulting holon and doublon excitations are mobile, inducing a nonequilibrium metallic state.

We choose to study photodoping via a pump characterised by a vector potential component
\begin{align}
\label{eq:pumpprotocol}
    A(t) = A_0 e^{-(t-t_p)^2/(2\sigma_p^2)}\sin[\Omega(t-t_p)],
\end{align}
where $A_0$ and $\Omega$ are the amplitude and frequency of the pump respectively, $\sigma_p$ is the pump width and $t_p$ is a time offset that sets the centre of the pump.
The coupling between electrons and the applied field during the optical pump is achieved through the Peierls substitution,
\begin{align*}
    t_{\parallel}c^\dagger_{i,l,\sigma}c_{i+1,l,\sigma} +\text{h.c.}&\to t_{\parallel} e^{i A_{\parallel}(t)} c^\dagger_{i,l,\sigma}c_{i+1,l,\sigma}+\text{h.c.},\\
    t_{\perp}c^\dagger_{i,1,\sigma}c_{i,2,\sigma}+\text{h.c.} &\to t_{\perp} e^{i A_{\perp}(t)} c^\dagger_{i,1,\sigma}c_{i,2,\sigma}+\text{h.c.}.
\end{align*}
We consider two different pump polarizations, the `leg pump' corresponding to
\begin{align*}
    A_\parallel(t) =A(t) \quad \text{with} \quad A_\perp=0,
\end{align*}
and the `rung pump',
\begin{align*}
     A_\parallel=0 \quad \text{with} \quad A_\perp(t) =A(t). 
\end{align*}
The complex phase factors in the hopping terms lead to a time-dependent Hamiltonian, $H(t)$.

We are interested in systems that have strong magnetic correlations in equilibrium, so we take our initial state to be the ground state of $H_0$ (Eq.~\ref{eq:hubbardladder}) at half filling, with $t_\perp=t_\parallel=t_h=1$ and $U=8 t_h$.
The pump is then applied, either in the leg or rung{-}direction, and the resulting time-dependent Hamiltonian $H(t)$ causes the state to evolve in a non-trivial manner.

We employ MPS techniques to study the ensuing nonequilibrium behaviour, computing the ground state of the equilibrium system  $H_0$ using the density matrix renormalization group (DMRG) technique, and applying time evolving block decimation (TEBD) to evolve the system under the influence of $H(t)$.
We study the thermodynamic limit, i.e.  infinitely long ladders, using infinite DMRG (iDMRG) and infinite TEBD (iTEBD).
Our simulations are performed using ChainAMPS~\cite{chainamps}{, a set of MPS algorithms developed to deal with systems of many coupled integrable chains \cite{james2013understanding,james2015quantum,james2018non,james2019nonthermal}. In the present case each rung is treated as a short `chain' and the computational requirements of the MPS representation are reduced by }taking into account the conservation of the total number of electrons and the total $z$-component of spin.
For leg pumps we also use parity (reflection symmetry between the two legs) to improve the numerical efficiency of the MPS representation.
This is not applicable for rung pumps because they break the parity symmetry.
This can be seen by applying the parity operator which entails $l=1,2 \to 2,1$: for a leg pump $H(t)$ remains invariant under this operation, but for a rung pump the $\perp$ hopping parameters are complex and applying the parity operator leads to
\begin{align*}
    &t_\perp (e^{i A(t)} c^\dagger_{i,1,\sigma}c_{i,2,\sigma} + e^{-i A(t)} c^\dagger_{i,2,\sigma}c_{i,1,\sigma}) \\
    &\rightarrow t_\perp (e^{i A(t)} c^\dagger_{i,2,\sigma}c_{i,1,\sigma}+e^{-i A(t)} c^\dagger_{i,1,\sigma}c_{i,2,\sigma}),
\end{align*}
so that the parameters for hopping in the rung{-}direction are conjugated.
{ Thus, during and after a rung{-}direction pump, the system is no longer guaranteed to be in a state of definite parity.}

Breaking the parity symmetry also has consequences for charge density expectation values.
For leg pumps on infinite ladders the occupation number $\langle n_{i,l} \rangle =1$ by translational invariance and parity symmetry.
For rung pumps translational invariance along the legs is maintained, so $\langle n_{i,l}\rangle = \langle n_{j,l}\rangle$, but the broken parity symmetry during the pump means that in general
$\langle n_{i,1} \rangle \ne \langle n_{i,2}\rangle$.
{ and it is only the total occupation of a rung that is conserved, $\langle n_{i,1} +n_{i,2} \rangle =2$.}

Following earlier work on the Hubbard chain~\cite{wang2017producing,ejima2020photoinduced}, we choose a pump frequency $\Omega=6$ that is close to the Hubbard interaction scale $U=8$ (approximately the gap between the lower and upper Hubbard bands), so that significant magnetic and charge effects should be observed in the nonequilibrium state {(we discuss the effect of altering $\Omega$ in Appendix~\ref{app:frequency})}.
We fix the pump width to be $\sigma_p=1$, centred at $t_p=\pi/2$, and apply the pump from $t=0$ to $t=\pi$.
These values result in the pump switching on and off when its instantaneous amplitude is zero, so that the effects of an abrupt change in $H$ are diminished and we can improve numerical efficiency by using a time independent Hamiltonian for the majority of the evolution.
{
As well as their relevance to previous theoretical works, these parameter choices are similar to those for studies of the ladder material SCCO, for which $t_\parallel \approx t_\perp$, $U=8 t_\parallel$ and $t_\parallel$ is on the order of $10^{-1}\,\text{eV}$~\cite{Fukaya2015,Padma}).
For example, if we set $t_\parallel \approx 0.2\,\text{eV}$ in our work then the pump we use has a length of $\approx 10\,\text{fs}$ and a wavelength of $\approx 1000\,\text{nm}$ (i.e. infrared).}

\section{Dynamics in real time and space}
\label{sec:realtime}
{ In this section we use finite and infinite MPS methods to compute equal time correlations during and after pumping, using the model described in Section~\ref{sec:model}.
We examine local observables (the energy density, magnetization, and doublon/holon density) and the spatial correlation functions for spin and charge in real space.
The response in energy and momentum space, taking into account the finite lifetime of the probe pulse, will be considered in Section~\ref{sec:dsf}.}
\subsection{Local observables}
\label{sec:local}
\begin{figure}
    \centering
    \includegraphics[scale=0.45]{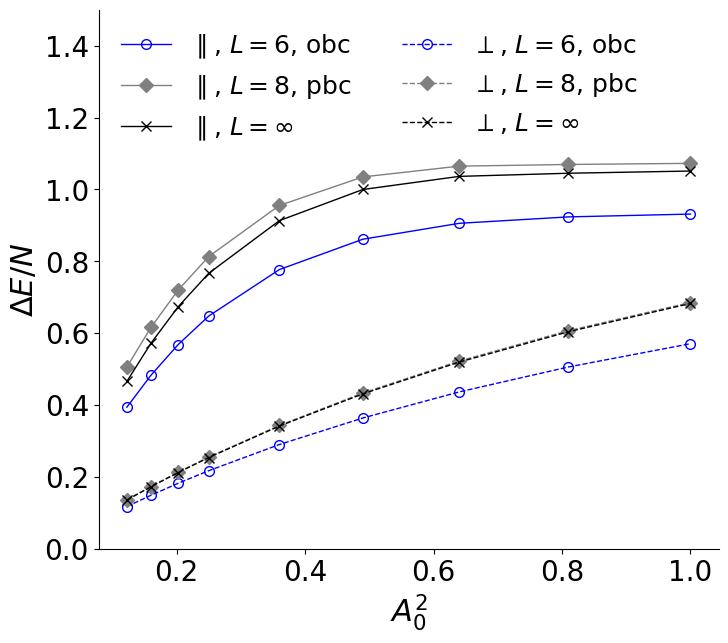}
    \caption{Energy density of the Hubbard ladder, relative to its ground state, after pumps of the type described in Eq.~\ref{eq:pumpprotocol} are applied along either the leg $(\parallel)$ or rung $(\perp)$ directions.
    Data are shown for infinitely long ($L=\infty$) and finite ladders {with $L=8$ and periodic boundary conditions (pbc) and $L=6$ and open boundary conditions (obc).
    The pump} parameters are $\Omega = 6.0$ and $\sigma_{p} = 1.0$.
    Lines are a guide to the eye.}
    \label{fig:energy-density}
\end{figure}

In order to compare the effects of pumping along the legs versus along the rungs, we first determine the energy imparted to the system.
{ The final photoinjected energy $\Delta E$ is given by} the difference between the energy after the pump ends and the energy of the initial equilibrium state (the ground state of Eq.~\ref{eq:hubbardladder} at half filling).
%The direction of the pump significantly affects the energy transferred to the system for a given fluence.
Fig.~\ref{fig:energy-density} shows the {photoinjected energy per site, $\Delta E/N$, as a function of $A_0^2$ (which is proportional to the fluence) for leg and rung{-}direction pumps.
We include results for infinite ladders (using iTEBD) and finite ladders (using ED).
For a given value of $A_0$, leg{-}direction pumping (denoted $\parallel$) leads to a larger photoinjected energy than the corresponding rung{-}direction pump (denoted $\perp$).
For the range of $A_0$ shown there is a plateau starting at $A_0 \sim 0.8$ (or $A_0^2 \approx 0.64$) for leg{-}direction pumping, but not for rung{-}direction pumping.

Leg and rung-direction pumping should be expected to yield qualitative differences: the direct coupling between the pump and the motion of the electrons is only along the pump direction, and the redistribution of energy and momenta along the other direction occurs through the scattering of excitations due to interactions.
The first (direct coupling) stage of this process is highly anisotropic, because the (infinite) leg direction supports continuous values of momenta $k_x$, but the (finite) rung direction admits only two discrete momenta $k_y$.

A short time expansion (see Appendix~\ref{app:shorttime}) shows that at lowest order the photoinjected energy scales as $\Delta E \sim z_\alpha t_\alpha A_0^2 \Omega^2$, where the index $\alpha=\parallel,\perp$ indicates the pump direction, and $z_\parallel=2$ and $z_\perp=1$ are the number of nearest neighbours in the leg and rung directions respectively.
This scaling confirms that the photoinjected energy is initially be governed by the different spatial properties in the two directions.}
%in qualitative agreement with the behaviour observed for the Hubbard chain~\cite{wang2017producing}.
%For fixed $A_0=1$, we find that the leg{-}direction pump (denoted $\parallel$) of the infinite ladder imparts approximately $40\%$ more energy than the corresponding rung{-}direction pump (denoted $\perp$).
%Adjusting the amplitude of the leg pump to $A_0=0.45$ {(or $A_0^2\approx0.2$)} results in a similar $\Delta E$ to that of the $A_0=1$ rung pump.
In the following we consider the rung pump with amplitude $A_0=1$ as our reference, and compare it with leg pumps at different $A_0$ corresponding to either similar fluence or similar photoinjected energy. 
%%%%

Next we consider the magnetization, and the density of doubly occupied and vacant sites.
Specifically we compute the expectation of the squared $z$-component of local spin, $\langle (S^z_{i,l})^2 \rangle$, where in our units $S^z_{i,l} = (1/2) c^\dagger_{i,l,a} \sigma^z_{ab} c_{i,l,b}= (n_{i,l,\uparrow}- n_{i,l,\downarrow})/2$.
This quantity can be recast in terms of the density of doubly occupied sites $d_{i,l}=n_{i,l,\uparrow}n_{i,l,\downarrow}$ and density of holes $h_{i,l}=(1-n_{i,l,\uparrow})(1-n_{i,l,\downarrow})$:
\begin{align*}
    \langle (S^z_{i,l})^2 \rangle &=\frac{1}{4}(1-\langle d_{i,l}+h_{i,l} \rangle).
\end{align*}
For half filling in equilibrium, $\langle d_{i,l}\rangle=\langle h_{i,l} \rangle$ because of the constraint {$\langle n_{i,l}\rangle=1 $}.
However, when the parity symmetry between the legs is broken (as it is for a rung{-}direction pump) this constraint is relaxed to {$\langle n_{i,1}+n_{i,2}\rangle=2$} and it is only true that $\sum_l\langle d_{i,l}\rangle=\sum_l\langle h_{i,l} \rangle$.
We define $\Delta d=\sum_l\langle d_{i,l}(t) - d_{i,l}(0)\rangle/2$ as the difference between the instantaneous doublon density and the equilibrium value, averaged over the two legs.

\begin{figure}
    \centering
    \includegraphics[scale=0.45]{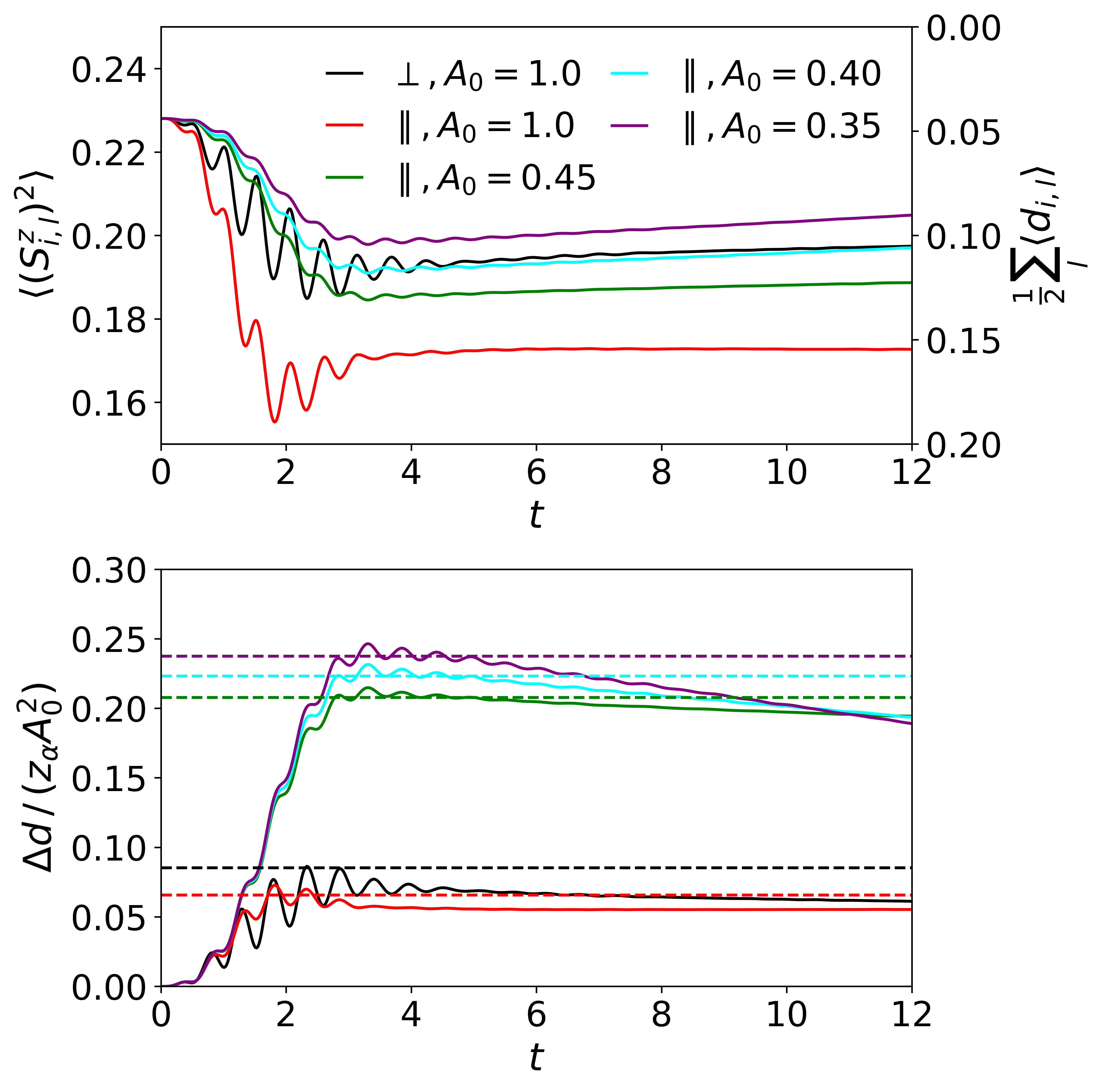}
    \caption{Upper panel: magnetization (left axis) and doublon number (right axis) as a function of time for the infinite Hubbard ladder when pumped along its legs $(\parallel)$ or along its rungs $(\perp)$ from $t=0$ to $t=\pi$.
    The pump parameters are $\Omega = 6.0$ and $\sigma_{p}=1.0$.
    Lower panel: The change in doublon number relative to the initial state $\Delta d=\sum_l(\langle d_{i,l}(t)-d_{i,l}(0)\rangle)/2$, scaled by { $z_\alpha A_0^2$ with $z_\parallel=2$ and $z_\perp=1$}, showing collapse onto a single curve at short times.
    Dashed lines show the values of $\Delta E/ (N U z_\alpha A_0^2)$ for the various pumps.
    {For convenience when comparing with Fig.~\ref{fig:energy-density}, the squared values of $A_0$ are, $(0.45^2)\approx0.2$, $(0.4)^2\approx 0.16$ and $(0.35)^2\approx0.12$.}
    }
    \label{fig:magnetization_iTEBD}
\end{figure}

%Photodoping during the pump decreases magnetisation and leads to oscillations in magnetisation and doublon number with double the pump frequency.
For $U\to \infty$, the ground state of the half-filled, translationally invariant, Hubbard ladder has $\langle (S^z_{i,l})^2 \rangle=1/4$ due to the presence of only one electron per site $\langle d_{i,l}\rangle=\langle h_{i,l} \rangle=0$.
At finite $U$ the actual magnetization is reduced due to quantum fluctuations (virtually doubly occupied and vacant sites).
For our parameters $\langle (S^z_{i,l})^2 \rangle\approx 0.23$ and {$\langle d_{i,l}\rangle \approx 0.05$}, as seen at time $t=0$ in the upper panel of Fig.~\ref{fig:magnetization_iTEBD}.
Photodoping during the pump further suppresses the magnetization (though not necessarily monotonically) due to the enhanced density of doubly occupied and vacant sites.
The suppression of magnetization (or equivalently the enhancement of doublon density) during the pump is accompanied by oscillations with a time scale $T=\pi/\Omega$, i.e. with a frequency double that of the pump.
{The oscillatory time dependence is consistent with the short time expansion in Appendix~\ref{app:shorttime}, which suggests that the change in magnetization should scale as $\sim z_\alpha t_\alpha A^2_0 \sin^2 \Omega t$.}
After the pump ends at $t=\pi$ the oscillations persist with damping { due to many-body interactions}, and the magnetisation either plateaus {(e.g. the leg pump with $A_0=1.0$)} or appears to recover slowly (e.g. the leg pump with $A_0=0.35$).

%Comparison shows that the amplitude of oscillations is governed by $A_0$ (consistent with a small $t$ expansion), whilst the overall trend and post pump magnetization is related to the total photoinjected energy.
%This can be seen in the lower panel of Fig.~\ref{fig:magnetization_iTEBD}, which plots the change in doublon density $\Delta d$ scaled by { $z_\alpha A_0^2$} the fluence, showing that $\Delta d/A_0^2$ is the same for all pumps at short times.
{The lower panel of Fig.~\ref{fig:magnetization_iTEBD} plots the change in doublon density $\Delta d$ divided by $z_\alpha A_0^2$, demonstrating that the results obey the short time scaling, collapsing onto a single curve for small $t$ compared to the characteristic hopping timescale $t_h^{-1}=1$. 
At later times interactions and higher order processes become important and the curves separate.
Then} $\Delta d$ approaches $\Delta E/(NU)$ (indicated by dashed horizontal lines on the plot) which is the expected result for the change in doublon density when injecting energy $\Delta E/N$ per site into a model with no hopping.
In the post pump period $\Delta d$ falls below $\Delta E/(NU)$.
This can be interpreted as the creation of increasing numbers of approximately stationary doublons during the pump, followed by the decay of some of them and an associated increase in kinetic energy.
The profile for the most energetic pump we consider, the $A_0=1$ leg pump, shows some decay of doublons before the pump ends at $t=\pi$.
This qualitative difference to the other pumps is consistent with the saturation shown in Fig.~\ref{fig:energy-density}.

The leg{-}direction pumps cause more disruption of magnetic correlations (and correspondingly a higher density doublons) than the rung pump, even if the energy transferred to the ladder is the same.
%The behaviour of $\Delta d/A_0^2$ for the rung pump is similar to that of the leg pumps at very short times, but the significantly different value of $\Delta E$ means that it separates from the other curves at a time $t<1$.
The photoinjected energy for the rung pump is $\Delta E \approx 0.7$, close to that of the $A_0=0.45$ leg pump.
However, the upper panel of Fig.~\ref{fig:magnetization_iTEBD} shows that the magnetization (and doublon) profile for the rung pump at later times is most similar to that of the leg pump with $A_0=0.4$, which has a lower photoinjected energy, $\Delta E \approx 0.6$. 

\subsection{Equal time correlations}
\label{sec:equaltimecor}
\begin{figure}
\centering
\includegraphics[width=\linewidth]{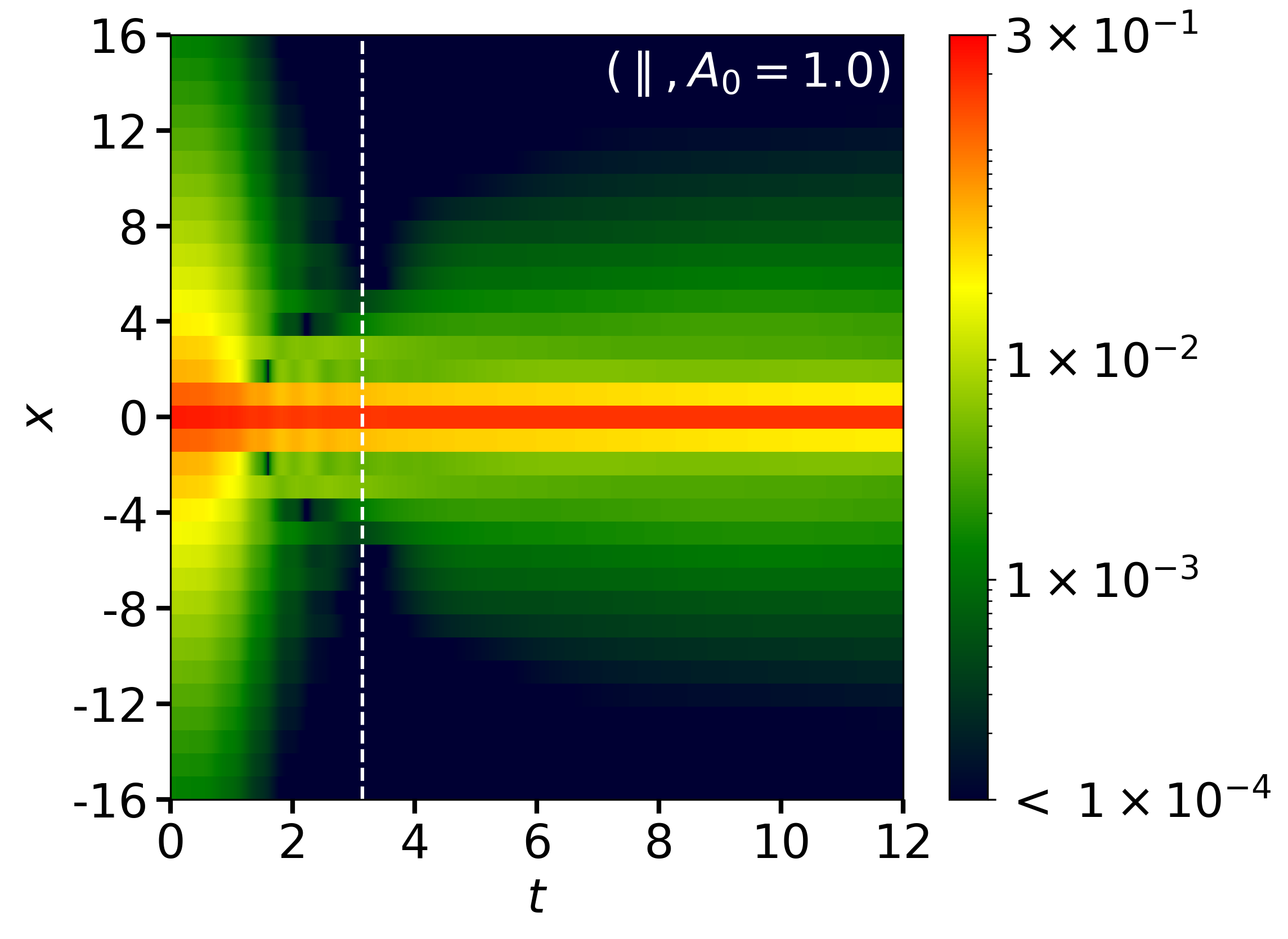} 
\includegraphics[width=\linewidth]{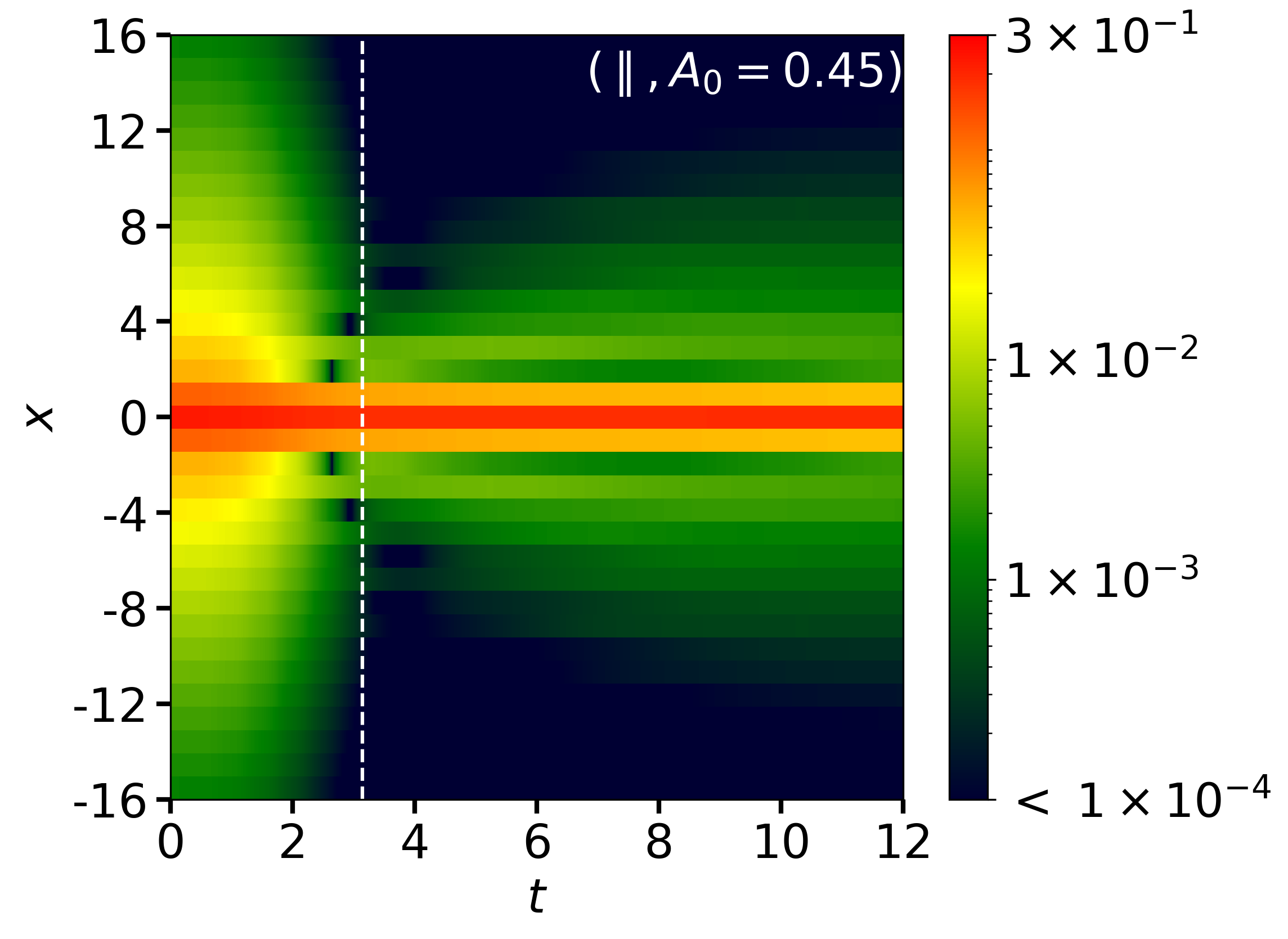} 
\includegraphics[width=\linewidth]{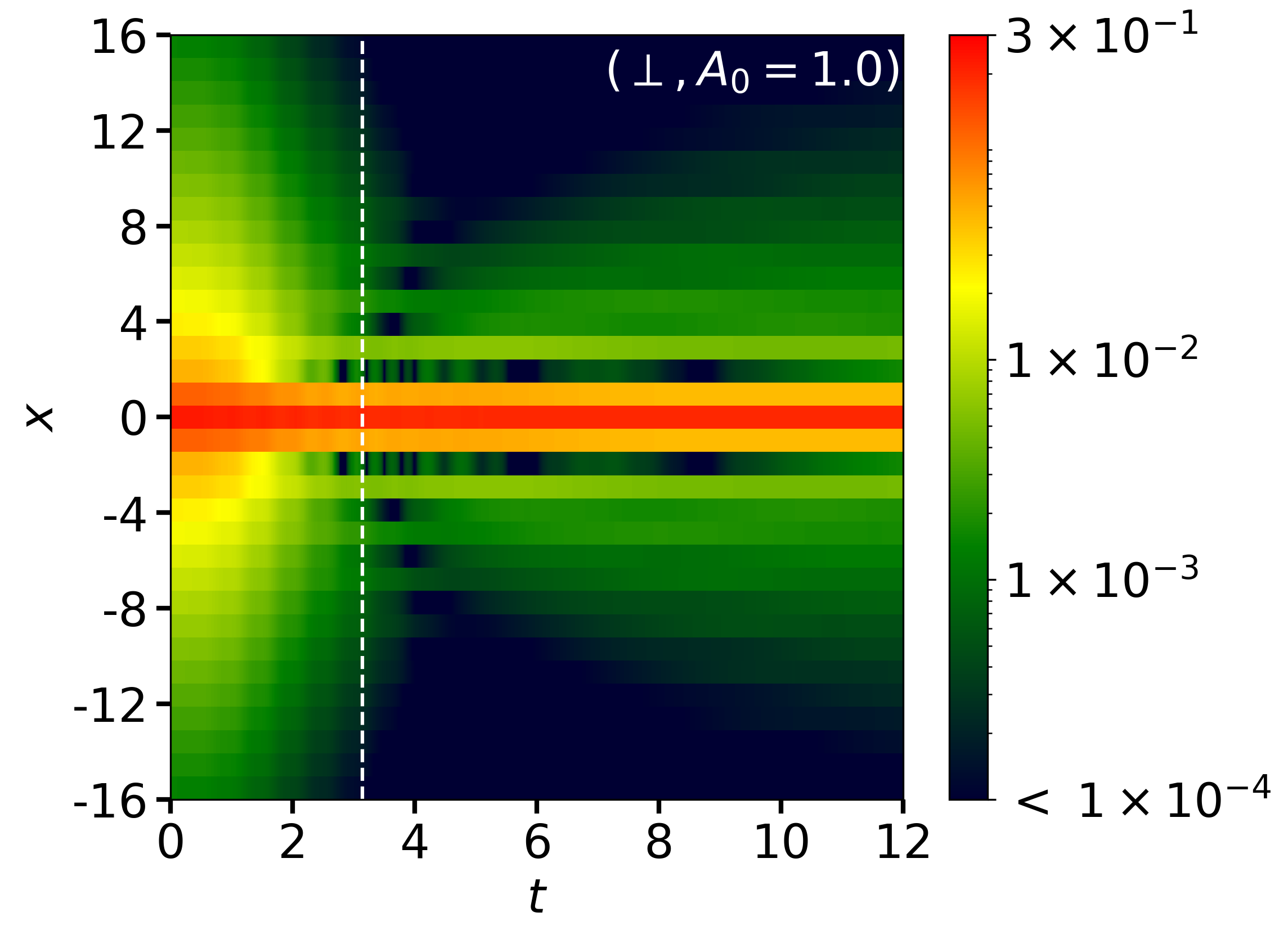}
\caption{Staggered spin-spin correlations $(-1)^x \langle S^z_{i,l} S^z_{i+x,l}\rangle$ along a leg { of an infinite ladder} as a function of separation $x$ and time $t$.
    The white dashed line indicates the time at which the pump is turned off.
    Upper panel: leg{-}direction pump with $A_0=1$. Middle panel: leg{-}direction pump with $A_0=0.45$. Lower panel: rung{-}direction pump with $A_0=1$.}
    \label{fig:spin_spatial_plots_same_leg}
\end{figure}

Spin-spin correlations are strongly suppressed by the pump, with the extent of the suppression and the details of the relaxation afterwards depending on both the energy imparted and the pump direction.
In Fig.~\ref{fig:spin_spatial_plots_same_leg} we plot the staggered spin-spin correlations along one leg $(-1)^x \langle S^z_{i,l} S^z_{i+x,l}\rangle$ { of an infinite ladder}, as a function of separation $x$ and time $t$, for three different pump protocols.

The ground state of the $U=8$ half filled Hubbard ladder is a Mott insulator with strong antiferromagnetic correlations and a spin-spin correlation length $\xi_s \approx 4.2$ (in agreement with Ref.~\cite{noack1996ground}). 
This is reflected at $t=0$ in each panel of Fig.~\ref{fig:spin_spatial_plots_same_leg}, where the spin-spin correlations extend to large distances {(as depicted using a logarithmic colour scheme)}.

Note that the results for $x=0$ correspond to those in Fig.~\ref{fig:magnetization_iTEBD}, but the suppression of onsite magnetization is less visible on a log scale.
All three panels of Fig.~\ref{fig:spin_spatial_plots_same_leg} show suppression of spatial correlations during the pump.
This is effect is fastest for the leg{-}direction, $A_0=1$ pump (Fig.~\ref{fig:spin_spatial_plots_same_leg} upper panel), where correlations beyond nearest neighbour are suppressed by an order of magnitude by the midpoint of the pump.
The plots for the less energetic, $A_0=0.45$ leg{-}direction and $A_0=1$ rung{-}direction, pumps in the middle and lower panels of Fig.~\ref{fig:spin_spatial_plots_same_leg} show suppression of longer-range spin correlations that continue past the point of peak pump amplitude.
All three panels display a limited recovery of spin correlations sometime after the pump has peaked, with correlations growing outwards from $x=0$.
The recovered spin-spin correlations are significantly weaker than in equilibrium, an effect that is seen clearly at next-nearest neighbour and beyond, $\vert x\vert \ge 2$.
%Where the inward-moving region of suppression meets the outward-moving recovery region there is a local minimum in the staggered spin correlations at even $x$.
%This is particularly clear for the rung pump shown in the lower panel of Fig.~\ref{fig:spin_spatial_plots_same_leg}, which shows a local minimum (for even $x$) moving outwards from $t \approx \pi$ with a constant velocity $v_m\approx 4 t_h$. 
In addition, for the $A_0=0.45$ leg{-}direction and $A_0=1$ rung{-}direction pumps there is a local minimum in the spin correlations at $x=2$ in the post pump region.
%%%
\begin{figure}
\centering
\includegraphics[width=\linewidth]{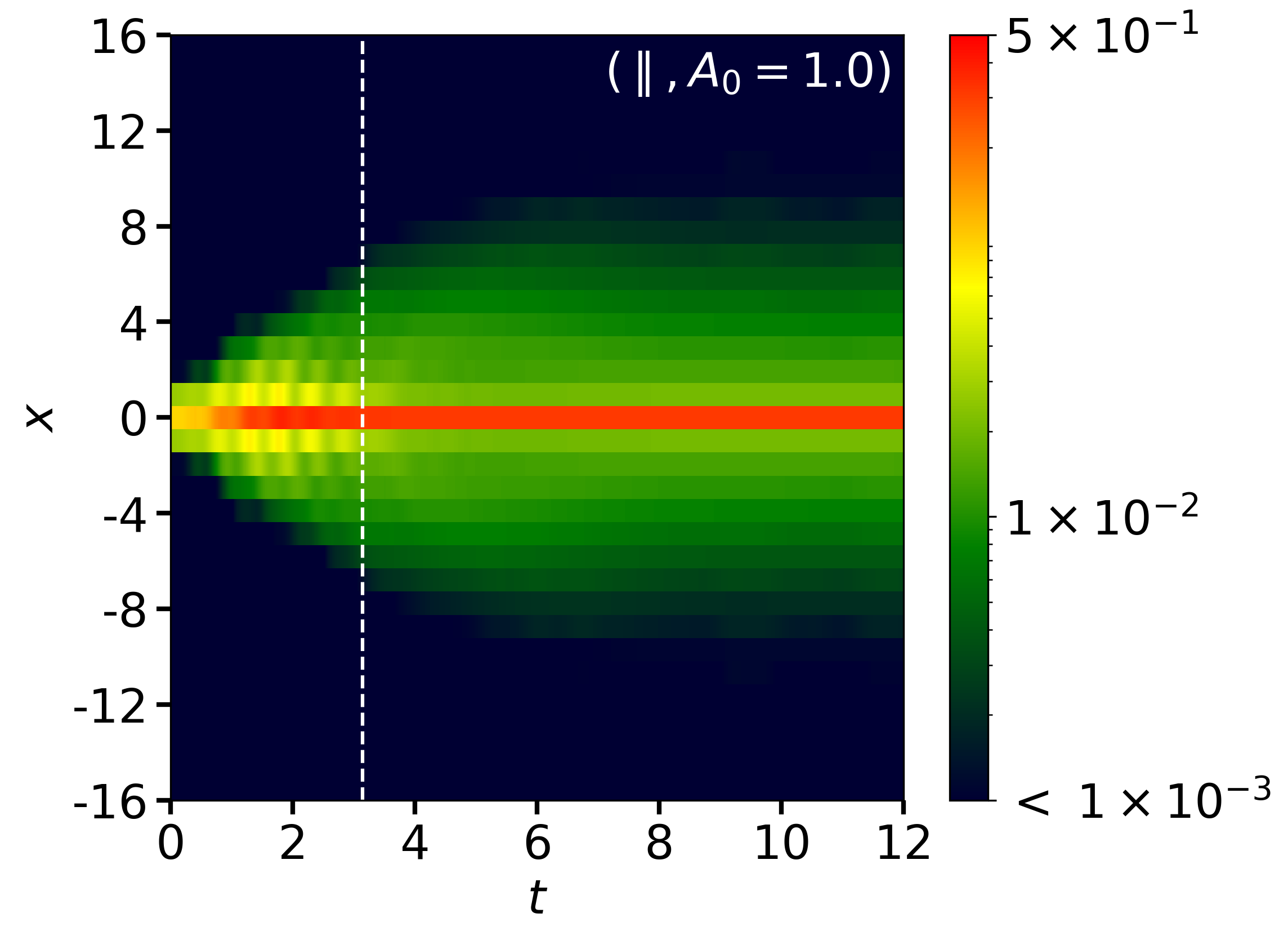} 
\includegraphics[width=\linewidth]{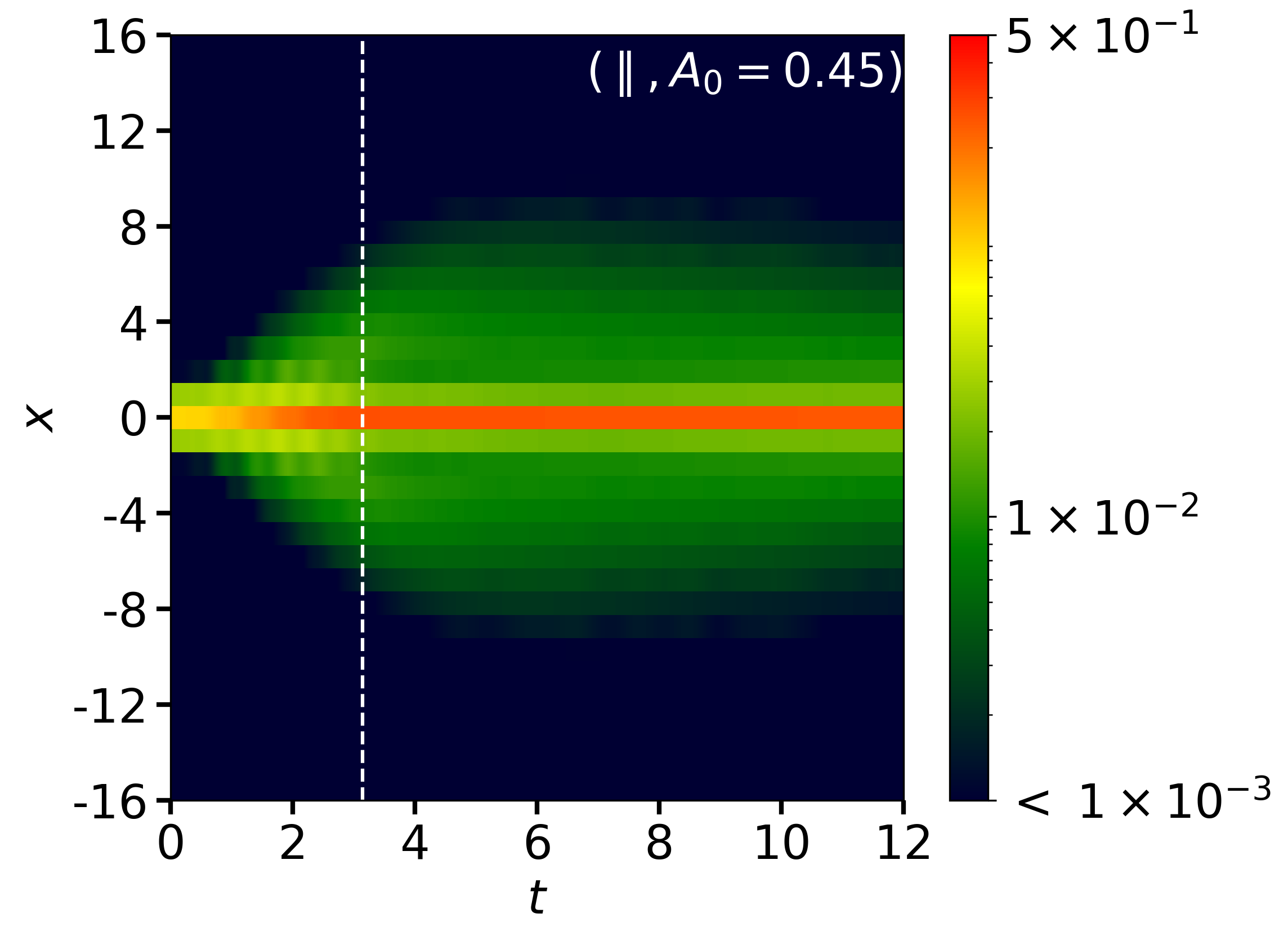} 
\includegraphics[width=\linewidth]{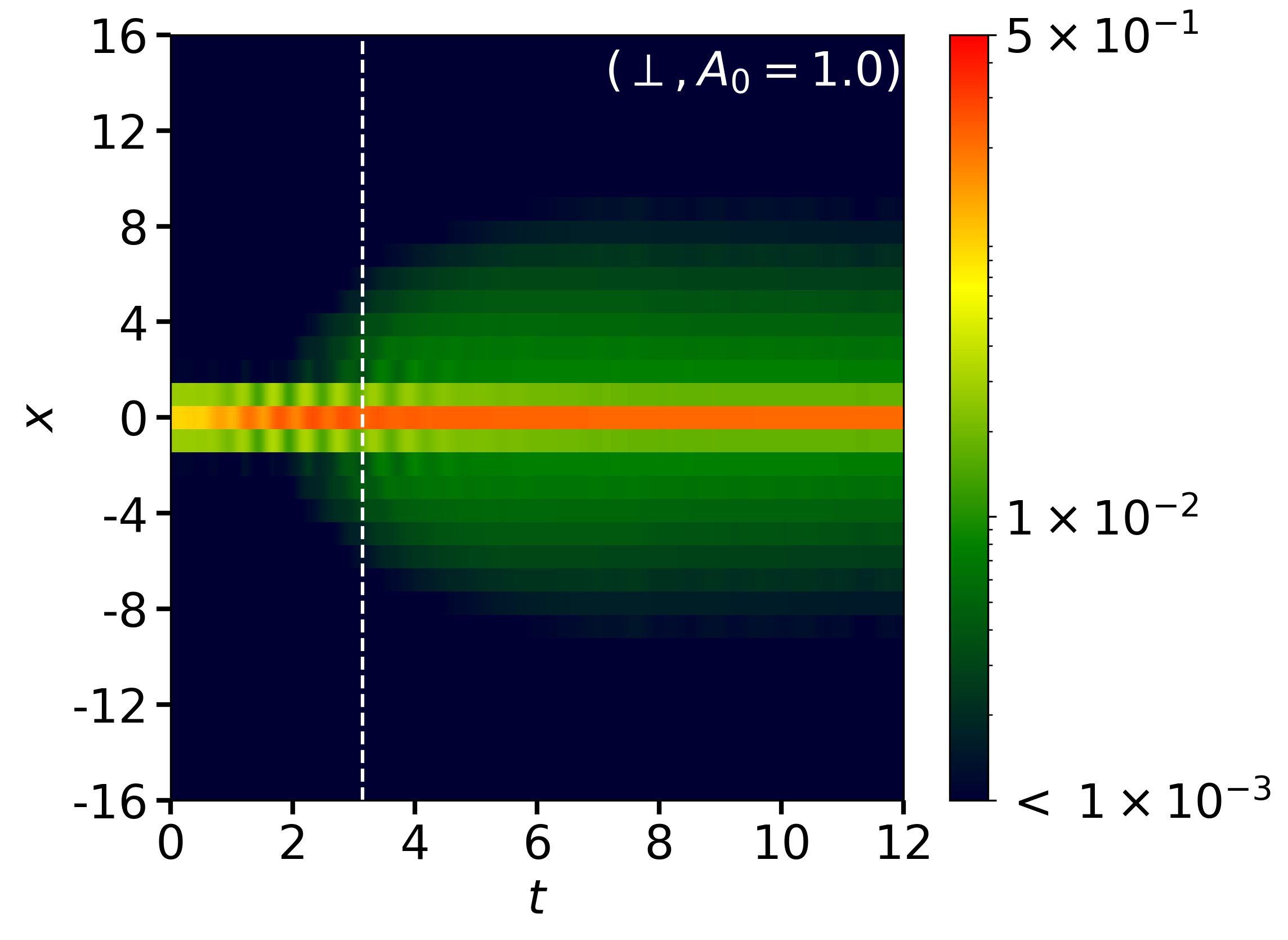} 
\caption{Connected charge (density-density) correlations $\vert\langle n_{i,l} n_{i+x,l}\rangle_c\vert$ along a leg { of an infinite ladder} as a function of separation $x$ and time $t$.
The white dashed line indicates the time at which the pump is turned off.  Upper panel: leg{-}direction pump with $A_0=1$. Middle panel: leg{-}direction pump with $A_0=0.45$. Lower panel: rung{-}direction pump with $A_0=1$.}
\label{fig:density_spatial_plots_same_leg}
\end{figure}

We now consider the charge correlations.
These are dominated by an infinite range component due to charge conservation and translational invariance which entail the constraint $\langle n_{i,l}\rangle=1$ for the leg pump, and the similar condition $\langle n_{i,1}+n_{i,2}\rangle=2$ for the rung pump.
In order to remove this infinite range component, we plot the absolute value of the connected charge correlation function $\vert \langle n_{i,l} n_{i+x,l}\rangle_c\vert =\vert \langle n_{i,l}n_{i+x,l}\rangle-\langle n_{i,l}\rangle\langle n_{i+x,l}\rangle \vert$ in Fig.~\ref{fig:density_spatial_plots_same_leg}.
%The infinite range component is a result of charge conservation and translational invariance, leading to the constraint $\langle n_{i,l}\rangle=1$ for the leg pump, and a similar condition $\langle n_{i,1}+n_{i,2}\rangle=2$ for the rung pump.
Note that no such distinction is needed for the spin correlation function as $\langle S^z_{i,l}\rangle=0$, independent of the pump protocol.
Taking the absolute value accounts for the fact that the connected charge correlations for $x=0$ are positive, but those for $\vert x\vert \ge 1$ are negative (this is also a result of charge conservation, because a fluctuation that locally enhances the charge density must be accompanied by a region of charge depletion).

The charge correlation length in the ground state of the half filled Hubbard ladder with $U=8$ is much shorter than the spin correlation length, and this is evident in Fig.~\ref{fig:density_spatial_plots_same_leg} at $t=0$.
In contrast to the spin correlations, the spatial charge correlations only grow with time during the pump.
This growth is strongest for the $A_0=1$ leg{-}direction pump (top panel), and weakest for the pump along the rung{-}direction.
The growth proceeds without any apparent sharp features associated with the time at which the pump amplitude peaks or the time at which the pump ends.
Unlike the spin correlations, the charge correlations within the `light cone' appear to decrease monotonically for small $x$, and there is no distinct minimum at $x=2$ in the charge correlations for the less energetic pump protocols.
Hence, there are distinct qualitative differences between the spatial spin and charge correlations, despite the simple relation between the local magnetisation and doublon number.
\begin{figure}
\centering
\includegraphics[width=\linewidth]{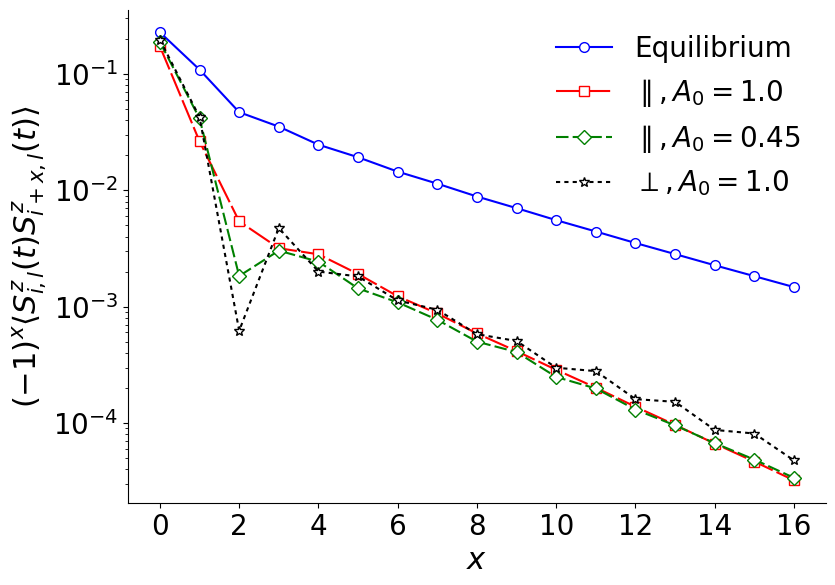} 
\includegraphics[width=\linewidth]{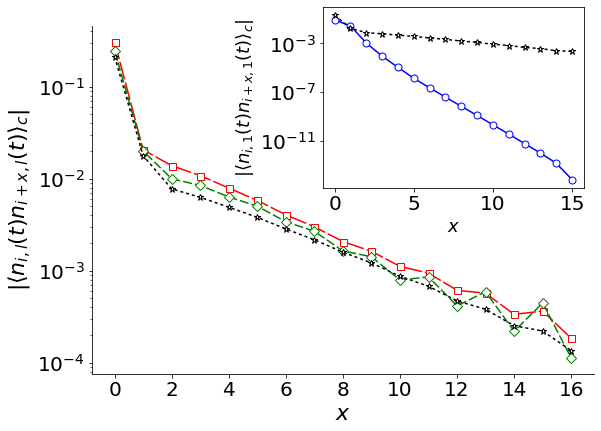} 
\caption{Upper panel: staggered spin correlations in the unperturbed, equilibrium system and at $t=10$ for the three different pump protocols discussed in the text.
Lower panel: as above, but the connected charge correlations.}
\label{fig:cuts_spin_charge_equal_time}
\end{figure}

Both spin and charge correlations decay exponentially at larger separations, and the correlation lengths are affected by photodoping, with the charge correlation length growing significantly relative to its equilibrium value.
Fig.~\ref{fig:cuts_spin_charge_equal_time} shows the spin and charge correlations in equilibrium (i.e. with $A_0=0$, $t<0$) and at $t=10$ after pumping.
The correlation lengths obtained by fitting an exponential
form to the data between $x=6$ and $x=12$ are given in Table~\ref{tab:xi}.
\begin{table}
    \centering
    \caption{Correlation lengths for spin, $\xi_s$, and charge, $\xi_c$, in equilibrium and at $t=10$ after pumping.
    We fit the correlation length between $x=6$ and $x=12$ using the data shown in Fig.~\ref{fig:cuts_spin_charge_equal_time}. 
    We include 2 significant figures, fit uncertainties are $<10^{-2}$.}
    \label{tab:xi}
    \begin{tabular}{|l|c|c|}
    \hline\hline
         & $\xi_s$ & $\xi_c$ \\
         \hline
        Equilibrium & 4.2 & 0.57 \\
        $\parallel, A_0=1$ &2.8 &3.2\\
        $\parallel, A_0=0.45$ &2.8 &3.1\\
        $\perp, A_0=1$ &3.3 &3.5 \\
        \hline
        \hline
    \end{tabular}
    \label{tab:my_label}
\end{table}
The spin correlation length $\xi_s$ is reduced by approximately one lattice unit after the pump.
However, the most notable difference between the equilibrium and nonequilibrium spin correlations is the very fast decay of the latter at short range, before the exponential decay characterised by $\xi_s$ takes over.
This behaviour is evident for $x \le 3$ in the upper panel of Fig.~\ref{fig:cuts_spin_charge_equal_time} where, although the spin correlations at $x=0$ (i.e. the squared magnetizations) are of similar magnitude for the equilibrium and nonequilibrium data, by $x=3$ there is an order of magnitude difference between the equilibrium and nonequilibrium cases{, showing strong suppression of magnetic correlations by the pump}. 
This effect is even more pronounced for the $A_0=0.45$ leg{-}direction pump and the rung{-}direction pump because of the local minimum at $x=2$.
Thus, the value of $\xi_s$ alone does not capture the extreme weakness of correlations beyond nearest neighbour in the nonequilibrium state.

The inset of the lower panel of Fig.~\ref{fig:cuts_spin_charge_equal_time} and the data in Table~\ref{tab:xi} show that connected charge correlations after pumping are strongly enhanced relative to equilibrium.
Although the post pump spin and charge correlation lengths are similar, their short range behaviours are different.
For charge correlations there is no additional strong suppression on going from $x=1$ to $x=2$, and there is no local minimum at $x=2$.
The increase in $\xi_c$ can be interpreted as a consequence of the conservation of charge: the pump increases the doublon occupancy of a site, which must also reduce the {surrounding} charge density, and thus leads to an enhanced anticorrelation.

\begin{figure} 
    \centering
    \includegraphics[scale=0.5]{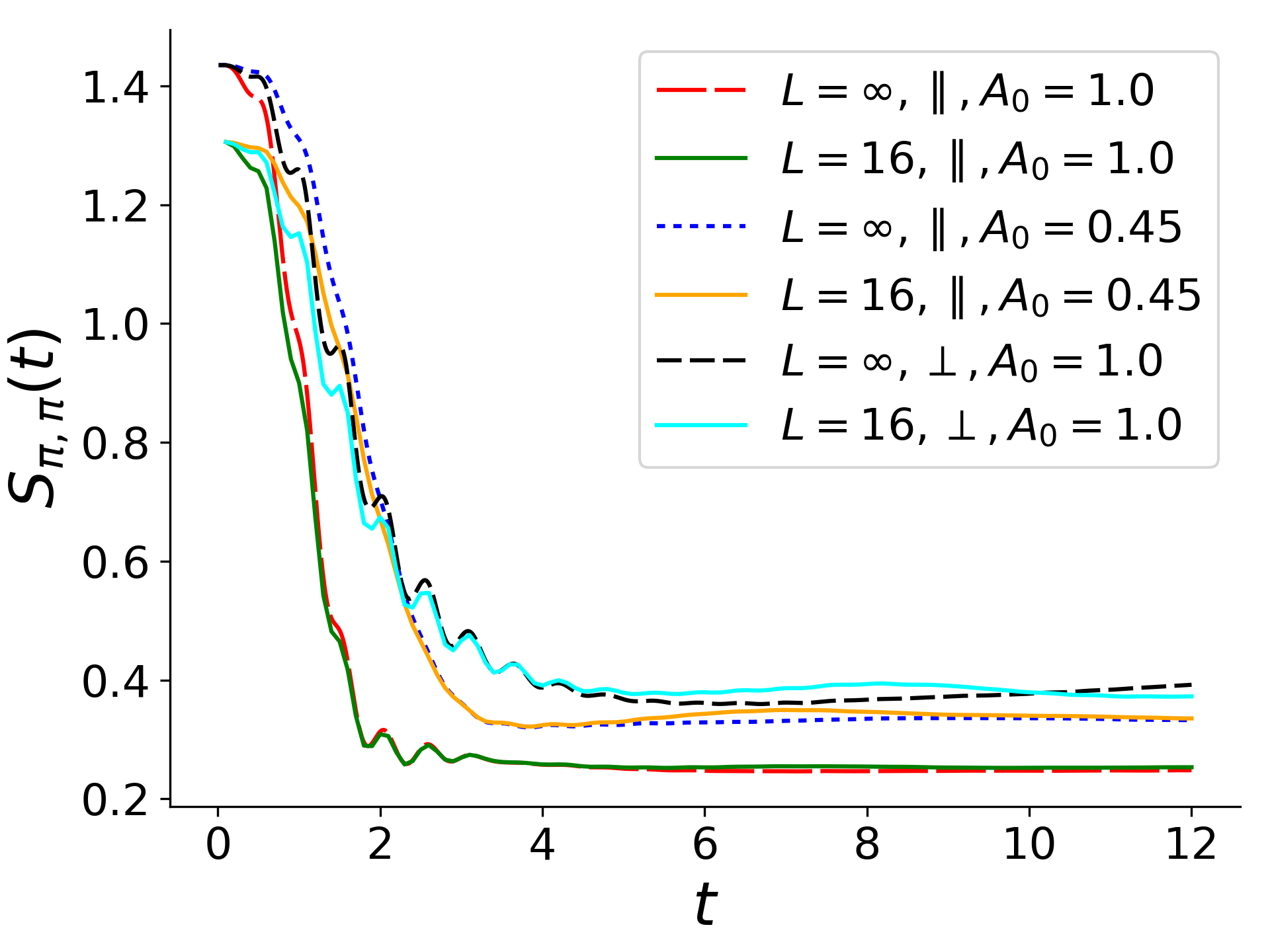}
    \caption{Equal time spin-spin correlation function at $\mathbf{k}=(\pi,\pi)$ for three different pump protocols as a function of time.
    Data is shown for both infinite and 16 rung ladders.}
    \label{fig:spin_PIPI_16vs_infty}
\end{figure}

We end this section by examining the equal time spin correlations in momentum space, 
\begin{align}
    S^{zz}(k_x,k_y,t)= \frac{1}{2L}\sum_{i,l,x,y} e^{-i (k_x x + k_y y)} \langle S^z_{i,l}(t) S^z_{i+x,l+y}(t) \rangle.
    \label{eq:szz}
\end{align}
%The sum over $i$ is trivial due to translational invariance, but the sum over $l$ is required for rung{-}direction pumping, because of the broken parity symmetry between the legs.
In Fig.~\ref{fig:spin_PIPI_16vs_infty} we show the correlations at the antiferromagnetic wave vector $(\pi,\pi)$.
We also include data for finite, $L=16$, rung ladders with open boundary conditions, using the two central rungs as a proxy for the two rung unit cell of approximate infinite system.
The leg{-}direction $A_0=1$ pump shows the fastest suppression of the antiferromagnetic peak, with $S^{zz}(\pi,\pi,t)$ essentially reaching its late time value by $t\approx 2$, during the pump.
For the lower fluence, $A_0=0.45$, leg{-}direction pump the suppression of $S^{zz}(\pi,\pi,t)$ is slower, reaching its late time value close to $t=\pi$ when the pump ends.
The rung{-}direction pump leads to the slowest suppression of the three protocols, with $S^{zz}(\pi,\pi,t)$ decreasing until $t\approx 4$ after the pump has ended.
At $t=0$ the $L=16$ rung data differ from the $L=\infty$ value of $S^{zz}(\pi,\pi,t)$ due to finite size effects.
As the spin correlation length decreases, the difference between a finite and infinite system becomes less distinct, and for all three pump protocols there is good agreement between the $L=16$ and $L=\infty$ data by time $t \approx 1$.
This agreement continues to improve until $t\approx 5$, at which point small differences between the $L=16$ and $L=\infty$ curves become apparent again.
This is consistent with the partial recovery of longer range staggered spin correlations seen in Fig~\ref{fig:spin_spatial_plots_same_leg} which increases the influence of the boundary in the finite system.
We note that there is still very good agreement at $t=10$ between the $L=16$ and $L=\infty$ data for the three protocols.
As $\xi_c \approx \xi_s$, these results suggest that an $L=16$ system is a good approximation for an infinite ladder during and after the pump.

\section{Nonequilibrium dynamical structure factor}
\label{sec:dsf}
In tr-RIXS experiments, a system is probed using ultrashort X-ray pulses, yielding a time-dependent scattering cross section.
For weak probes the related nonequilibrium dynamical structure factor has the form of a double time integral~\cite{freericks2009theoretical},
 \begin{align}
     I(\mathbf{k},\omega,t) = \iint \text{d}t_1 \text{d}t_2 e^{i\omega(t_1-t_2)}g(t_1,t)g(t_2,t)C(\mathbf{k},t_1,t_2).
     \label{eq:Idoubletime}
 \end{align}
Here the $g(t_i,t)$ are windowing functions associated with the { probe pulse}, which we assume to have a Gaussian form,
\begin{align}
    g(t_i,t) = \frac{1}{\sqrt{2\pi}\sigma_{X}}e^{-(t_i-t)^2/2\sigma_{X}^2},
    \label{eq:probewindow}
\end{align}
where $\sigma_{X}$ is the { probe's width in time.
The finite duration of the probe pulse limits the resolution of the time resolved response.}
$C(\mathbf{k},t_1,t_2)=\langle O_\mathbf{k}(t_1) O^\dagger_{-\mathbf{k}}(t_2) \rangle$ is a two time correlation function in momentum space.
For our purposes $O_\mathbf{k}=S^z_\mathbf{k}$ or $n_\mathbf{k}$, respectively the spin $z$-component or charge density operator in momentum space. 

Evaluating the unequal time correlations that are encoded in $C(\mathbf{k},t_1,t_2)$ necessitates applying a local operator at one time, which breaks the translational invariance of an infinite matrix product state (iMPS), and then evolving the system and applying another local operator at a later time.
Although techniques have been developed to embed a finite window with broken translational invariance within an iMPS~\cite{phien2012infinite,phien2013dynamical,milsted2013variational,zauner2015time,ejima2020photoinduced}, we note that the finite size effects for the $L=16$ rung system after the pump (e.g. as seen in Fig.~\ref{fig:spin_PIPI_16vs_infty}) are small at the latest times we simulate.
Therefore, we use an $L=16$ rung system to compute the nonequilibrium dynamical structure factor.
Furthermore, to improve efficiency and simulation time we approximate the double time integral by a single time integral:
\begin{align}
    I(\vec{k},\omega,t) \approx \int \text{d} \tau &\frac{1}{4\sqrt{\pi}\sigma_X}\exp[i\omega \tau -\tau^2/(4\sigma_X^2)]\nonumber\\
    &\times \left[C(\mathbf{k},t+\tau,t)+C(\mathbf{k},t,t-\tau) \right].
    \label{eq:Isingletime}
\end{align}
This approximation is reasonable if $C(\mathbf{k},t+\tau,t)$ varies slowly with $t$ relative to the scale $\sigma_X$.
In our results we set $\sigma_X=1$.
We provide more details in Appendix~\ref{app:integral} including a comparison that shows there is good agreement between the approximate form Eq.~\ref{eq:Isingletime} and the full form  Eq.~\ref{eq:Idoubletime}, even during the pump when the conditions for the approximation are not strictly met.
%%%
%%%
%%%
\subsection{Nonequilibrium sum rules}
In this subsection we show that the combined total spectral weight of the nonequilibrium dynamical spin and charge responses is time-independent.
{ Although we concentrate on the case of the two leg ladder, the derivation generalizes in a simple manner and the result also applies to single band Hubbard models on other lattices (and in other dimensions).}
Relations of this type, between { sums of} dynamical quantities and a constant, are called sum rules.

The total spectral weight in a channel is given by integrating Eq.~\ref{eq:Idoubletime} for $I(\mathbf{k},\omega,t)$ over all $\mathbf{k}$ and $\omega$,
\begin{align}
    W_O(t)&=\frac{1}{N}\sum_\mathbf{k} \int \frac{\text{d}\omega}{2\pi}  I(\mathbf{k},\omega,t)\nonumber\\
    &=\int \text{d} t_1 \,g^2(t,t_1) \frac{1}{N}\sum_{i,l} \langle O_{i,l}(t_1)O^\dagger_{i,l}(t_1)\rangle.
    \label{eq:IW}
\end{align}
In equilibrium, as a consequence of time translation invariance, the expectation is independent of $t_1$.
Then for reasonable probe windows (those that are a function of $t-t_1$ only, such as Eq.~\ref{eq:probewindow}) the total weight $W_O$ will be independent of time $t$,
\begin{align}
    W_O&=G \frac{1}{N} \sum_{i,l} \langle O_{i,l}O^\dagger_{i,l}\rangle.
    \label{eq:IWeq}
\end{align}
where $G=\int \text{d} t_1 \,g^2(t,t_1)$ is a constant that depends on the probe shape.

Returning to the general case, Eq.~\ref{eq:IW}, and setting $O=S^z$, gives the total weight in the spin channel,
\begin{align}
    W_{S^z}(t)&=\int \text{d} t_1 \,g^2(t,t_1) \frac{1}{N}\sum_{i,l} \langle [S^z_{i,l}(t_1)]^2\rangle.
    \label{eq:IWS1}
\end{align}

In cases where $\langle [S^z_{i,l}(t)]^2\rangle$ is conserved (e.g. as in Heisenberg models) Eq.~\ref{eq:IWS1} reduces to the standard spin sum rule and $W_{S^z}$ is a constant.

In contrast, for the Hubbard model $\langle [S^z_{i,l}(t)]^2\rangle$ is not necessarily conserved, due to the coupling of spin and charge degrees of freedom.
Instead, using the relation between the $S^z$ operator and occupation number gives,
% \begin{align}
%      \langle [S^z_{i,l}(t)]^2 \rangle &=\frac{1}{4}\left(1- \sum_l \left\langle d_{i,l}(t_1) \right\rangle\right),
% \end{align}
\begin{align*}
    W_{S^z}(t)&=\int \text{d} t_1 \,g^2(t,t_1)\frac{1}{4N}\sum_{i,l} \left\langle n_{i,l}(t_1)-2d_{i,l}(t_1) \right\rangle.
\end{align*}
As the total number of electrons, $\sum_{i,l} n_{i,l}=N_e$, is conserved, this becomes
\begin{align}
    W_{S^z}(t)&=\frac{G}{4}\rho - \frac{1}{4}\int \text{d} t_1 \,g^2(t,t_1)\frac{2}{N}\sum_{i,l} \left\langle d_{i,l}(t_1) \right\rangle,
    \label{eq:IWS2}
\end{align}
where the density $\rho=N_e/N$ is unity at half-filling.
Pumping leads to a nontrivial time-dependence for $\left\langle d_{i,l}(t) \right\rangle$ as shown in Fig.~\ref{fig:magnetization_iTEBD}.

Similarly, the total charge spectral weight can be written as
\begin{align}
    W_n(t)&=\int \text{d} t_1 \,g^2(t,t_1) \frac{1}{N}\sum_{i,l} \langle [n_{i,l}(t_1)]^2\rangle\nonumber\\
    &=G \rho+\int \text{d} t_1 \,g^2(t,t_1) \frac{2}{N}\sum_{i,l} \left\langle d_{i,l}(t_1) \right\rangle.\label{eq:IWn}
\end{align}
Combining Eq.~\ref{eq:IWS2} and Eq.~\ref{eq:IWn} yields a sum rule which remains valid for a nonequilibrium system,
\begin{align}
    4W_{S^z}(t)+W_n(t)=2G\rho.
    \label{eq:sumrule}
\end{align}
Hence, there is a direct relation between changes in the total spectral weights for spin and charge:
a change $\Delta W_{S^z}$ in the total spin response entails a change $\Delta W_{n}=-4 \Delta W_{S^z}$ in the total charge response.

Processes that alter the distribution of spectral weight can be separated into two types, those that only redistribute weight within each channel so that $\Delta W_{S^z}=\Delta W_{n}=0$, and those that transfer dynamical response from one degree of freedom to another $\Delta W_{n}=-4 \Delta W_{S^z} \ne 0$.
An example of the former would be raising the temperature by an amount $ \Delta_{S} < k_\text{B} T \ll U$, where $\Delta_{S}$ is the spin gap.
This would lead to a finite population of spin excitations, altering the distribution of spectral weight in the spin channel, but leaving the response in the charge channel essentially unaffected due to the large Mott gap.
In the following we show that photodoping of the Hubbard ladder is an example of the latter type of process, with significant transfer of spectral weight from the spin to the charge sector.

\subsection{Spin dynamical structure factor}
\label{sec:sdsf}
The spin dynamical structure factor in the ladder in equilibrium is dominated by the strong AF correlations, leading to a large response around the $(\pi,\pi)$ point.
\begin{figure*} 
  \centering
  \begin{minipage}[b]{0.32\textwidth}
    \centering
    \includegraphics[width=\linewidth]{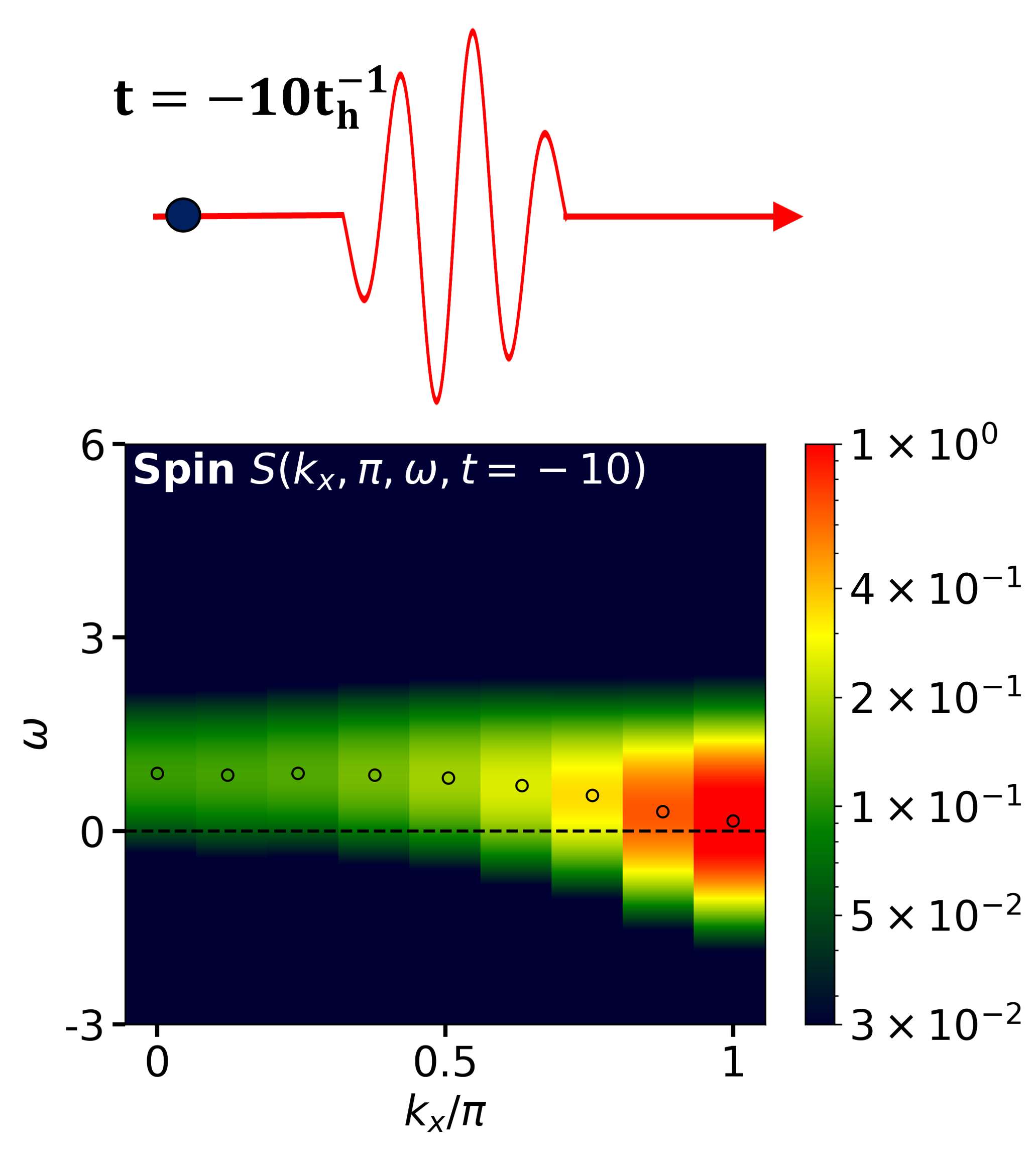}
  \end{minipage}
  \hfill
  \begin{minipage}[b]{0.32\textwidth}
    \centering
    \includegraphics[width=\linewidth]{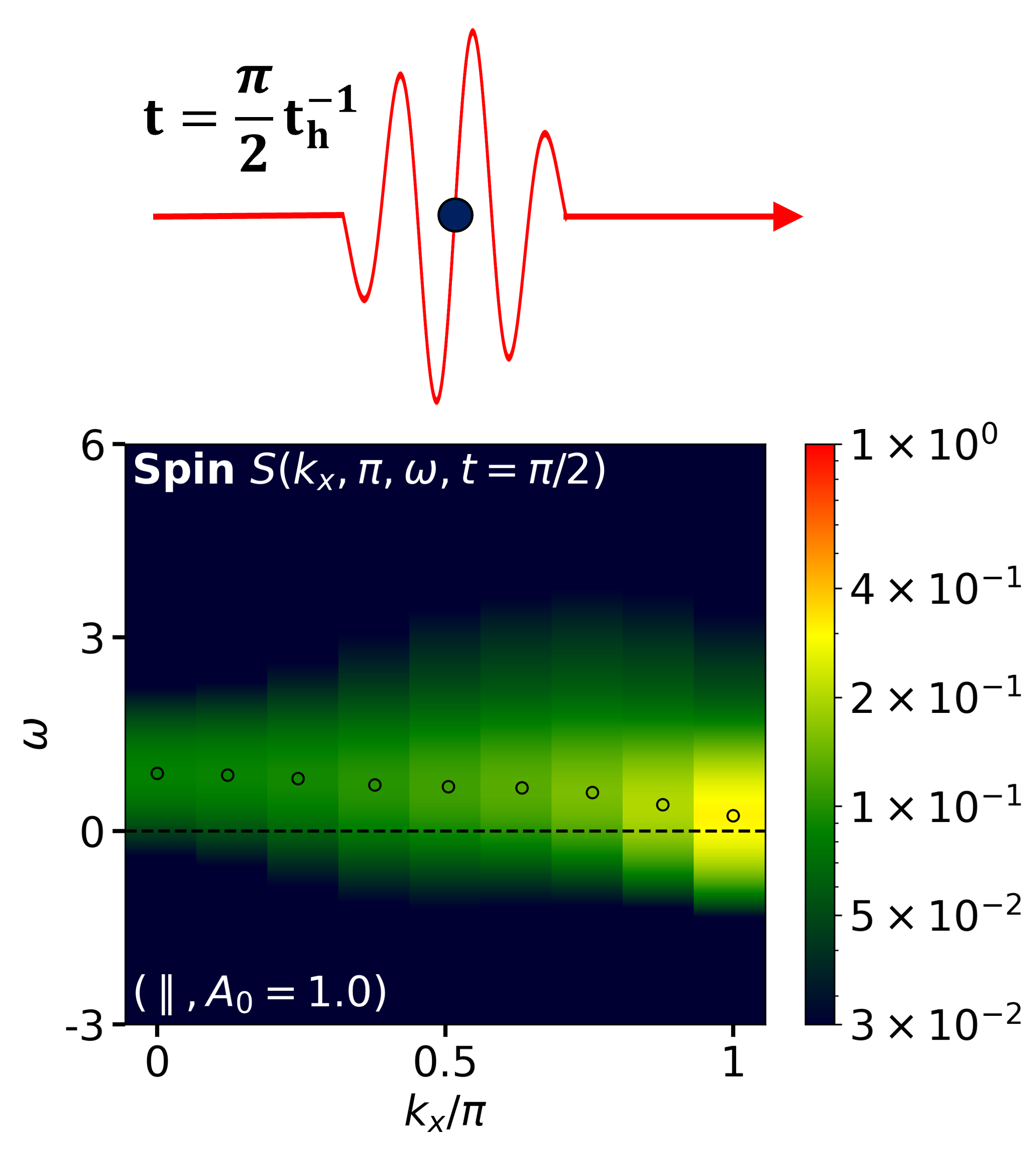}
  \end{minipage}
  \hfill
  \begin{minipage}[b]{0.32\textwidth}
    \centering
    \includegraphics[width=\linewidth]{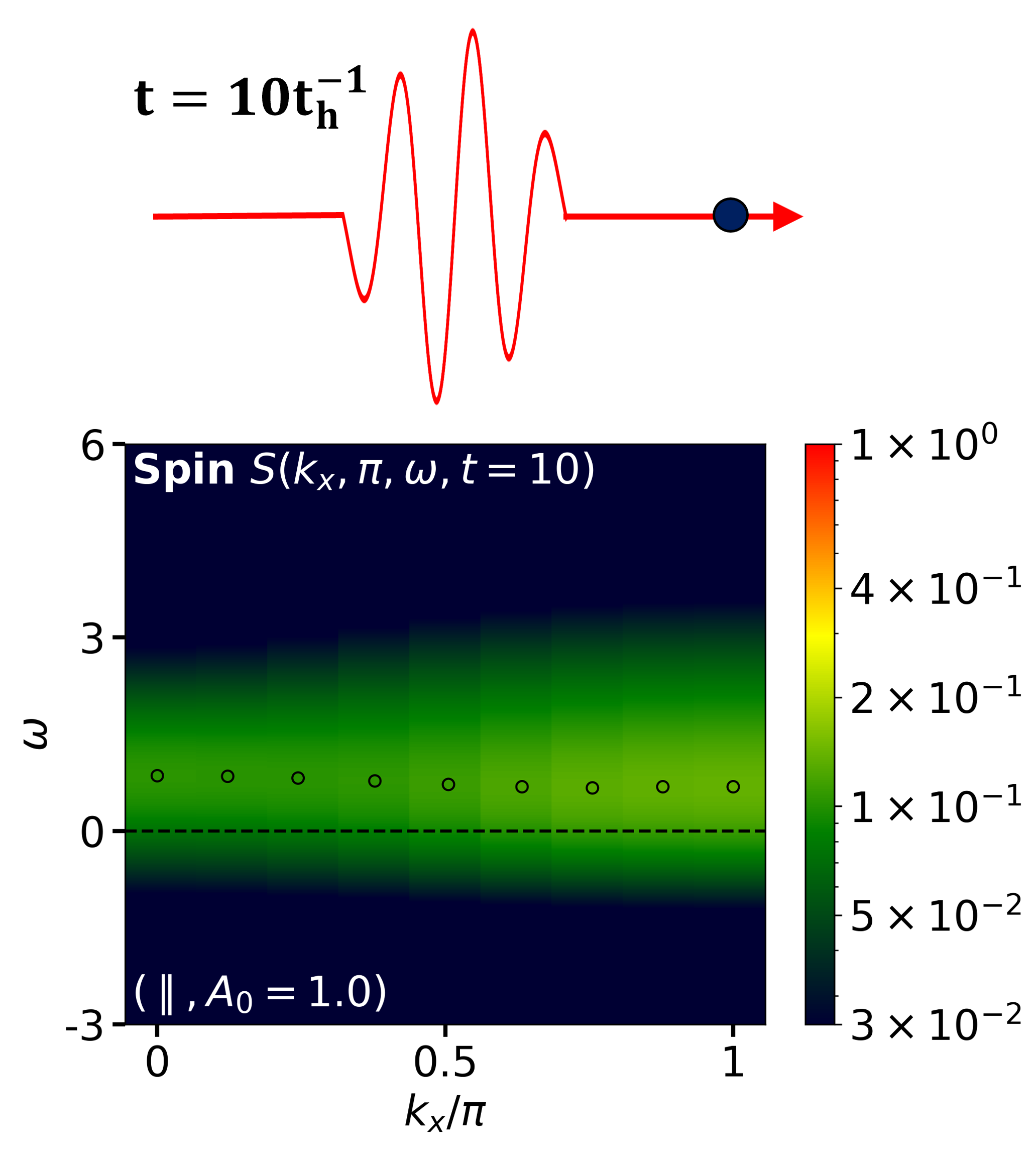}
  \end{minipage}
  \caption{Nonequilibrium dynamical spin structure factor at $k_y = \pi$ for a pump with $A_0 = 1$ in the leg ($\parallel$) direction.
  The leftmost column shows the response in equilibrium before the pump.
  The middle and rightmost columns show results during and after the pump respectively.
  The black open circles mark maxima in the response at a given wavevector $k_x$.
  The corresponding plots for the leg{-}direction pump with $A_0 = 0.45$ and rung{-}direction pump with $A_0=1$ are qualitatively very similar.}
  \label{fig:struc_fac_ky_PI_legsquench}
\end{figure*}
We compute the dynamical spin response $S(k_x,k_y,\omega,t)$ using Eq.~\ref{eq:Isingletime} before pumping, mid way through the pump at $t=\pi/2$ and after the pump at $t=10$.

In equilibrium, the spin excitations of the Hubbard ladder at half filling and large $U$ are well described by those of a Heisenberg ladder with antiferromagnetic exchange interactions.
This two-leg spin ladder has a ground state characterised by singlet formation on rungs~\cite{giamarchi2003quantum}.
Spin excitations are formed by breaking rung singlets, leading to massive (gapped) dispersing triplets.
There is a spin gap and no true long range order, and the system is sometimes described as a spin liquid.

{ Parity selection rules mean that the spin response at different $k_y$ is related to the number of rung triplet excitations that are formed~\cite{bouillot2011statics}.
A rung singlet has negative parity (the wavefunction changes sign under interchange of the legs) whilst a rung triplet has positive parity (the wavefunction sign remains the same under interchange of the legs).
If the ladder is in a state of well defined total parity then breaking an odd number of singlets to form triplets will change the total parity of the state from positive to negative or vice-versa. 
In contrast, breaking an even number of singlets to form triplets leaves the total parity unchanged.
The spin operator $S^z_{k_x,k_y}$ with $k_y=0$ preserves parity but for $k_y=\pi$ it alters parity.
Accordingly, the spin response for $k_y=\pi$ probes the formation of odd numbers of triplets (as the parity is altered) and for $k_y=0$ it probes the formation of even numbers of rung triplets (as the total parity is unchanged).
As the triplets are massive the minimum energy spin response is expected for $k_y=\pi$ which allows for the creation of a single rung triplet.
}
However, because the spin gap {$\Delta_s \approx 0.12 t_\parallel$} is small and the antiferromagnetic correlations are somewhat long ranged, the low energy dynamics are approximately those of a spin-1 magnon, with a sharply defined mode that has an almost linear dispersion relation as it approaches zero energy near $\mathbf{k}=(\pi,\pi)$~\cite{fabrizio1993spin,endres1996dynamical}.

{Fig.~\ref{fig:struc_fac_ky_PI_legsquench} shows the spin response for $k_y=\pi$ as a function of $\omega$ and $k_x$ for a leg{-}direction pump with $A_0=1$ at three different times: $t=-10$ before pumping (equivalent to the equilibrium result), during the pump at $t=\pi/2$, and at $t=10$ after pumping.
We do not show the results for the $A_0=0.45, \parallel$ and $A_0=1, \perp$ pumps as they are qualitatively very similar when plotted in this way.
The equilibrium behaviour (convolved with the resolution due to the probe pulse's temporal width}) is seen in the left-hand panel of Fig.~\ref{fig:struc_fac_ky_PI_legsquench}.
The middle panel of the same figure shows that during the pump the spin response near $\mathbf{k}=(\pi,\pi)$ is reduced in intensity, in agreement with the real time results of the previous section.
There is also a notable broadening of the response.
At later times after pumping (right-hand panel of Fig.~\ref{fig:struc_fac_ky_PI_legsquench}) the antiferromagnetic peak is substantially reduced in intensity, the response is broader for all $k_x$, and the dispersion has flattened.
%The suppression of the antiferromagnetic peak is strongest for the $A_0=1$ leg{-}direction pump, although all three protocols lead to a similar response near $\mathbf{k}=(0,\pi)$.
\begin{figure*}
  \centering
  \begin{minipage}[b]{0.32\textwidth}
    \centering
    \includegraphics[width=\linewidth]{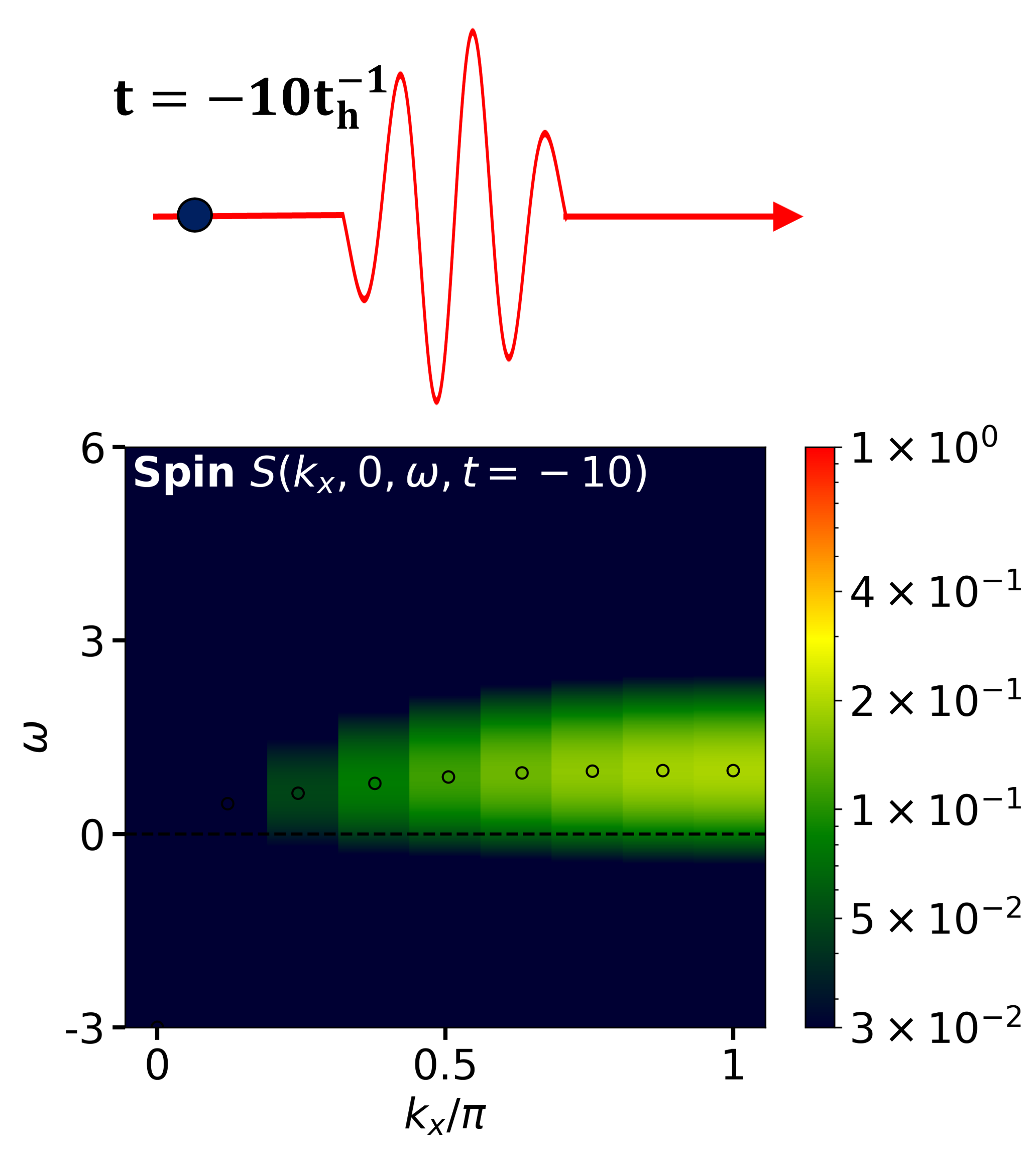}
  \end{minipage}
  \hfill
  \begin{minipage}[b]{0.32\textwidth}
    \centering
    \includegraphics[width=\linewidth]{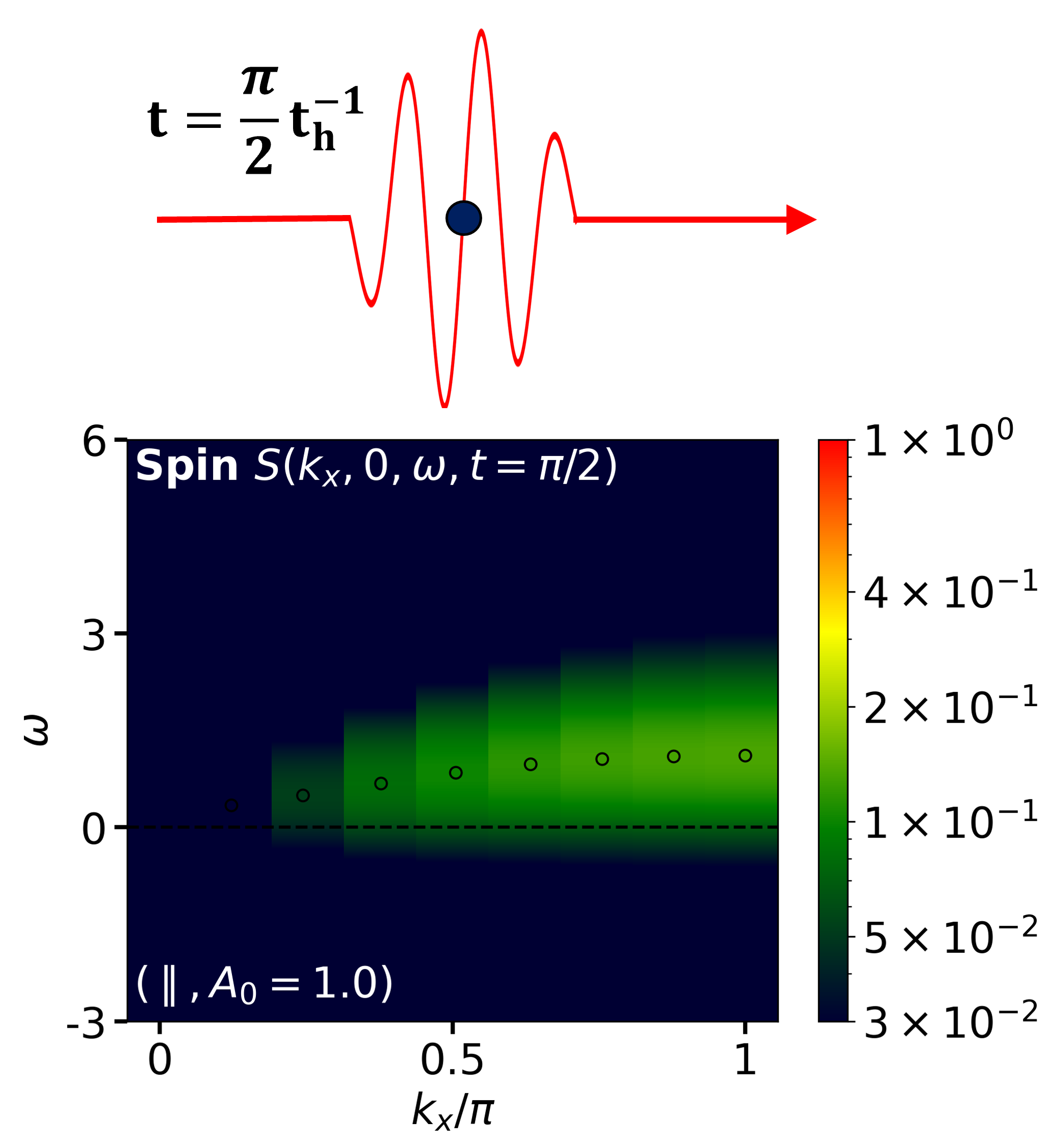}
  \end{minipage}
  \hfill
  \begin{minipage}[b]{0.32\textwidth}
    \centering
    \includegraphics[width=\linewidth]{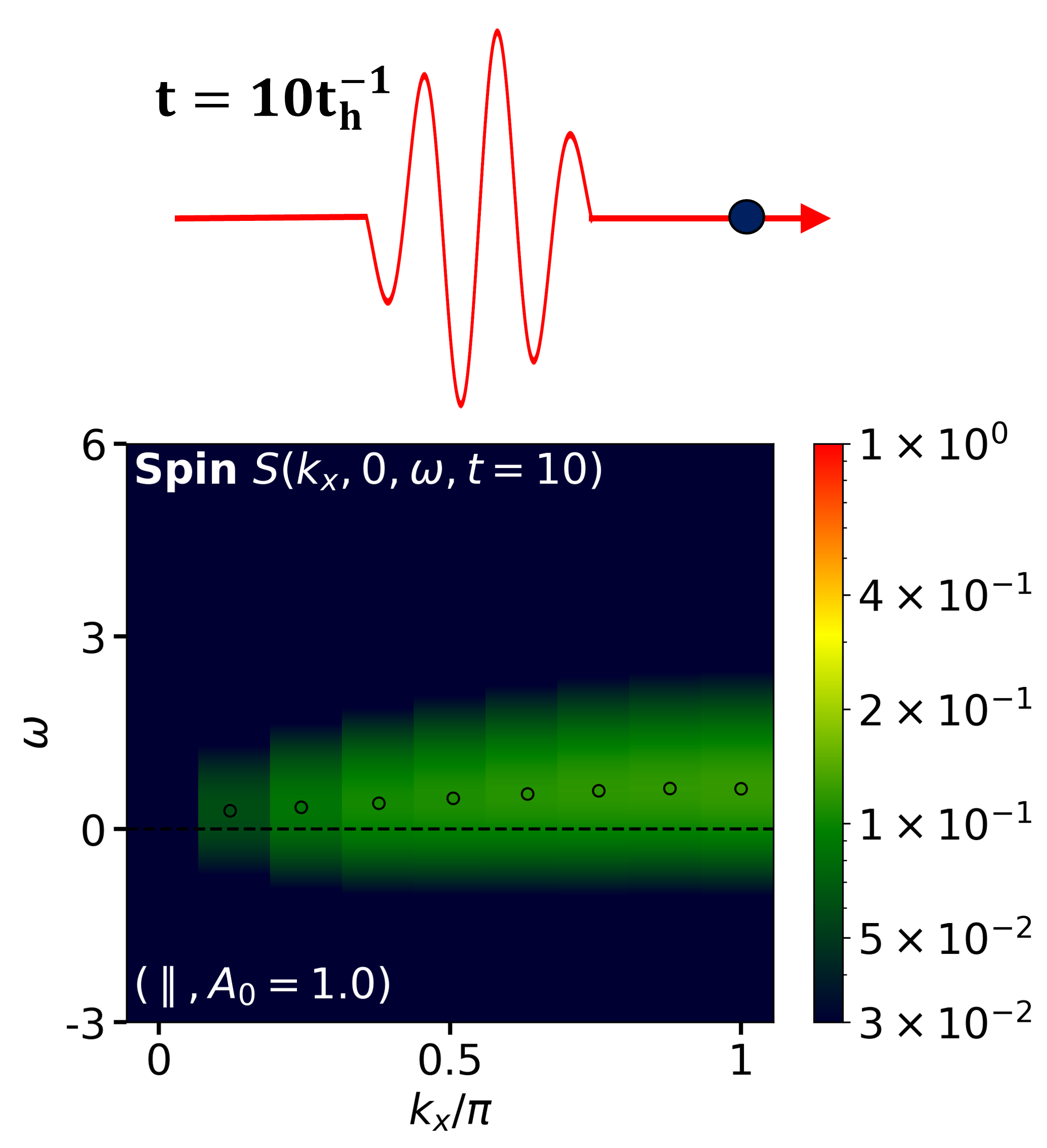}
  \end{minipage}
  \caption{Nonequilibrium dynamical spin structure factor at $k_y = 0$ for the $A_0 = 1$ leg{-}direction pump, before, during, and after the pump.
  The black open circles mark maxima in the response at a given wavevector $k_x$.
  The corresponding plots for the leg{-}direction pump with $A_0 = 0.45$ and rung{-}direction pump with $A_0=1$ are qualitatively very similar.}
  \label{fig:struc_fac_ky_0_legsquench}
\end{figure*}
Fig.~\ref{fig:struc_fac_ky_0_legsquench} shows the equivalent dynamical spin response for the $A_0=1$ leg{-}direction pump along $k_y=0$ ({ the other two pump protocols give qualitatively similar results}).
The response along $k_y=0$ vanishes as $k_x$ approaches $0$ (because $\sum_{i,l} S^z_{i,l}=0$ is a conserved quantity) and is strongest at $k_x=\pi$.
After the pump the response at $(\pi,0)$ is weaker, although this suppression is less significant than at $(\pi,\pi)$.
There is also some \emph{enhancement} of the spectral weight close to $k_x=0$ after pumping, relative to the equilibrium case.

\begin{figure}
    \centering
    \includegraphics[scale=0.4]{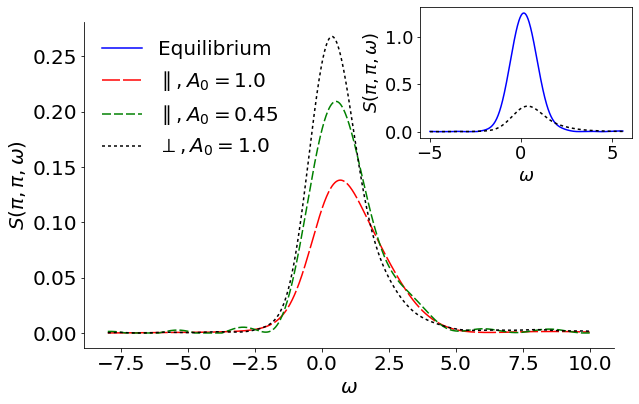}
    \caption{Magnetic correlations with $\mathbf{k}=(\pi,\pi)$, at $t = 10$ after pumping along the legs $(\parallel)$ or rungs $(\perp)$.
    The inset shows the strong antiferromagnetic peak in equilibrium.}   \label{fig:compare_rungs_vs_legs_quench_cutsPIPI}
\end{figure}
\begin{figure}
    \centering
    \includegraphics[scale=0.4]    {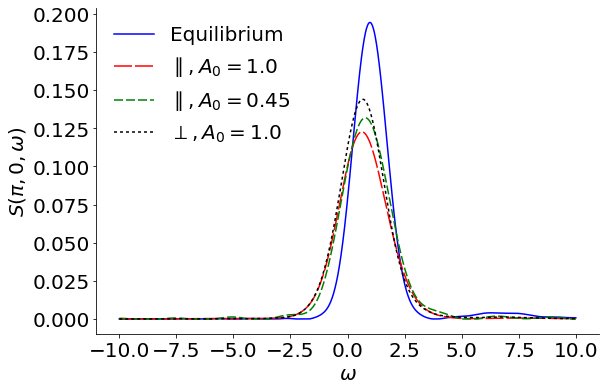}
     \caption{Magnetic correlations with $\mathbf{k}=(\pi,0)$, at $t = 10$ after pumping along the legs $(\parallel)$ or rungs $(\perp)$.}
  \label{fig:compare_rungs_vs_legs_quench_cutsPI0}
  \end{figure}
More precise information on changes in the energy profile of the spin response is provided by taking cuts at fixed wave vectors. 
In Figs.~\ref{fig:compare_rungs_vs_legs_quench_cutsPIPI} and~\ref{fig:compare_rungs_vs_legs_quench_cutsPI0} we compare the equilibrium and $t=10$ post pump responses for $\mathbf{k}=(\pi,\pi)$ and $\mathbf{k}=(\pi,0)$ respectively~\footnote{The very small amount of negative spectral weight visible for two of the curves in Figs.~\ref{fig:compare_rungs_vs_legs_quench_cutsPIPI} and~\ref{fig:compare_rungs_vs_legs_quench_cutsPI0} is a consequence of the finite window used in the frequency Fourier transform.
This negative weight could be removed by using a larger time window at a significant computational cost.}.
The equilibrium response at both wave vectors is essentially a { Gaussian peak, i.e. a single mode that is resolution limited due to the windowing effect of the probe pulse.
The equilibrium response for $k_y=\pi$ can be interpreted in the rung triplet picture as showing interband transitions from the ground state to a band of single triplet excitations (or a band of double triplet excitations for $k_y=0$).

The most notable result of the pump at $\mathbf{k}=(\pi,\pi)$, other than the overall weakening of the response,} is that the broadening of the peak occurs in an asymmetric manner with a tail towards higher energies, in contrast to the symmetric broadening { of a single particle mode} that is expected in a simple scattering lifetime approximation.

Asymmetric broadening of a single particle mode in the dynamical spin response has been described previously in gapped spin systems with triplet excitations and hard-core interactions, at low but finite temperatures~\cite{essler2008finite,james2008finite,goetze2010low,tennant2012anomalous,quintero2012asymmetric}.
At zero temperature the single particle mode in these systems is a result of interband transitions from the ground state { i.e. breaking a singlet and creating a state} with a single propagating triplet.
As temperature increases, triplet states become thermally populated and due to strong interactions they dress the single particle mode in a nontrivial manner, leading to an interband response with an asymmetric lineshape.
{ Broadening of the single particle response occurs due to interband transitions from a thermally populated single particle state to a state with two particles.
The precise form of the joint density of states for occupied single particle modes and unoccupied two particle states determines the direction of the asymmetry.}
Moreover, the existence of thermally populated triplets reduces the joint density of states for interband transitions, so the spectral weight associated with the single mode response decreases.
These effects are accompanied by the emergence of a new spin response at $\omega\sim 0$ due to intraband transitions (scattering of the probe from thermally occupied states) { which can occur for vanishing values of energy and momentum}.
This increased spectral weight at low energies is known as a Villain mode~\cite{villain1975propagative}, and it compensates for the decrease in the single mode weight.
The total integrated spin response then remains a conserved quantity as expected for a pure spin model.

Photodoping of the Hubbard ladder injects energy, breaking rung singlets and producing a finite density of excitations, suggesting an analogy with thermal behaviour of gapped spin systems.
The application of this description to the spin response to the Hubbard ladder is complicated by two factors (although see Ref.~\cite{nocera2018finite} for a treatment of the Hubbard chain in the $U\to\infty$ limit).
First, the spin gap for the ladder is very small, preventing the identification of a separate intraband response for $\omega < \Delta_s$, and confounding the use of the analytical methods in Refs.~\cite{essler2008finite,james2008finite,goetze2010low}.
Second, the presence of charge excitations means that Eq.~\ref{eq:sumrule} applies and the total spin response need not be conserved, with the possibility of weight being transferred into the charge sector instead of into a Villain mode. 
Determining the existence of a separate Villain mode is not possible for our simulations, due to the resolution $\sigma_X$ and the small spin gap.
However, the emergence of a Villain mode would be consistent with the increased low energy response after the pump at $t=10$, visible as increased weight at and below $\omega=0$ for wave vectors away from the AF peak (in particular close to $\mathbf{k}=(0,0)$, see the rightmost column of Fig.~\ref{fig:struc_fac_ky_0_legsquench}).

The total spin spectral weight $W_{S^z}$ can be determined using Eq.~\ref{eq:IWS2} from the doublon number computed with iTEBD, or by integrating the dynamical spin response for the 16 rung system shown in Figs.~\ref{fig:struc_fac_ky_PI_legsquench} and ~\ref{fig:struc_fac_ky_0_legsquench} over all $\omega$ and $\mathbf{k}$.
We find excellent agreement between both these methods, which provides further confirmation of the accuracy of the approximation Eq.~\ref{eq:Isingletime} and the use of the 16 rung system.
For the $A_0=1$ leg{-}direction pump the ratio of the spin weight after the pump to that in equilibrium is $W_{S^z}(t=10)/W_{S^z}(t=-10)=0.76$.
For the $A_0=1$ rung{-}direction pump the corresponding figure is $W_{S^z}(t=10)/W_{S^z}(t=-10)=0.86$.
Therefore, the integrated spin response decreases significantly after the optical pump, indicating a substantial transfer of spectral weight to the charge sector.
%This confirms that charge excitations are a key feature of the nonequilibrium dynamics, and interpretations of photodoping experiments that only consider magnetic behaviour should be treated with caution.

\subsection{Charge dynamical structure factor}
\label{sec:cdsf}
\begin{figure*} % Use figure* for two-column wide figures
  \centering
  \begin{minipage}[b]{0.32\textwidth}
    \centering
    \includegraphics[width=\linewidth]{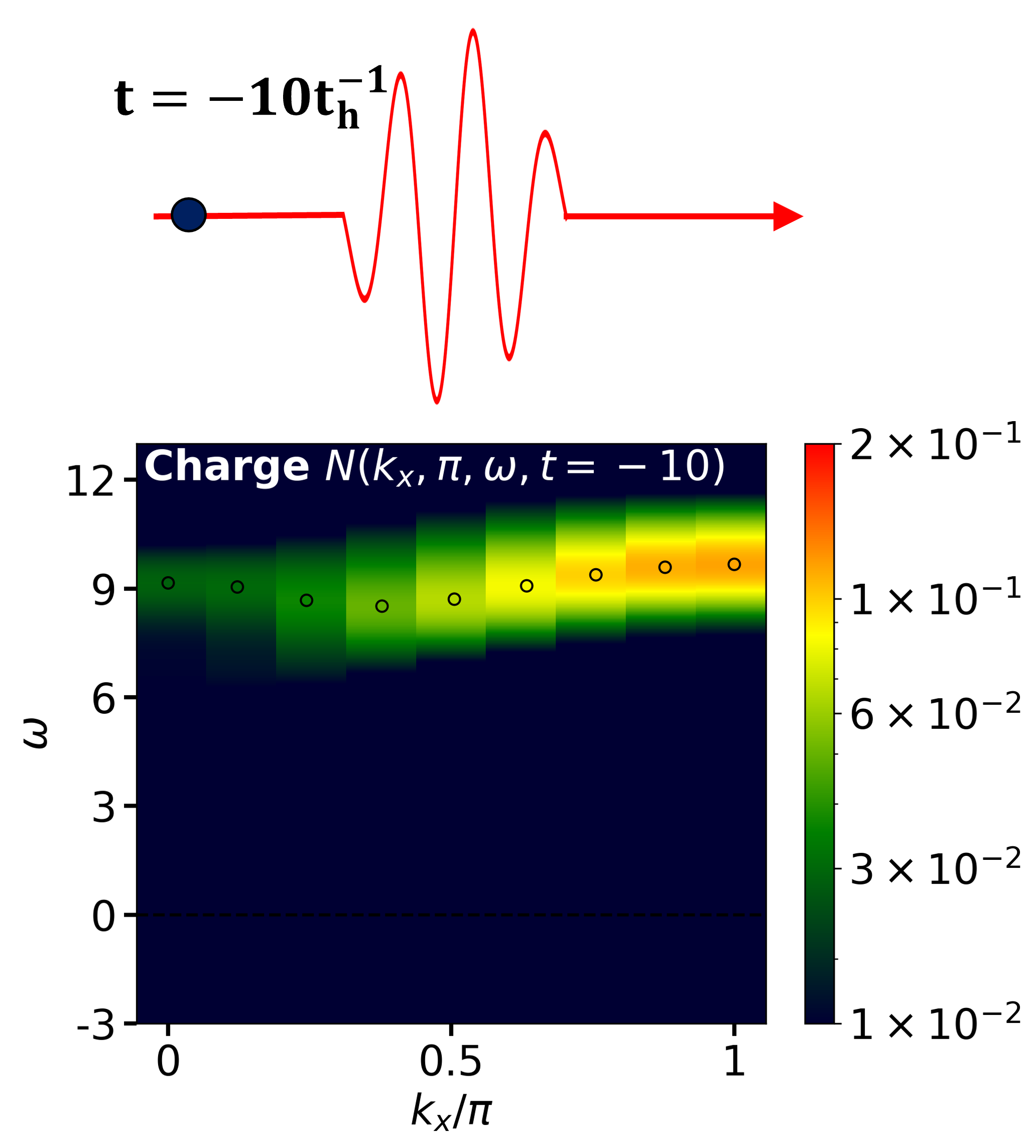}
  \end{minipage}
  \hfill
  \begin{minipage}[b]{0.32\textwidth}
    \centering
    \includegraphics[width=\linewidth]{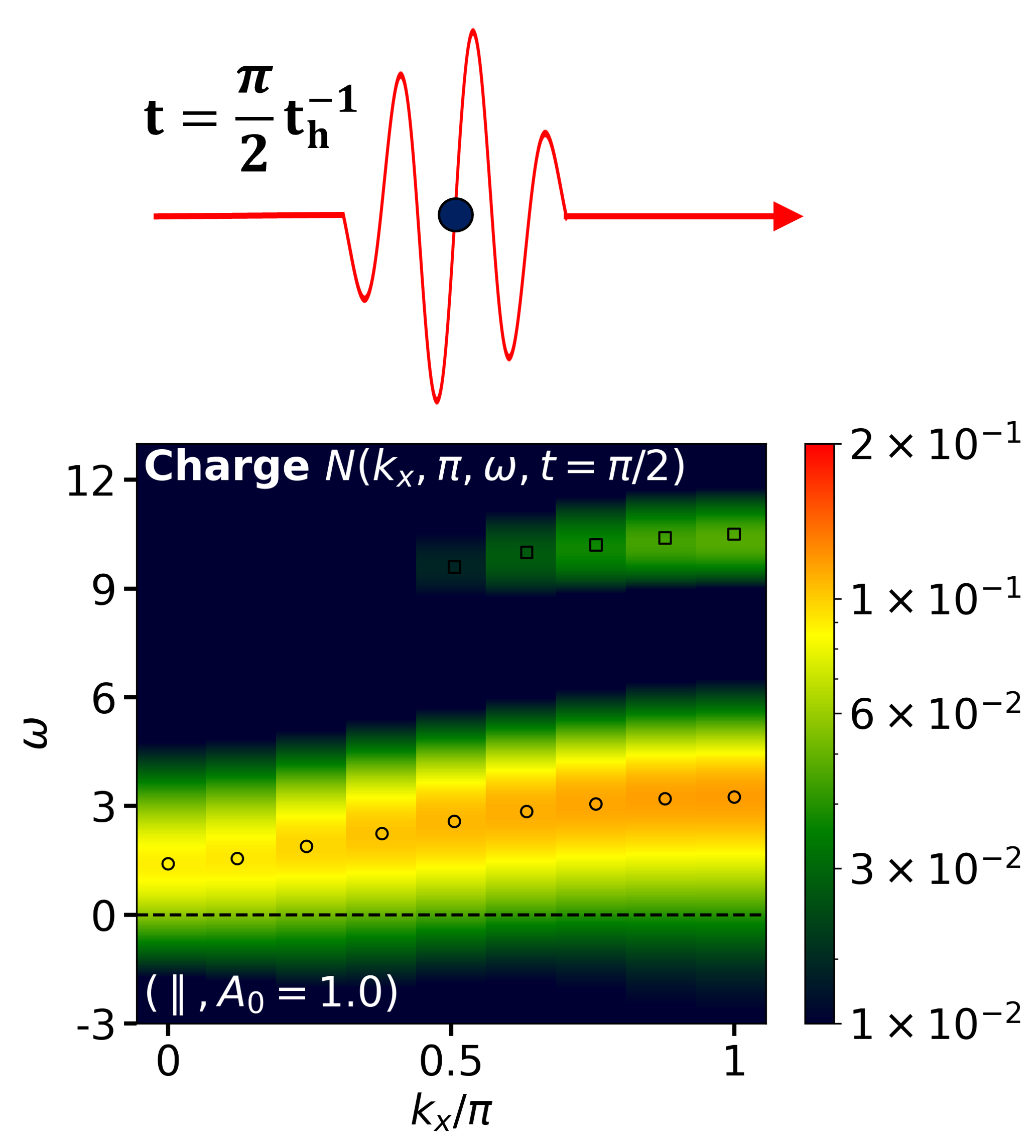}
  \end{minipage}
  \hfill
  \begin{minipage}[b]{0.32\textwidth}
    \centering
    \includegraphics[width=\linewidth]{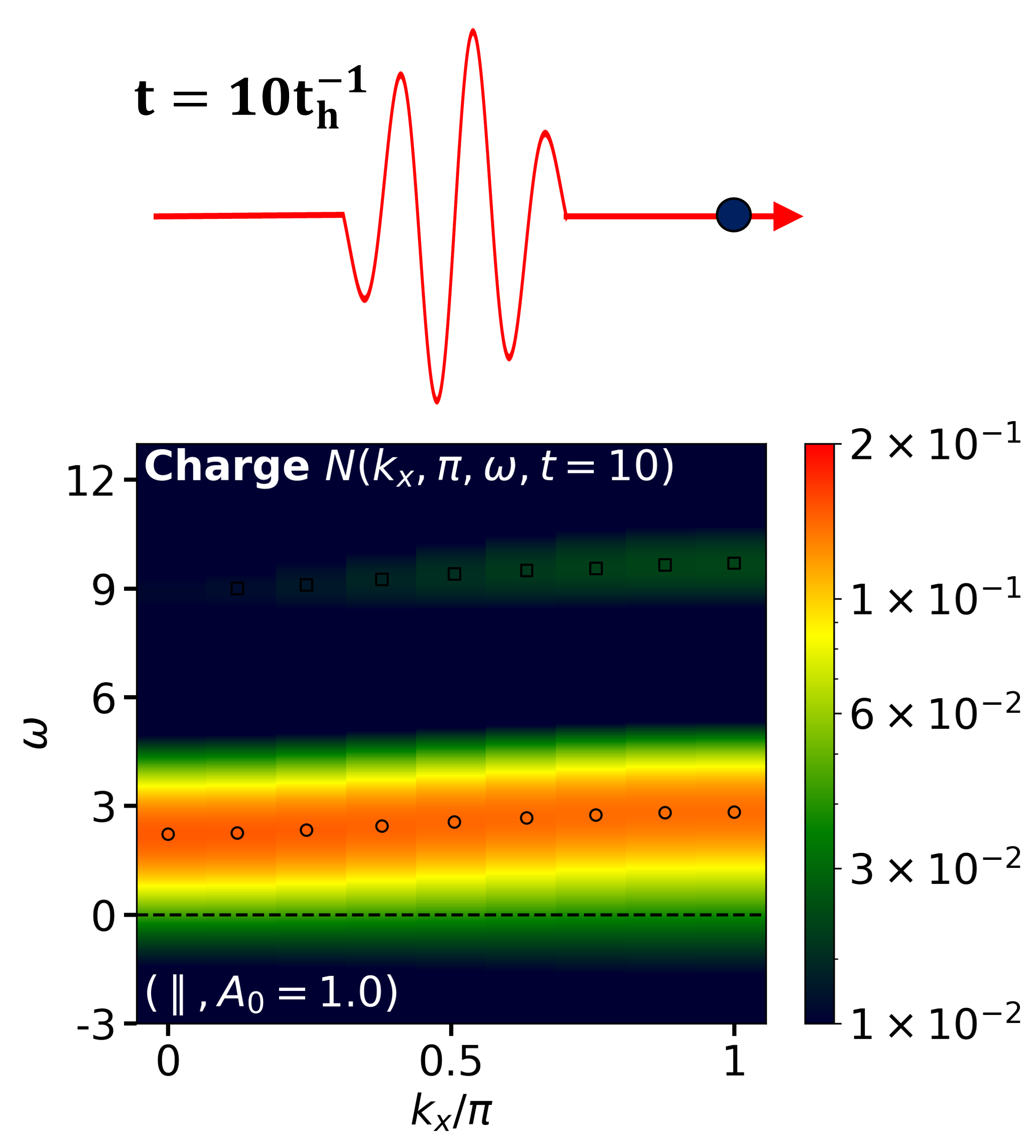}
  \end{minipage}
  \vspace{\baselineskip} % Add some vertical space between the two sets of plots
  \begin{minipage}[b]{0.32\textwidth}
    \centering
    \includegraphics[width=\linewidth]{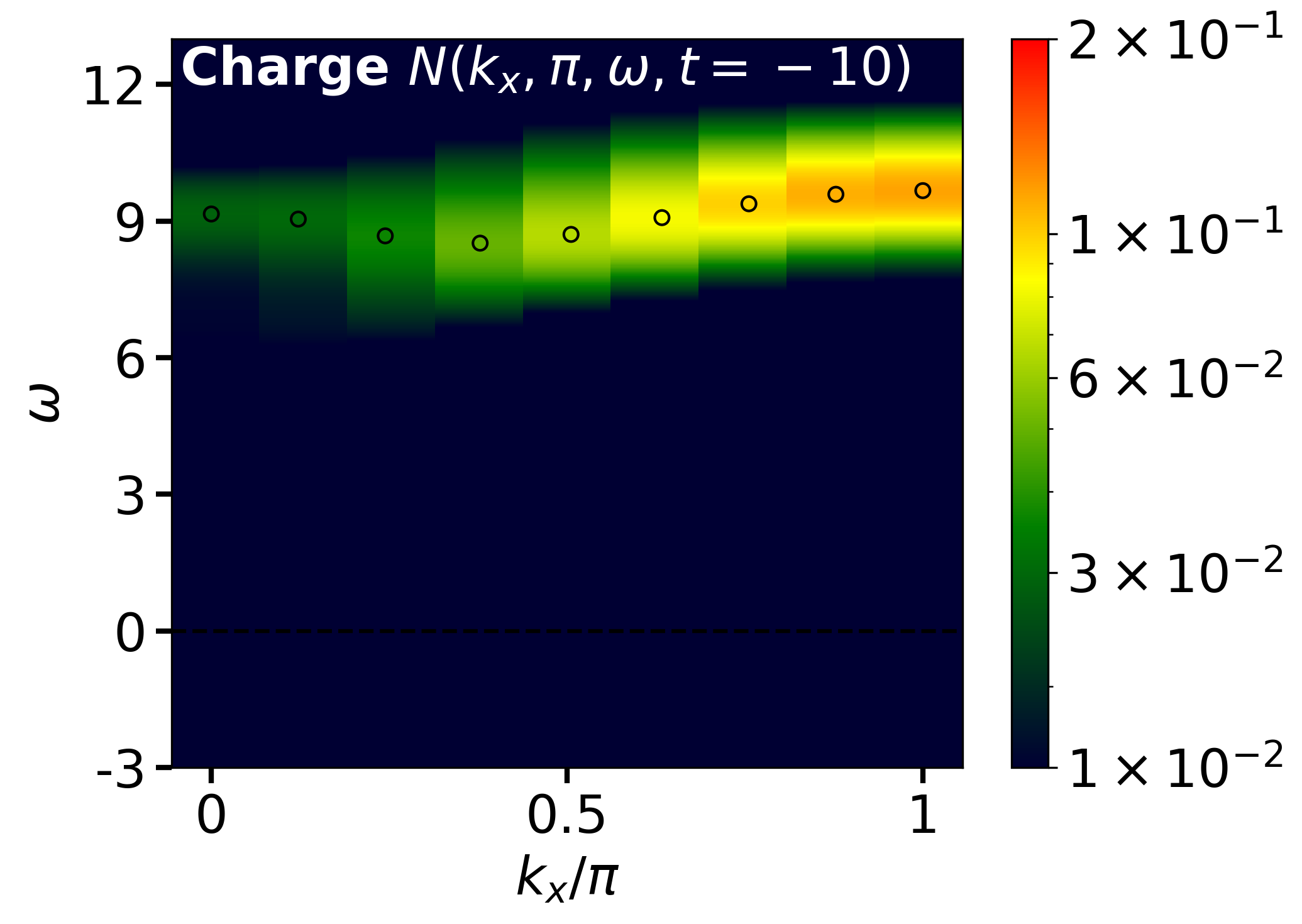}
  \end{minipage}
  \hfill
  \begin{minipage}[b]{0.32\textwidth}
    \centering
    \includegraphics[width=\linewidth]{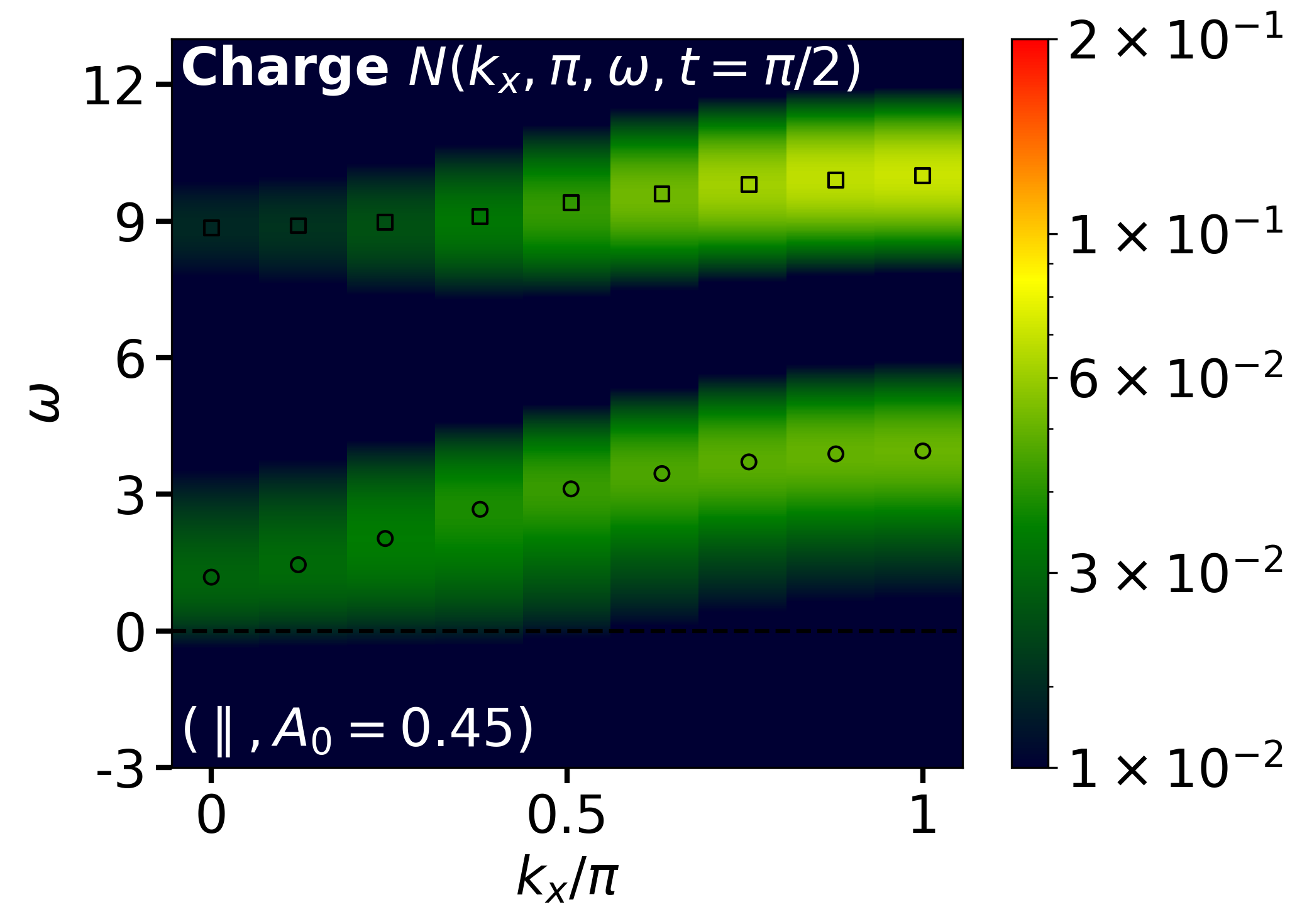}
  \end{minipage}
  \hfill
  \begin{minipage}[b]{0.32\textwidth}
    \centering
    \includegraphics[width=\linewidth]{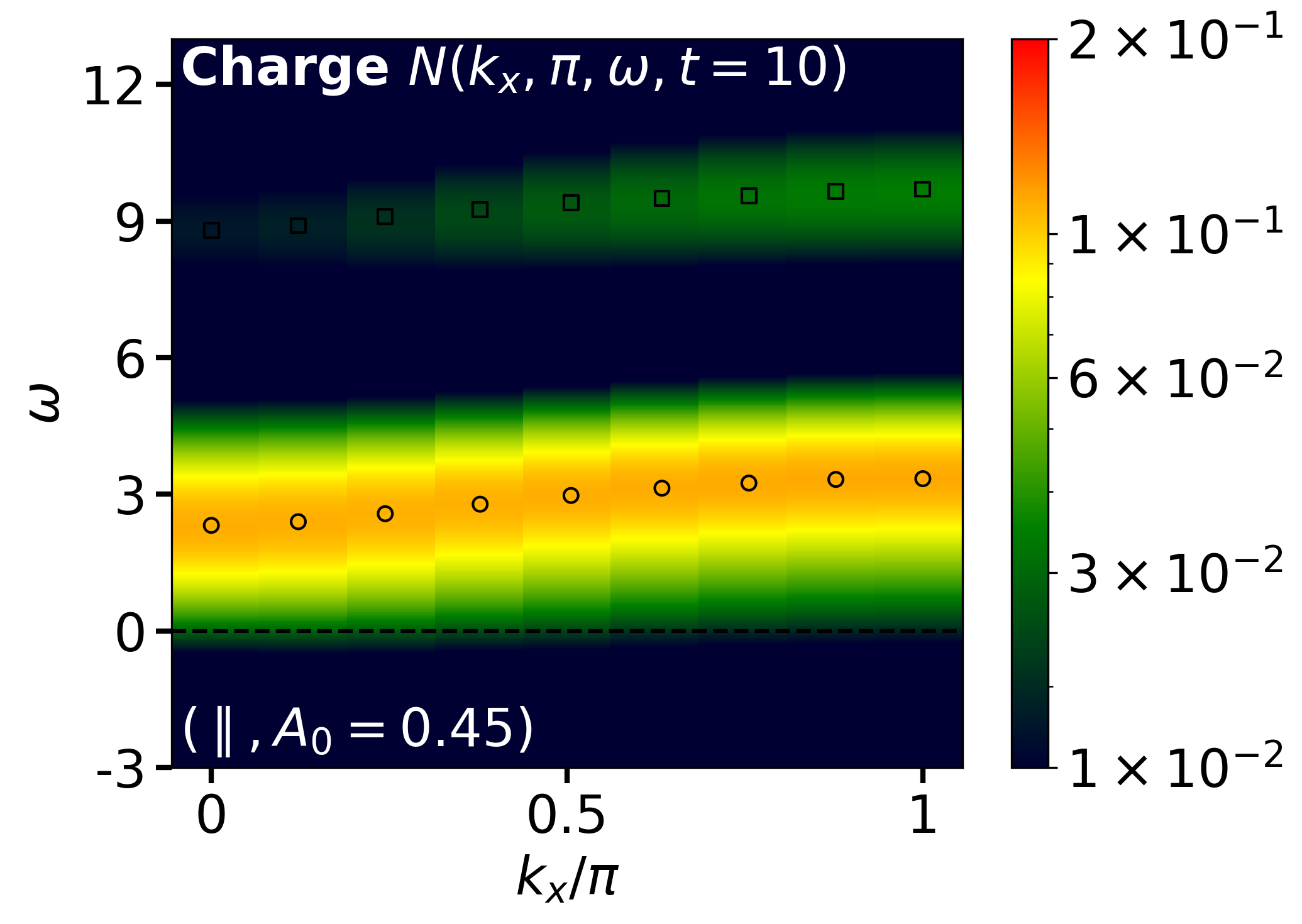}
  \end{minipage}
  \vspace{\baselineskip} % Add some vertical space between the two sets of plots
  \begin{minipage}[b]{0.32\textwidth}
    \centering
    \includegraphics[width=\linewidth]{charge_ky_PI_equilibrium}
  \end{minipage}
  \hfill
  \begin{minipage}[b]{0.32\textwidth}
    \centering
    \includegraphics[width=\linewidth]{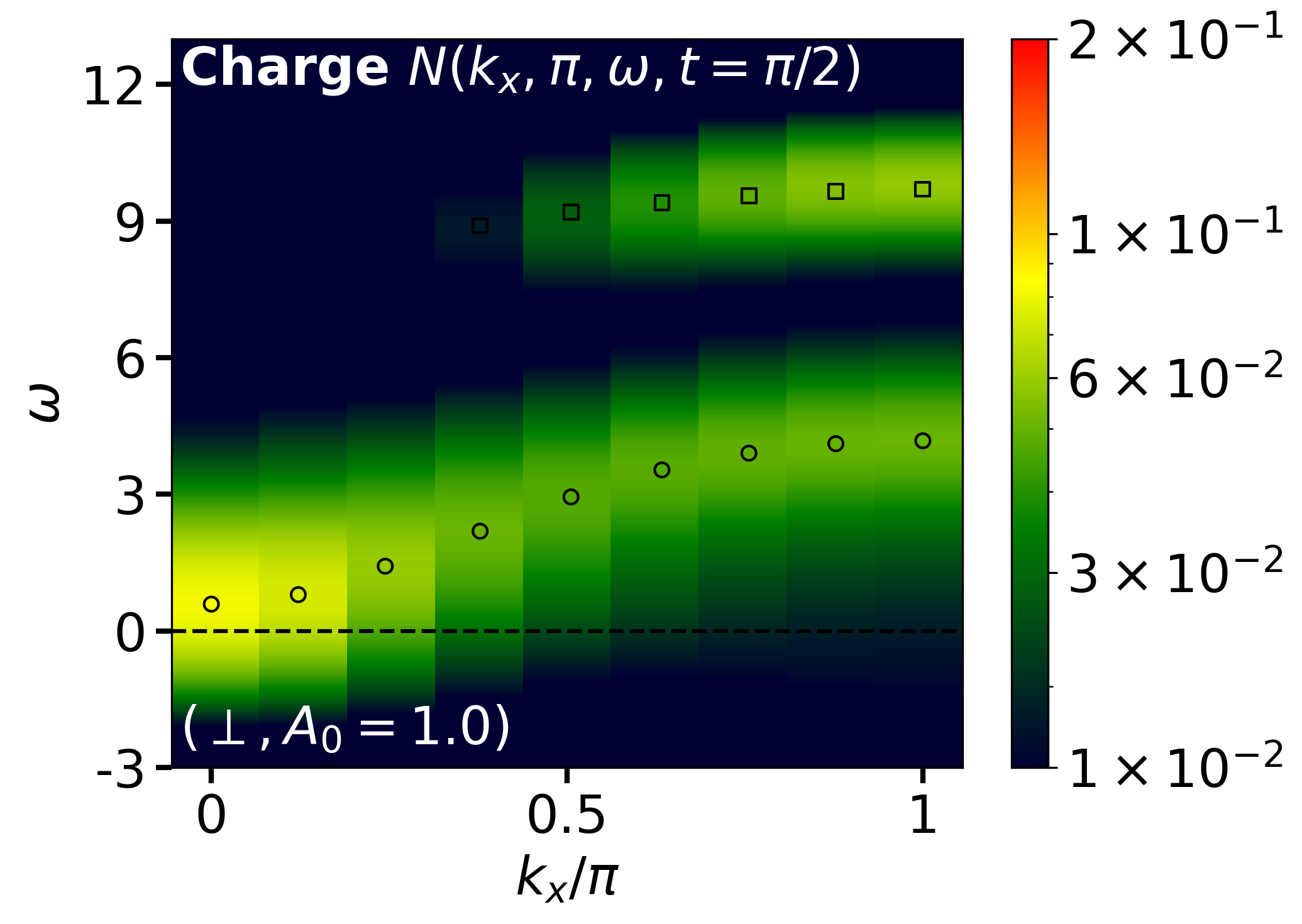}
  \end{minipage}
  \hfill
  \begin{minipage}[b]{0.32\textwidth}
    \centering
    \includegraphics[width=\linewidth]{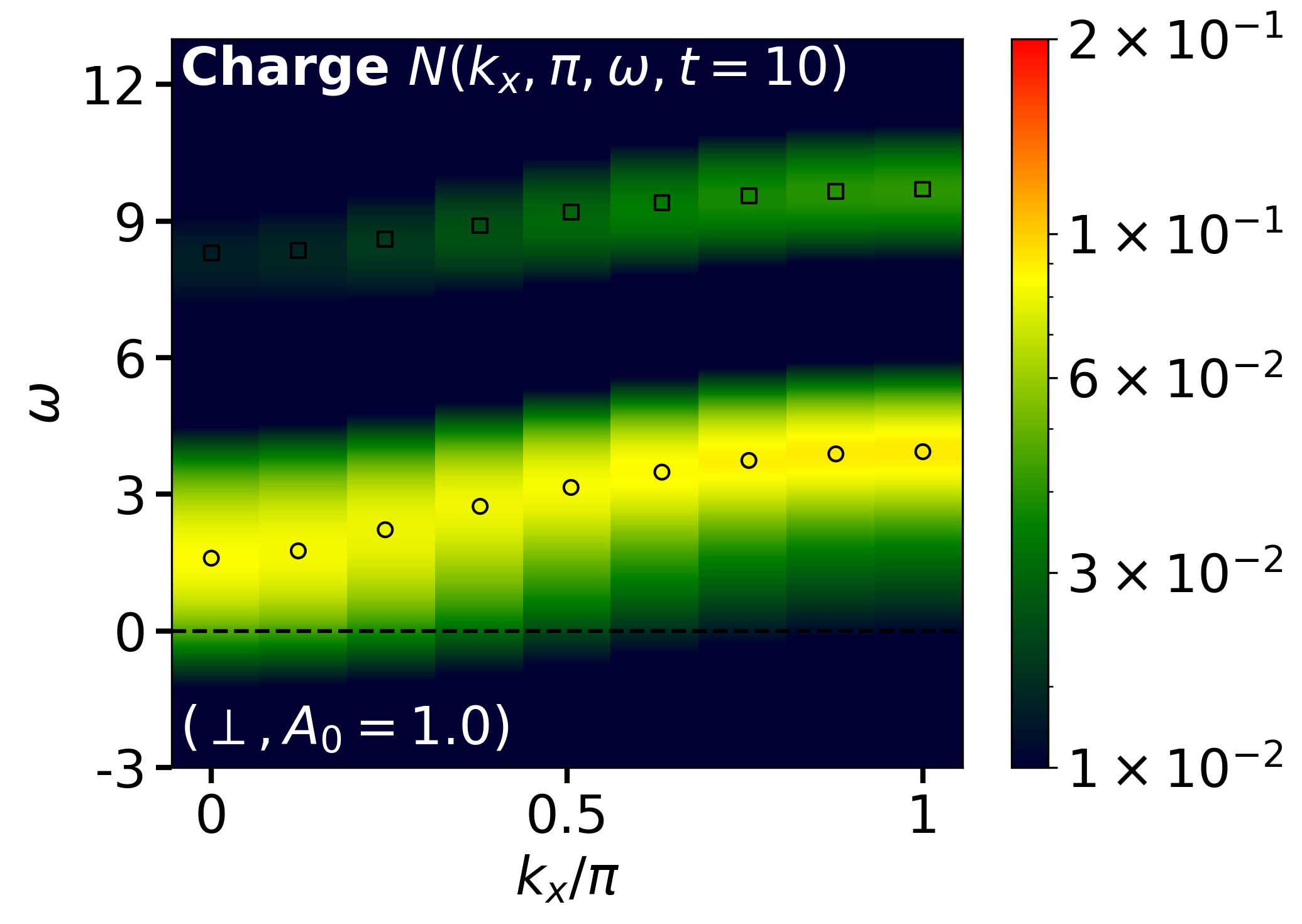}
  \end{minipage}
  \caption{Nonequilibrium dynamical charge structure factor at $k_y = \pi$.
  In equilibrium the response consists of excitations across the Mott gap from the lower to the upper Hubbard band ($\omega\sim U$).
  After photodoping the dominant feature is a low energy response, due to scattering from excited states within the upper Hubbard band.
  The three panels in the leftmost column are the same, but are provided for ease of comparison across the rows.}
  \label{fig:charge_structure_factor_ky_PI}
\end{figure*}
\begin{figure*} % Use figure* for two-column wide figures
  \centering
  \begin{minipage}[b]{0.32\textwidth}
    \centering
    \includegraphics[width=\linewidth]{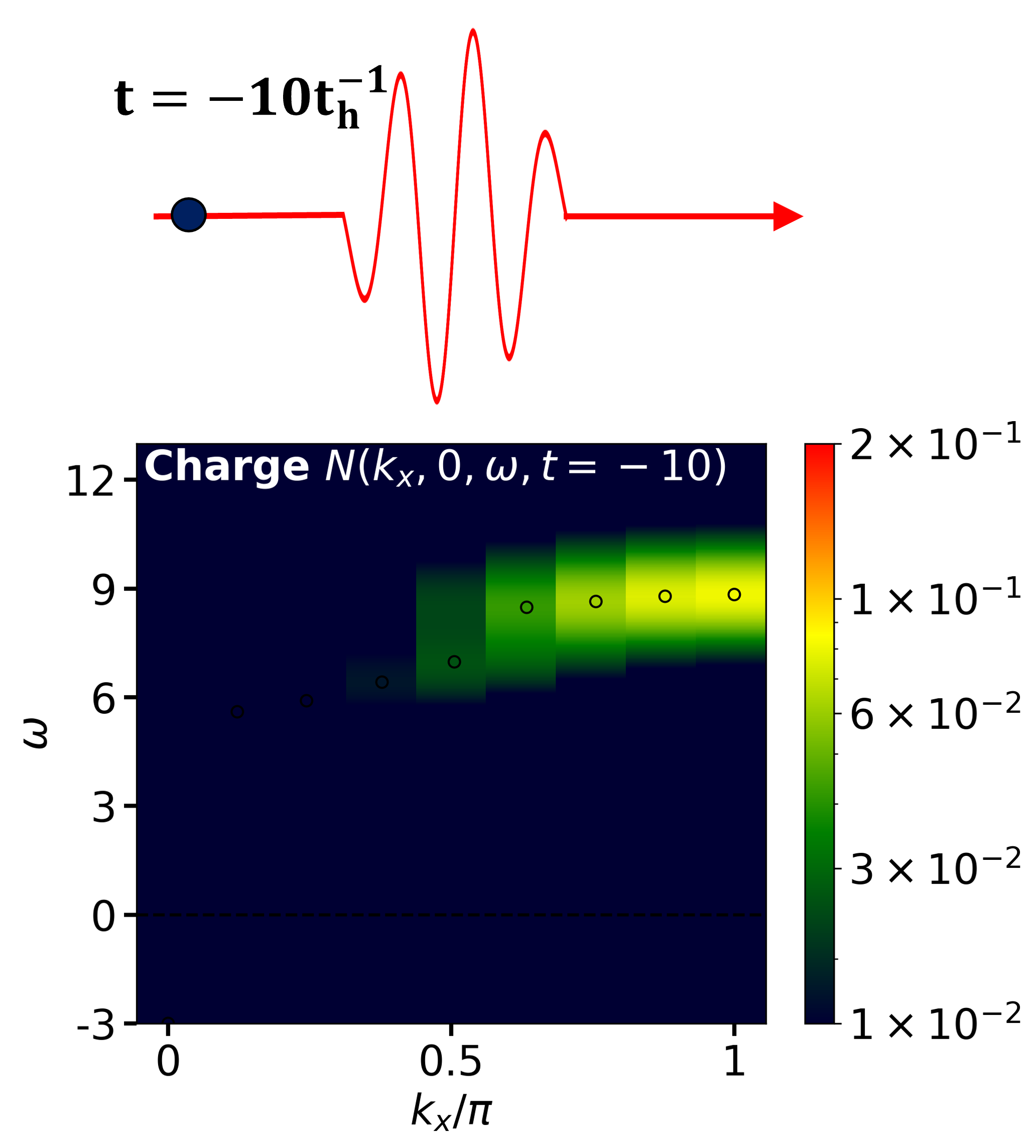}
  \end{minipage}
  \hfill
  \begin{minipage}[b]{0.32\textwidth}
    \centering
    \includegraphics[width=\linewidth]{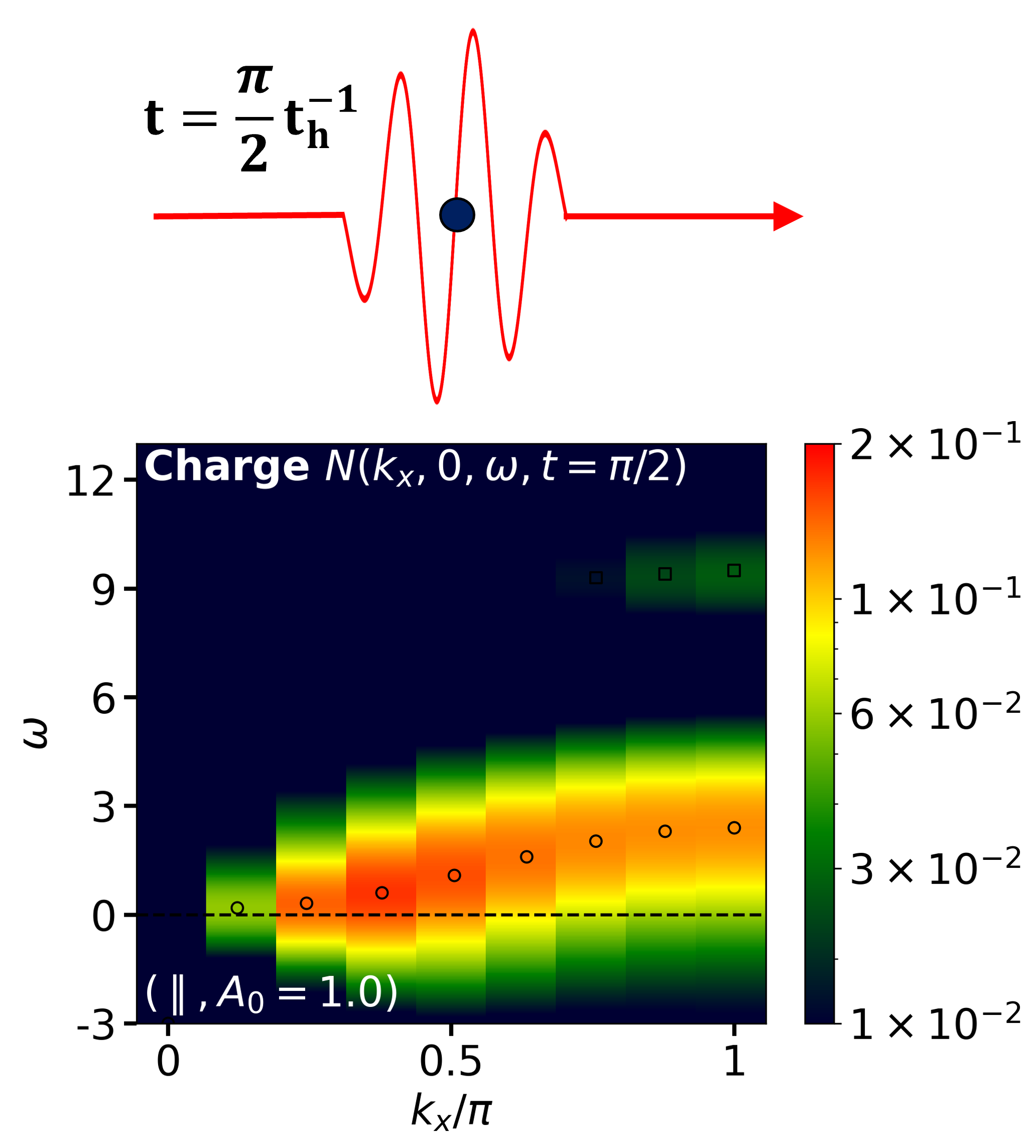}
  \end{minipage}
  \hfill
  \begin{minipage}[b]{0.32\textwidth}
    \centering
    \includegraphics[width=\linewidth]{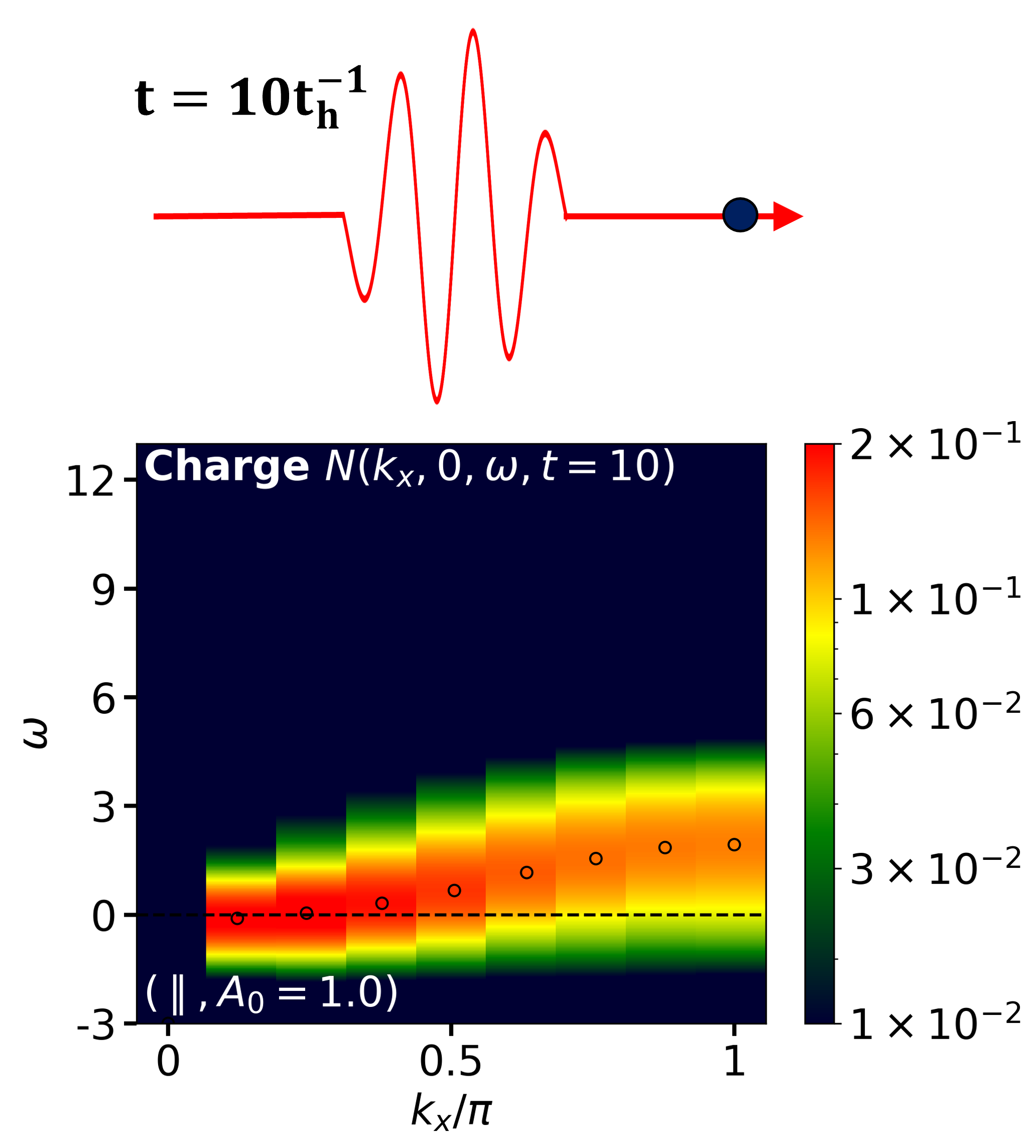}
  \end{minipage}
  \vspace{\baselineskip} % Add some vertical space between the two sets of plots
  \begin{minipage}[b]{0.32\textwidth}
    \centering
    \includegraphics[width=\linewidth]{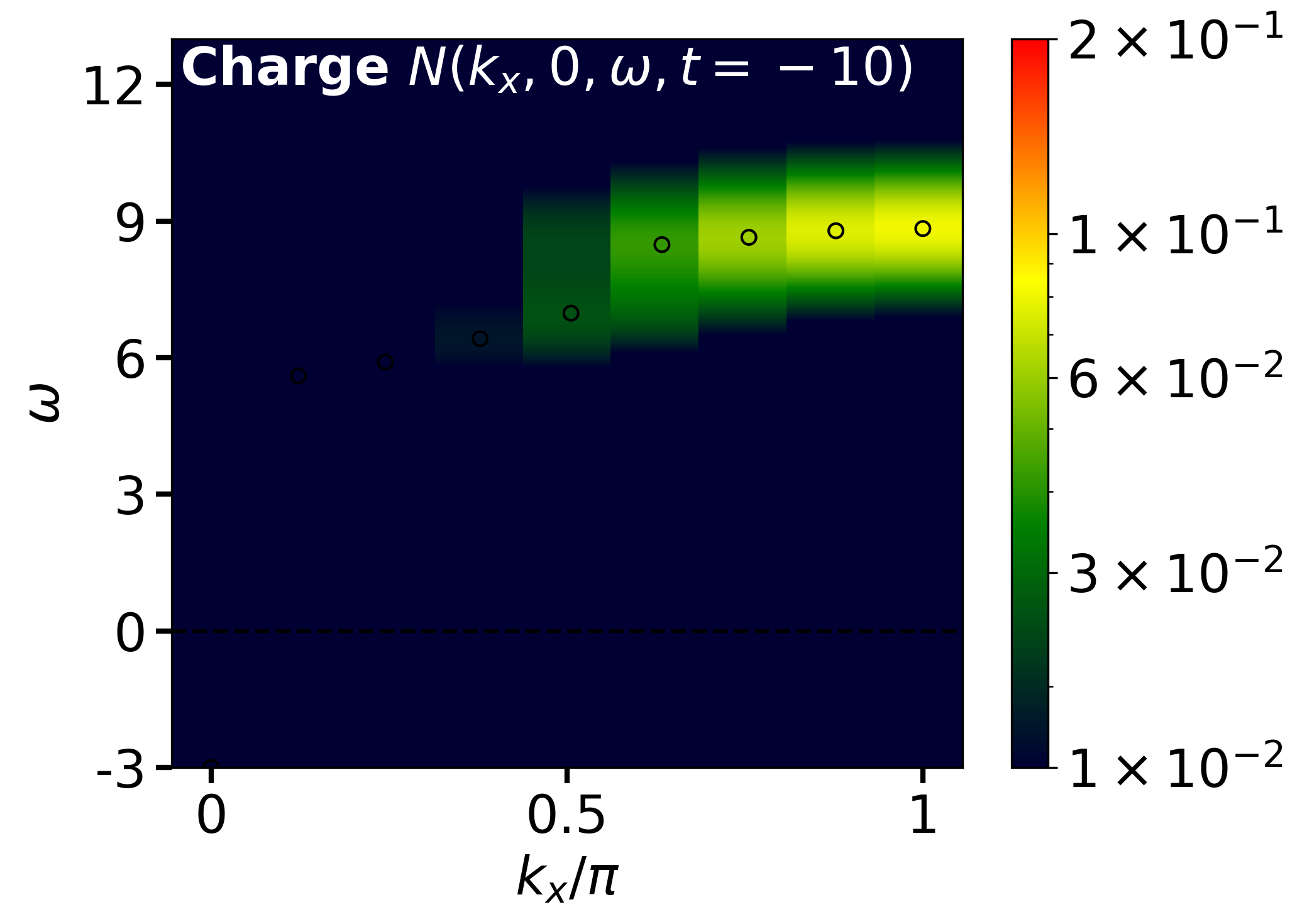}
  \end{minipage}
  \hfill
  \begin{minipage}[b]{0.32\textwidth}
    \centering
    \includegraphics[width=\linewidth]{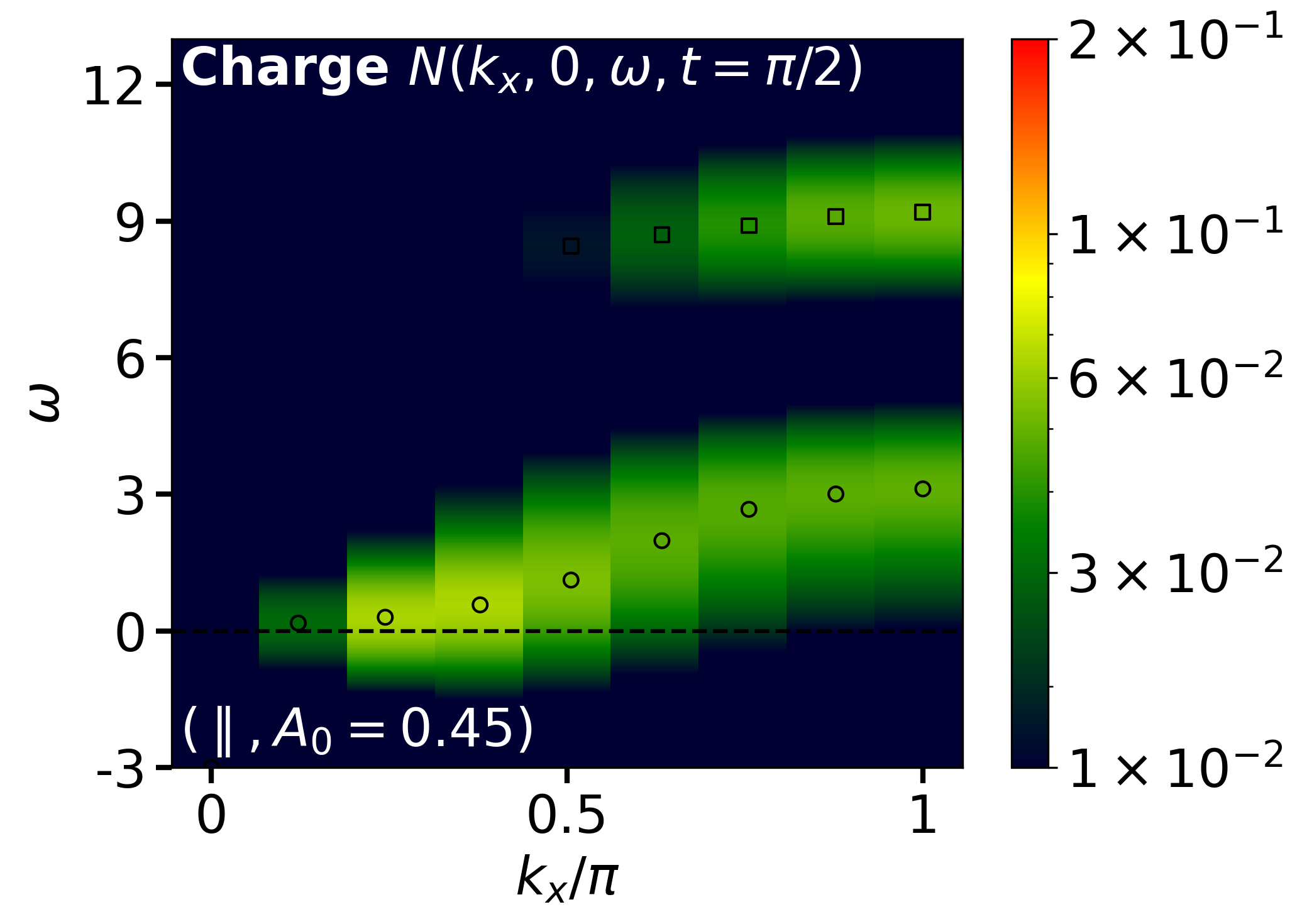}
  \end{minipage}
  \hfill
  \begin{minipage}[b]{0.32\textwidth}
    \centering
    \includegraphics[width=\linewidth]{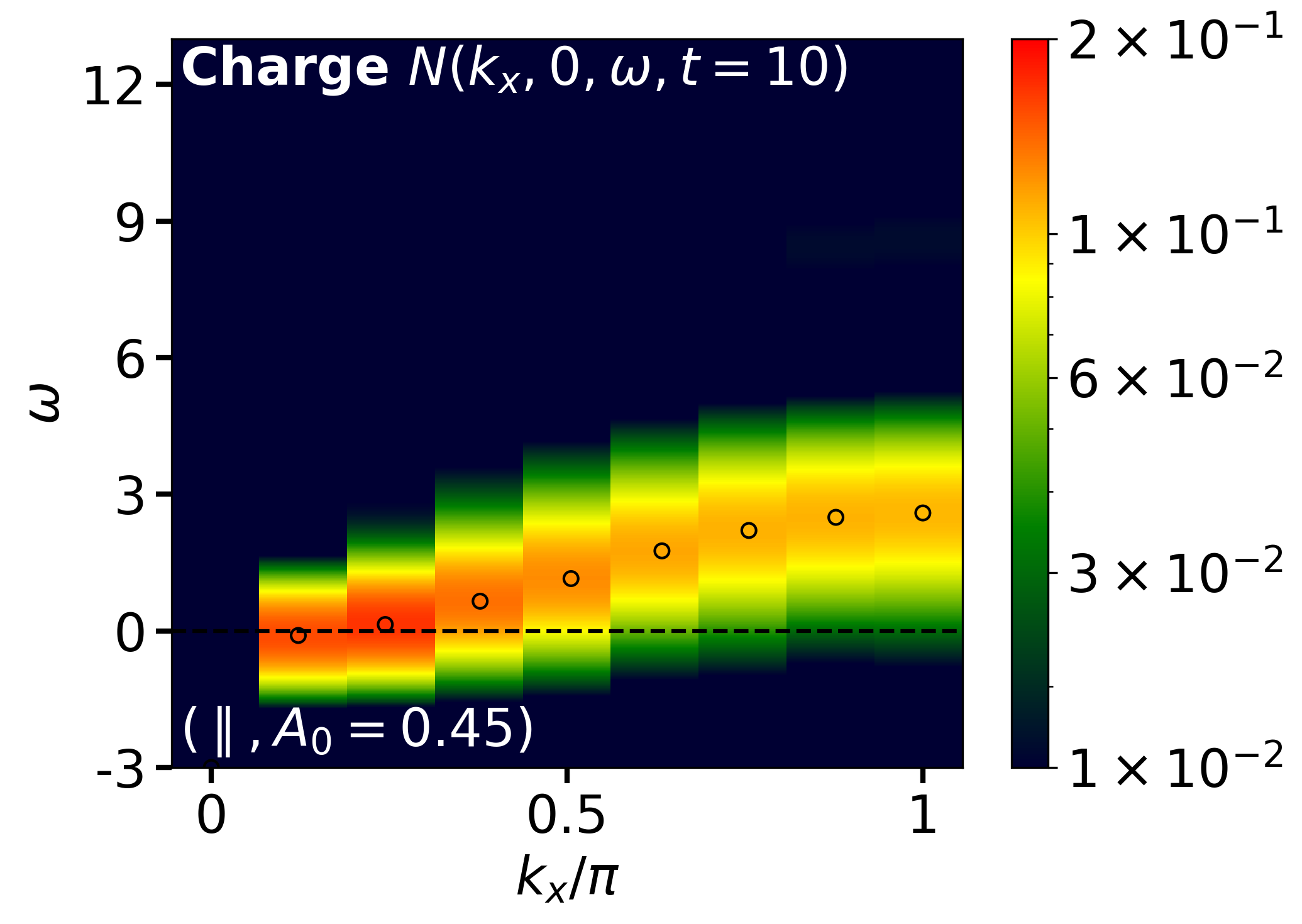}
  \end{minipage}
  \vspace{\baselineskip} % Add some vertical space between the two sets of plots
  \begin{minipage}[b]{0.32\textwidth}
    \centering
    \includegraphics[width=\linewidth]{charge_ky_0_equilibrium.png}
  \end{minipage}
  \hfill
  \begin{minipage}[b]{0.32\textwidth}
    \centering
    \includegraphics[width=\linewidth]{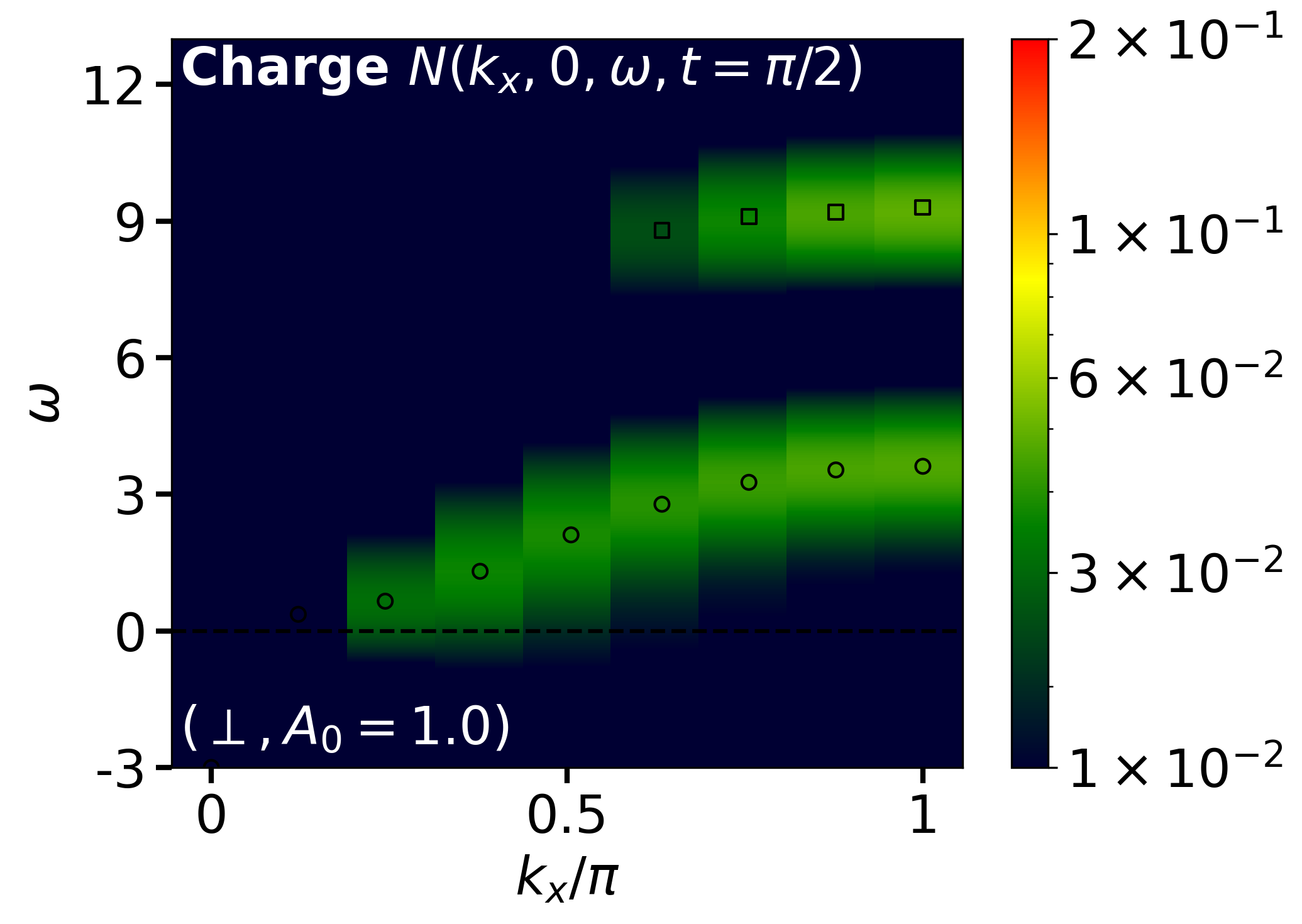}
  \end{minipage}
  \hfill
  \begin{minipage}[b]{0.32\textwidth}
    \centering
    \includegraphics[width=\linewidth]{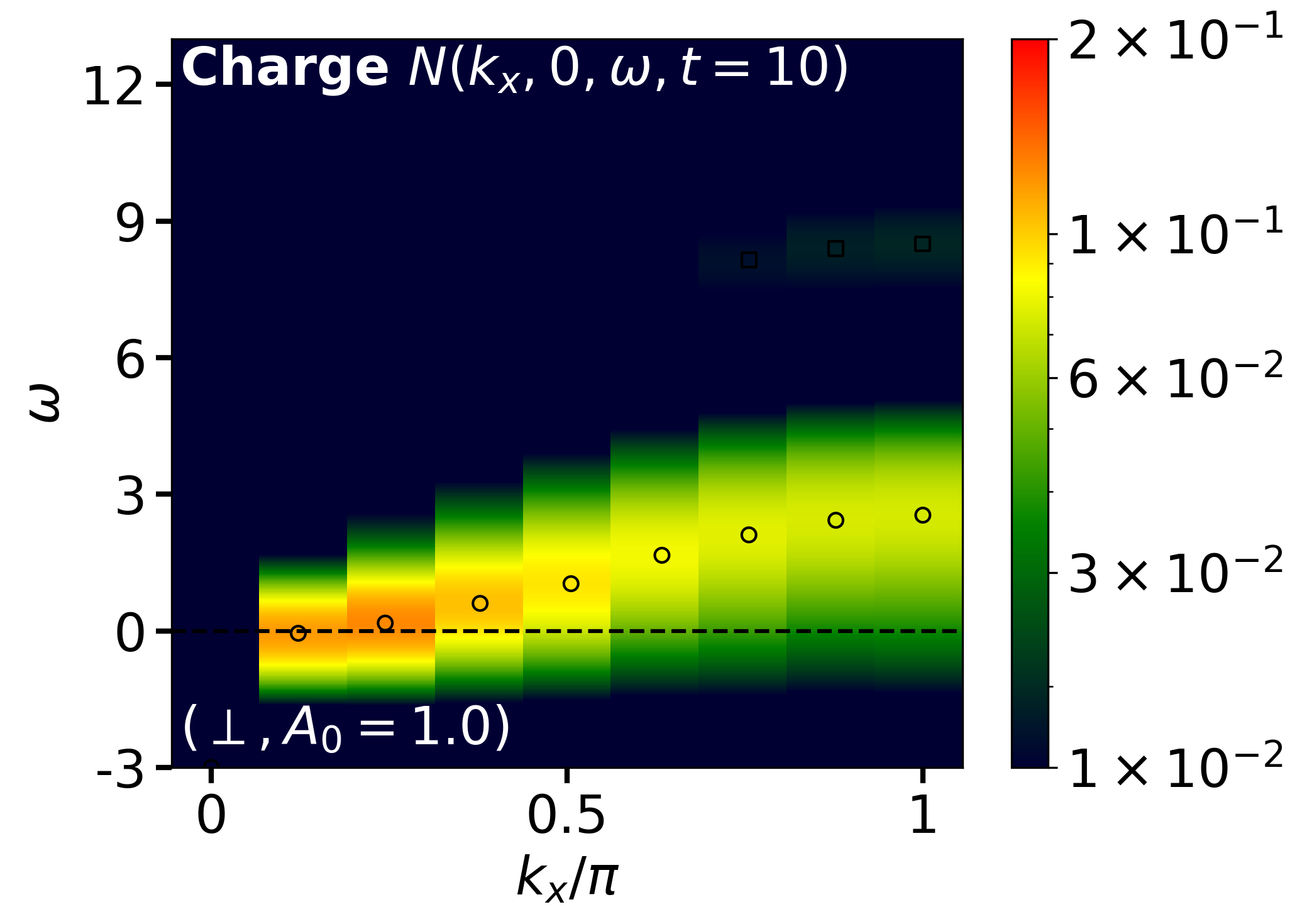}
  \end{minipage}
  \caption{Nonequilibrium dynamical charge structure factor for $k_y = 0$.
 The static peak at $\mathbf{k}=(0,0)$ and $\omega\sim 0$ due to charge conservation has been subtracted off.
  The three panels in the leftmost column are the same, but are provided for ease of comparison across the rows.}
  \label{fig:rungsquench_structure_factor_Ky_0}
\end{figure*}

The dynamical charge response reflects Mott insulator physics, displaying the Mott gap in equilibrium, but showing an increasing low energy response as electrons in the lower Hubbard band are excited into the upper Hubbard band by the pump.
We compute the nonequilibrium charge dynamical structure factor, using Eq.~\ref{eq:Isingletime}, and show the results for $k_y=\pi$ in Fig.~\ref{fig:charge_structure_factor_ky_PI} and for $k_y=0$ in Fig.~\ref{fig:rungsquench_structure_factor_Ky_0}.
%Although the spin response is mostly sensitive to pump direction through its effect on the total injected energy, the charge response shows some pump direction dependence even when the total injected energy is the same.

In equilibrium the charge dynamics are characterised by an interband response at the energy required to create a doublon, $\omega \sim U$ (left column of Figs.~\ref{fig:charge_structure_factor_ky_PI} and~\ref{fig:rungsquench_structure_factor_Ky_0}).
Our equilibrium results are in broad agreement with the charge response calculated using quantum Monte Carlo in Ref.~\cite{endres1996dynamical}.
The strongest response in Fig.~\ref{fig:charge_structure_factor_ky_PI} is at $\mathbf{k}=(\pi,\pi)$, reflecting the anticorrelation between nearest neighbour sites -- if the occupation of a site is enhanced then the occupation of its neighbouring sites is lowered.
In Fig.~\ref{fig:rungsquench_structure_factor_Ky_0} we have subtracted the strong response at $\mathbf{k}=(0,0)$ and $\omega\sim 0$ that results from charge conservation and the finite probe width $\sigma_X$.
This response behaves as $\sim N e^{-\sigma_X^2 \omega^2}$, where $N=2L$ is the total number of electrons.

Photodoping creates doublons, and as only a finite number of doublons can be created this process reduces the joint density of states for interband transitions in the charge channel.
This suppresses the response at $\omega \sim U$, as shown in the middle and right columns of Figs.~\ref{fig:charge_structure_factor_ky_PI} and~\ref{fig:rungsquench_structure_factor_Ky_0}.

In contrast to the spin sector, the emergence due to pumping of an intraband charge response at low energies $\omega \sim 0$ is unambiguous because of the large Mott gap.
{ After pumping the intraband charge response is a strong feature, with more spectral weight than the interband charge response in the original equilibrium system.
The intraband response arises due to scattering from pump induced doublons.
The lowest panels of Fig.~\ref{fig:charge_structure_factor_ky_PI} and Fig.~\ref{fig:rungsquench_structure_factor_Ky_0} show the results for the rung pump, which produces fewer doublons than the two leg-direction pumps.
Accordingly, the rung pump results show a relatively weaker intraband response (due to fewer photoinduced doublons) and stronger interband response (reflecting a larger density of states for the creation of new doublons by a probe).
For the rung-direction pump the intraband response has a pronounced dispersion, becoming gapless on approaching $\mathbf{k}=(0,0)$.
This can be interpreted as an effective dispersion for the propagating photoinduced doublons.
The leg-direction pumps produce more doublons and the resulting intraband response is flatter: as doublons are hardcore they interact strongly and impede each other, renormalizing the effective charge hopping parameters and reducing the bandwidth.
}
%This intraband response, due to scattering from pump induced doublons, has a pronounced dispersion and becomes gapless on approaching $\mathbf{k}=(0,0)$.

%For the $A_0=1$ leg pump the ratio of the integrated charge response (including the suppressed static peak at $\mathbf{k}=(0,0)$ and $\omega \sim 0$) after pumping to that in equilibrium is $W_n(t=10)/W_n(t=-10) \approx 1.2$.
%For the $A_0=1$ rung pump the same ratio gives $\approx 1.1$.
\begin{figure}
    \centering
    \includegraphics[scale=0.4]{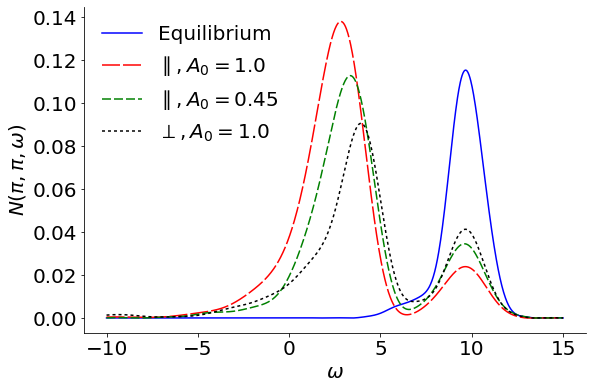}
    \caption{Charge correlations with $\mathbf{k}=(\pi,\pi)$, at $t = 10$ after pumping along the legs $(\parallel)$ or rungs $(\perp)$.}
  \label{fig:compare_rungs_vs_legs_quench_cuts_charge_PIPI}
\end{figure}
\begin{figure}
    \centering
    \includegraphics[scale=0.4]{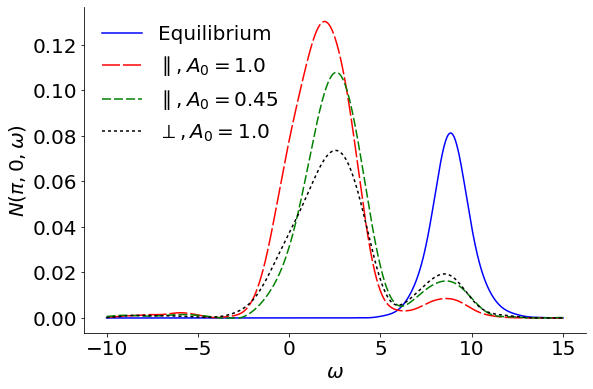}
       \caption{Charge correlations with $\mathbf{k}=(0,\pi)$, at $t = 10$ after pumping along the legs $(\parallel)$ or rungs $(\perp)$.}
  \label{fig:compare_rungs_vs_legs_quench_cuts_charge_PI0}
\end{figure}

{ As in the case of the spin response, the detailed effects of the different pumps in the charge sector are best seen by considering cuts at fixed wave vectors.}
Figs.~\ref{fig:compare_rungs_vs_legs_quench_cuts_charge_PIPI} and~\ref{fig:compare_rungs_vs_legs_quench_cuts_charge_PI0} show cuts through the charge response at fixed $\mathbf{k}$.
Similar to the spin dynamics, the equilibrium charge response at both $(\pi,\pi)$ and at {$(\pi,0)$} appears as a resolution limited peak (except for a small feature close to $\omega=6$ for $k_y=\pi$).
Pumping suppresses this peak and makes it asymmetric, as well as producing a strong low-energy, intraband response.
The energetically similar $A_0=0.45$ leg{-}direction, and $A_0=1$ rung{-}direction pumps lead to { similar} suppression of the interband response at $\omega \sim 10$ { for $(\pi,\pi)$ (see Fig.~\ref{fig:compare_rungs_vs_legs_quench_cuts_charge_PIPI}).}
However, there is pump anisotropy visible in the intraband response that develops.
Figs.~\ref{fig:compare_rungs_vs_legs_quench_cuts_charge_PIPI} and~\ref{fig:compare_rungs_vs_legs_quench_cuts_charge_PI0} show that the rung pump produces a weaker response at {$\mathbf{k}=(\pi,0)$} than at $\mathbf{k}=(\pi,\pi)$ with an asymmetric energy profile that differs qualitatively between the two wave vectors.
In comparison, the $A_0=0.45$ leg{-}direction pump produces a very similar post pump response in terms of intensity and shape at both $k_y=\pi$ and $k_y=0$.

\subsection{Comparison {of numerics} with the sum rule}
We finish this section by checking that the integrated spin and charge responses obey Eq.~\ref{eq:sumrule}.
Table~\ref{tab:weight} displays the results of evaluating $W_s$ and $W_c$ numerically using two different methods, at three different times: $t=-10$ before the pump, $t=10$ after the pump, and $t=\pi/2$ half way through the pump.
The first method involves evaluating Eq.~\ref{eq:IWS1} and Eq.~\ref{eq:IWn} using iTEBD data for $\langle S^2_{i,l}(t) \rangle$ and $\langle d_{i,l}(t) \rangle$.
This method confirms the accuracy of the iTEBD results by satisfying the sum rule $4W_s+W_c=2G\rho$ to high precision ($<10^{-9}$ missing weight).

The second method we use is to integrate the dynamical response $I(\mathbf{k},\omega,t)$, determined using TEBD for the 32 site ladder, over $\omega$ and $\mathbf{k}$.
The results of this method for $W_s$ and $W_c$ are expected to differ from those for iTEBD due to finite size effects and errors introduced by the single integral approximation, Eq.~\ref{eq:Isingletime}.
The latter can also cause departures from the sum rule.
Concomitantly, in Table~\ref{tab:weight} the worst agreement with the sum rule occurs at $t=\pi/2$ where the single integral approximation is least accurate.
However, the largest error, seen for the leg{-}direction $A_0=0.45$ pump, is still small and $<4\%$.

\begin{table*}[t]
\caption{Comparison of the spectral weight determined from iTEBD data using Eq.~\ref{eq:IWS1} and Eq.~\ref{eq:IWn}, with that found by integrating the nonequilibrium DSF computed for a 32 site ladder using TEBD.
Figures are shown to 4 decimal places. For our parameters $G=(2\sqrt{\pi})^{-1}=0.2821...$}
    \centering
    \begin{tabular}{|l|l|l|c|c|c|}
    \hline \hline
         \multicolumn{3}{|c|}{} & $W_s$ & $W_c$ & $\frac{4W_s+W_c}{2G\rho}$ \\ \hline
       \multirow{2}{*}{Equilibrium}& \multirow{2}{*}{$t=-10$} & iTEBD & 0.0643 & 0.3069 & 1\\
       & & integrated TEBD DSF & 0.0643 & 0.3073 & 1.0009 \\
       \hline
       \multirow{4}{*}{$\parallel, A_0=1$} & \multirow{2}{*}{$t=10$} & iTEBD & 0.0487 & 0.3695 & 1\\
       & & integrated TEBD DSF & 0.0486 & 0.3684 & 0.9976 \\
       \cline{2-6}
       & \multirow{2}{*}{$t=\pi/2$} & iTEBD & 0.0495 & 0.3662 & 1\\
       & & integrated TEBD DSF & 0.0480 & 0.3684 & 0.9929 \\
       \hline
       \multirow{4}{*}{$\parallel, A_0=0.45$} & \multirow{2}{*}{$t=10$} & iTEBD & 0.0530 & 0.3521 & 1\\
       & & integrated TEBD DSF & 0.0527 & 0.3664 & 1.0228 \\
       \cline{2-6}
       & \multirow{2}{*}{$t=\pi/2$} & iTEBD & 0.0594 & 0.3267 & 1\\
       & & integrated TEBD DSF & 0.0599 & 0.3496 & 1.0444 \\
       \hline
       \multirow{4}{*}{$\perp, A_0=1$} & \multirow{2}{*}{$t=10$} & iTEBD & 0.0550 & 0.3442 & 1\\
       & & integrated TEBD DSF & 0.0553 & 0.3449 & 1.0033 \\
       \cline{2-6}
       & \multirow{2}{*}{$t=\pi/2$} & iTEBD & 0.0592 & 0.3276 & 1\\
       & & integrated TEBD DSF & 0.0586 & 0.3494 & 1.0346 \\
       \hline
       \hline
    \end{tabular}
    \label{tab:weight}
\end{table*}

\section{Conclusions}
\label{sec:discussion}
We have studied the nonequilibrium dynamics of the half-filled Hubbard ladder when it is pumped by an idealised optical pulse, using MPS techniques.
We find that the antiferromagnetic correlations that dominate the equilibrium state are significantly reduced in the post pump state: equilibrium AF correlations are strongly suppressed at nearest and next nearest neighbour separations { relative to the equilibrium state}, before exponential decay sets in with a slightly reduced correlation length.
Charge correlations in the initial, equilibrium, state are very weak, but they are strongly enhanced by the pump until they rival the magnetic correlations.
We have shown that the anisotropic nature of the ladder means that the energy imparted by the pump is dependent on the pump direction.
In general, pumping with a polarisation that alters the hopping parameter along the legs of the ladder leads to greater energy transfer and disruption of spin correlations than pumping polarised in the rung{-}direction.

{ A small $t$ expansion and the numerical results show that at short times the response of the system to the pump scales with the number of nearest neighbours and hopping parameters as $\sim z_\alpha t_\alpha A_0^2$, where $\alpha=\parallel,\perp$ labels the pump direction.
The small $t$ expansion does not rely on the ladder geometry.
Hence, our results suggest this scaling will also apply to the short time behaviour of 2D (or 3D) systems.
Once $t$ exceeds the characteristic hopping time $t_h^{-1}$ the scaling breaks down, and the response of the system is instead more specifically dependent on the pump characteristics (for example, pumps in different directions that have almost equal total photoinjected energies nonetheless produce different numbers of doublon excitations).
}
%leg{-}direction pumping also leads to saturation of the transferred energy at lower fluence than for rung{-}direction pumping.

We have demonstrated that the spin and charge dynamical response functions obey a form of sum rule, with the total weight conserved even in the nonequilibrium state.
The sum rule shows that the photodoping causes spectral weight to be pumped from the magnetic to the charge degrees of freedom.
The overall effect is a sizeable shift of dynamical response to the low energy region of the charge sector, where a gapless dispersing mode develops, indicating that the post pump system is in a correlated metallic state.

We note that the separate spin and charge contributions to the sum rule are related to the density of doubly occupied sites (doublon number) which can be determined using infinite system techniques.
This suggests a method for benchmarking finite system results, by checking that the integrated spectral weight in each channel agrees with the result for the doublon number from an infinite system.

We interpret our results in terms of the hard-core nature of the doublon charge excitations of the system.
Photodoping excites electrons from the lower to upper Hubbard band, creating doublons.
As they are hard-core, each doublon reduces the available phase space for the creation of new doublons, and this leads to a reduced interband response at energies $\omega \sim U$.
%For pumping near saturation the interband response is strongly suppressed.
The photodoping induced doublons lead to a new low energy response due to intraband transitions.
This response is sharply defined and dispersing.
In spin systems the corresponding intraband behaviour is known as a Villain mode~\cite{villain1975propagative}.
A similar intraband charge response has been noted in the equilibrium Hubbard chain at temperatures that are finite, but small compared to the Mott gap~\cite{nocera2018finite}.
A consequence of the strongly correlated nature of excitations in gapped, hard-core, spin systems is that the thermal broadening of the interband response is strongly asymmetric~\cite{essler2008finite,james2008finite,goetze2010low}.
We see similar asymmetry in the interband { charge response} after pumping.
%spin and charge responses after pumping.

This analogy with thermal effects in strongly correlated spin systems raises the question of whether an effective temperature might be ascribed to the ladder after pumping.
An earlier work by Wang et al.~\cite{wang2017producing} compared the nonequilibrium post pump state of the Hubbard chain to the equilibrium system at finite temperatures and found no agreement.
However, it is possible that the Hubbard chain is a special case as it has a gapless spin sector with spin-$1/2$ spinon excitations.
We have attempted to extract an effective temperature $\beta^{-1}$ from our post pump results using a general property of the \emph{equilibrium} dynamical structure factor $I(\omega,\mathbf{k})$,
\begin{align}
    I(-\omega,\mathbf{k}) =e^{-\beta \omega} I(\omega,\mathbf{k}).
\end{align}
We have not found a robust value of $\beta^{-1}$ that is shared for both spin and charge structure factors and at different wave vectors (for details see appendix~\ref{app:effectiveT}).
We conclude that, like the chain, the post pump state cannot be simply described using a thermal distribution.
Indeed Ref.~\cite{murakami2022exploring} has used a generalised Gibbs ensemble (GGE) approach to characterise the optically pumped Hubbard chain at late times, with the conservation of doublon number leading to an additional generalised temperature.
Checking if the post pump Hubbard ladder can be described using a GGE would be an interesting subject for future work.

\begin{acknowledgments}
The authors thank Andreas Weichselbaum, { Mark Dean, Calum MacCormick and Chris Hooley} for useful discussions.
RMK was supported by the U.S. Department of Energy, Office of Basic Energy Sciences, under Contract No. DE-SC0012704.
This work was supported by the Engineering and Physical Sciences Research Council [grant number EP/L010623/1].
\end{acknowledgments}
\appendix
{
\section{Short time expansion}
\label{app:shorttime}
The directional dependence of a optical pump on the ladder system can be demonstrated by considering a short time expansion.
The ladder Hamiltonian in the absence of pumping can be written as
\begin{align}
    H_0&=H_{\text{leg}}+H_{\text{rung}}+H_U,\nn
    H_{\text{leg}}&=\sum_{i\ell\sigma} \left[-\tpar \left(c^\dagger_{i\l \sigma} c_{i+1\l \sigma}
    +c^\dagger_{i+1\l \sigma} c_{i\l \sigma}\right)\right],\nn
    H_{\text{rung}}&=\sum_{i\sigma} \left[-\tper \left(c^\dagger_{i,1,\sigma} c_{i,2,\sigma}
    +c^\dagger_{i,2,\sigma} c_{i,1,\sigma}\right)\right],\nn
    H_U&=\sum_{i\ell} n_{i\ell \uparrow} n_{i \ell \downarrow}.
\end{align}
The pump induces time-dependent hoppings in the leg and rung{-}directions, 
\begin{align}
    H_{\text{leg}}(t)&=\sum_{i\ell\sigma} \left[-\tpar \left(e^{\i\phipar} c^\dagger_{i\l \sigma} c_{i+1\l \sigma}
    +e^{-\i\phipar} c^\dagger_{i+1\l \sigma} c_{i\l \sigma}\right)\right],\nn
    H_{\text{rung}}(t)&=\sum_{i\sigma} \left[-\tper \left(e^{\i\phiper} c^\dagger_{i,1,\sigma} c_{i,2,\sigma}
    +e^{-\i\phiper} c^\dagger_{i,2,\sigma} c_{i,1,\sigma}\right)\right],
\end{align}
where $\phipar$ and $\phiper$ are the time-dependent phases given by Peierls substitution.
The full time-dependent Hamiltonian can then be written as
\begin{align}
    H(t)=H_{\alpha}(t)+H_0,
\end{align}
where for a parallel (leg{-}direction) pump $\alpha=\parallel$ and
\begin{align}
    H_{\parallel}(t)=H_{\text{leg}}(t)-H_{\text{leg}},
\end{align}
and for a perpendicular (rung{-}direction) pump $\alpha=\perp$, 
\begin{align}
    H_{\perp}(t)=H_{\text{rung}}(t)-H_{\text{rung}}.
\end{align}
Consider a pump applied for a very short time $\delta t$, so that the time evolution operator $U(t)$ is given approximately by 
\begin{align}
    U(t) \approx \exp[-\i H(\delta t) \delta t].
\end{align}
We assume that the pump is switched on suddenly at $t=0$ (as in our numerics) but suppressed by a Gaussian factor, so that $A_\alpha(\delta t) \approx A_0 \exp[-t_p^2/(2\sigma_p^2)] \Omega \delta t \ll 1$.

The energy that is imparted by the pump in this short time $\delta t$, relative to the energy of the initial state, is
\begin{align}
    \delta E = \langle e^{\i H(\delta t) \delta t} H_0 e^{-\i H(\delta t) \delta t} \rangle - \langle H_0 \rangle
\end{align}
where the expectation is evaluated with respect to the ground state $\vert \psi_0 \rangle$ of the initial Hamiltonian.
Expanding the exponentials gives
\begin{align}
    \delta E = &\left\langle \i \left[H(\delta t),H_0\right]\delta t  -\frac{1}{2}\left\{H^2(\delta t),H_0\right\}\delta t^2\right. \nonumber \\
    &\qquad \left.\phantom{\frac{1}{2}}+H(\delta t)H_0 H(\delta t)\delta t^2 +O(\delta t^3) \right\rangle
\end{align}
The commutator in the first term vanishes because the initial state is an eigenstate of $H_0$, that is, $H_0\vert \psi_0 \rangle=E_0 \vert \psi_0 \rangle$.
Writing $H=H_0+H_\alpha$, expanding and then collecting terms gives
\begin{align}
    \delta E &= \langle H_\alpha(\delta t) (H_0-E_0) H_\alpha(\delta t) \rangle \delta t^2 +O(\delta t^3)
    \label{eq-deltaEapprox}
\end{align}
Further explicit evaluation would require knowledge of the precise form of $\vert \psi_0 \rangle$.
However, the dependence of $\delta E$ can be extracted heuristically as follows.
Processes that contribute to Equation~\ref{eq-deltaEapprox} involve a hop with coefficient $ t_\alpha(e^{\pm \i A_\alpha}-1)\approx \pm \i t_\alpha A_\alpha  $, a double occupancy penalty, $H_0-E_0 \sim U$ and then a hop back $t_\alpha(e^{\mp \i A_\alpha}-1)\approx \mp \i t_\alpha A_\alpha $.
For leg{-}direction quenches this gives a contribution $\sim 2t_\parallel\phipar^2 U$ per site where the factor of two arises due to the two nearest neighbours in the leg{-}direction.
For rung{-}direction quenches the contribution is $\sim t_\perp\phiper^2 U$ per site, because each site only has one nearest neighbour in the rung{-}direction.
Hence, the anisotropic nature of the Hamiltonian results in an energy dependence on pump direction even at the shortest times, and when $t_\perp=t_\parallel$ energy transfer to the ladder is enhanced for pumps in the leg{-}direction relative to the rung{-}direction.

\section{Pump frequency}
\label{app:frequency}
%The response of the ladder to the pump will depend on both the pump fluence (the total energy that can be photoinjected) and the specific characteristics of the pump: the amplitude, frequency and duration.
%The effect on final energy density of varying the amplitude $A_0$ was shown in Fig.~\ref{fig:energy-density}.

The results in the main body of this work were obtained using a pump frequency $\Omega=6$.
This value was chosen to enable comparison with a previous nonequilibrium investigation of the Hubbard chain.
In this appendix we consider other values of $\Omega$ and their effect on the injected energy density.
\begin{figure}
    \centering
    \includegraphics[scale=0.4]{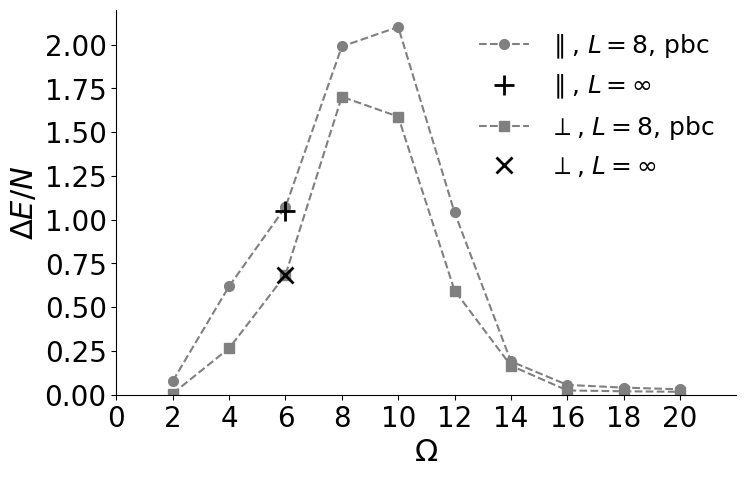}
    \caption{The change in energy density after the pump, as a function of the pump frequency $\Omega$ for $A_0=1$. Results are shown for a finite ladder with $L=8$ rungs and periodic boundary conditions (pbc). The infinite system results for $\Omega=6$ are also plotted for comparison.}
  \label{fig:energy_vs_omega}
\end{figure}
The response of the ladder to pumping depends on the pump frequency $\Omega$.
For very low frequencies, $\Omega \ll U$, the response is almost adiabatic: the state of the system will closely track the instantaneous ground state of the time-dependent Hamiltonian and at the end of the pump the system will be in a state that closely approximates the initial state, so $\Delta E \approx 0$.
At high frequencies, when $\Omega$ exceeds the inverse electron hopping timescale, the electrons cannot react fast enough to respond to the pump. This limits energy transfer and $\Delta E$ again approaches zero.
Between these extremes, a strong response is expected for frequencies $\Omega \sim U$, close to the Mott gap.

Fig.~\ref{fig:energy_vs_omega} shows the injected energy density as a function of pump (angular) frequency $\Omega$ for an infinite ladder and for an 8 rung ladder with periodic boundary conditions.
The form of the pump is given by Eq.~\ref{eq:pumpprotocol} with amplitude parameter $A_0=1$ and width $\sigma_p=1$.

For both pumping directions $\Delta E$ increases with frequency, reaching a maximum at $\Omega \approx U$ ($\Omega =10$ for the leg{-}direction and $\Omega=8$ for the rung{-}direction) before declining. 
This is consistent with the expectation that pumping with a frequency near the Mott gap should yield the strongest response.
}

\section{Effect of bond dimension}
\begin{figure}
    \centering
    \includegraphics[scale=0.4]{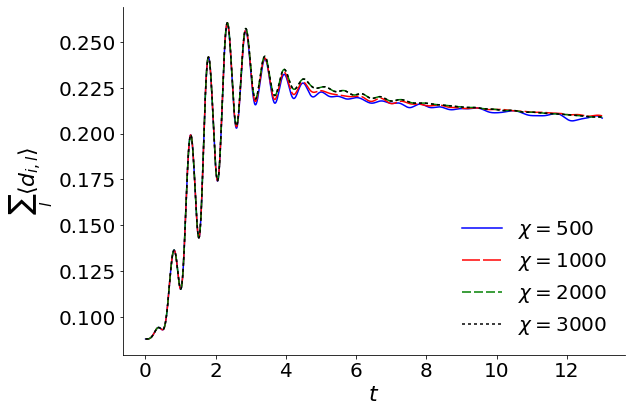}
    \caption{Doublon and holon density as the Hubbard ladder is pumped in the rung{-}direction, i.e. $(\perp , A_0 = 1.0)$, at different bond dimensions $\chi$.}
    \label{fig:doublon_convergence}
\end{figure}

Our simulations are carried out using infinite and finite MPS with bond dimensions of up to $\chi=3000$, and we perform time evolution using iTEBD and TEBD with a second order Trotter decomposition and time step, $\Delta t=0.01$.
In general we find truncation errors of order $10^{-6}$ at the latest times we report, but because correlations become very short ranged after the pump this does not significantly affect our results (especially once the experimental resolution $\sigma_X$ is taken into account).
As an example, in Fig.~\ref{fig:doublon_convergence} we plot $\langle d_{i,l}\rangle$ for the rung{-}direction pump with $A_0=1.0$ using iTEBD with different bond dimensions, $\chi$.
The rung pump case breaks parity symmetry between the legs leading to a less compact structure for the MPS matrices.
This makes it numerically more challenging to simulate than the leg{-}direction pumps, even those that inject greater energy.
The figure shows that even $\chi=500$ achieves good quantitative agreement for much of the simulation, but does develop an oscillatory error at late times.
The largest bond dimensions are all in excellent agreement for almost all times shown.
%A small departure occurs for $\chi=1000$ close to the end of the pump, but there is excellent agreement again by $t\approx 6$ for the larger $\chi$ simulations.
%
%
\section{Single time integral approximation}
\label{app:integral}
Eq.~\ref{eq:Idoubletime} is a two time integral that in principle requires computing $C(\mathbf{k},t_1,t_2)$ at many different values of $t_1$ and $t_2$.
This is numerically expensive because applying the operators within the correlator $C(\mathbf{k},t_1,t_2)$ at $t_1$ and $t_2$ alter the wave function, so a separate time evolution must be run for each $t_2$ (for example).
Some redundancy can be leveraged to reduce the numerical effort, by storing the wave function $\vert \psi(t) \rangle$ up to the maximum simulation time.
The recorded $\vert \psi(t) \rangle$ can then be used to form $\vert \psi(t_2) \rangle$ and $\langle \psi(t_1) \vert$, and then a separate evolution is needed from $t_2$ to $t_1$.
Note that this includes combinations of $t_1$ and $t_2$ that correspond to evolution both forward and backward in time.
To reduce the numerical overhead further we approximate Eq.~\ref{eq:Idoubletime} using a single time integral.
Defining $\bar{t}=(t_1+t_2)/2$ and $\tau=t_1-t_2$,
\begin{align}
  I(\mathbf{k},\omega,t) &= \iint \text{d}t_1 \text{d}t_2 e^{i\omega(t_1-t_2)}g(t_1,t)g(t_2,t)C(\mathbf{k},t_1,t_2)\nonumber\\
  &= \iint \text{d}\bar{t} \text{d}\tau \frac{\exp\left[i\omega \tau-\frac{\tau^2}{4\sigma_X^2}\right]\exp\left[-\frac{(t-\bar{t})^2}{\sigma_X^2}\right]}{2\pi \sigma_X^2}\nonumber\\
  &\quad\quad \times C(\mathbf{k},\bar{t}+\tau/2,\bar{t}-\tau/2)\nonumber\\
  &\approx\int  \text{d}\tau \frac{\exp\left[i\omega \tau-\frac{\tau^2}{4\sigma_X^2}\right]}{2\sqrt{\pi} \sigma_X}C(\mathbf{k},t+\tau/2,t-\tau/2).
\end{align}
The last line follows if $C(\mathbf{k},\bar{t}+\tau/2,\bar{t}-\tau/2)$ is slowly varying as a function of $\bar{t}$ for $\vert\bar{t}-t\vert \lesssim \sigma_X$.
Assuming this is true then
\begin{align*}
C(\mathbf{k},t+\tau/2,t-\tau/2)\approx \frac{1}{2}[C(\mathbf{k},t+\tau,t)+C(\mathbf{k},t,t-\tau)],
\end{align*}
and
\begin{align}
  I(\mathbf{k},\omega,t) &\approx \int  \text{d}\tau \frac{\exp\left[-\frac{\tau^2}{4\sigma_X^2}\right]}{4\sqrt{\pi} \sigma_X}
  \text{Re}\left[e^{i \omega t}C(\mathbf{k},t+\tau,t)\right],
  \label{eq:Iapprox}
\end{align}
which is equivalent to Eq.~\ref{eq:Isingletime} in the main text.

\begin{figure*} 
  \centering
  \begin{minipage}[b]{0.32\textwidth}
    \centering
    \includegraphics[width=\linewidth]{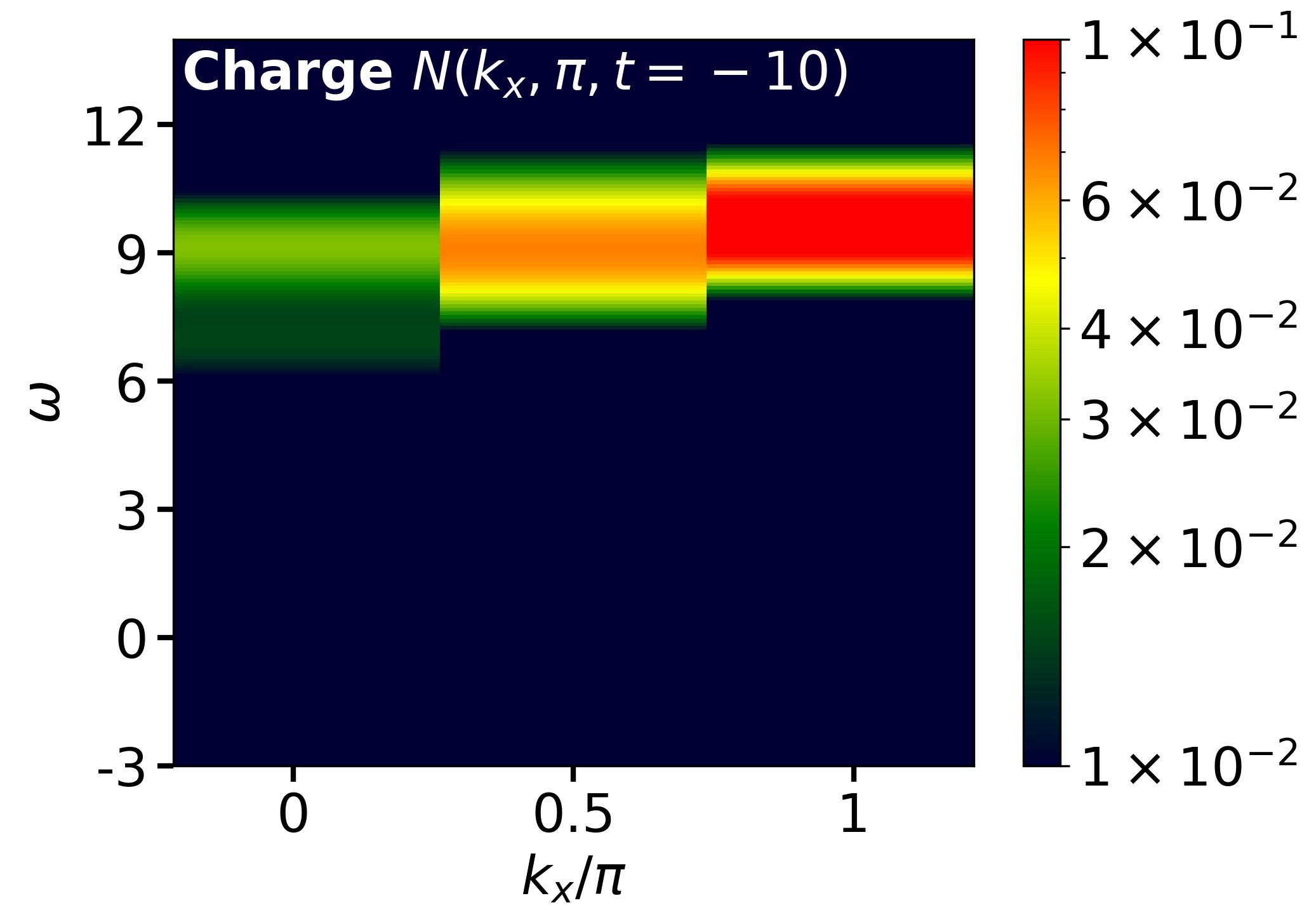}
  \end{minipage}
  \hfill
  \begin{minipage}[b]{0.32\textwidth}
    \centering
    \includegraphics[width=\linewidth]{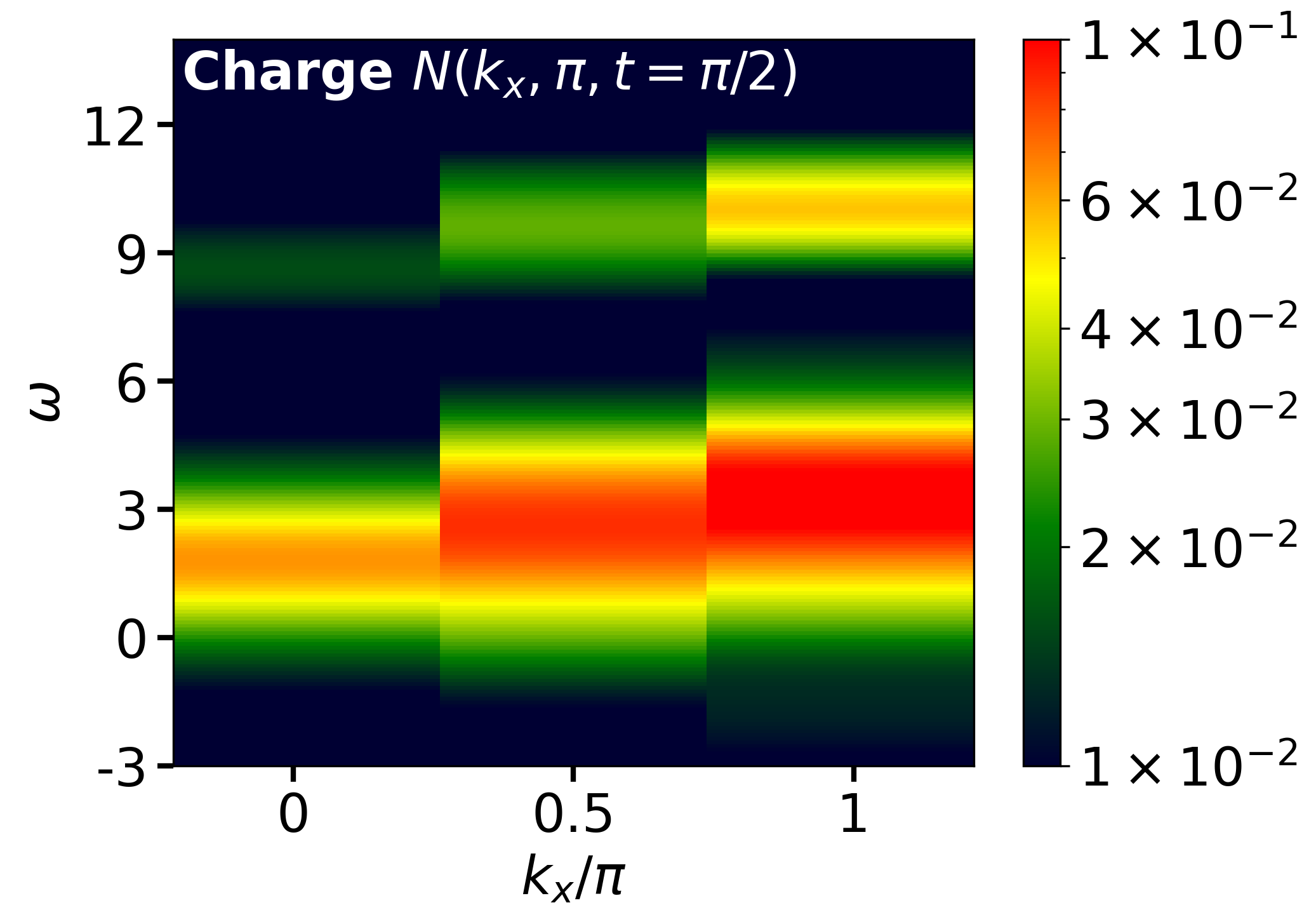}
  \end{minipage}
  \hfill
  \begin{minipage}[b]{0.32\textwidth}
    \centering
    \includegraphics[width=\linewidth]{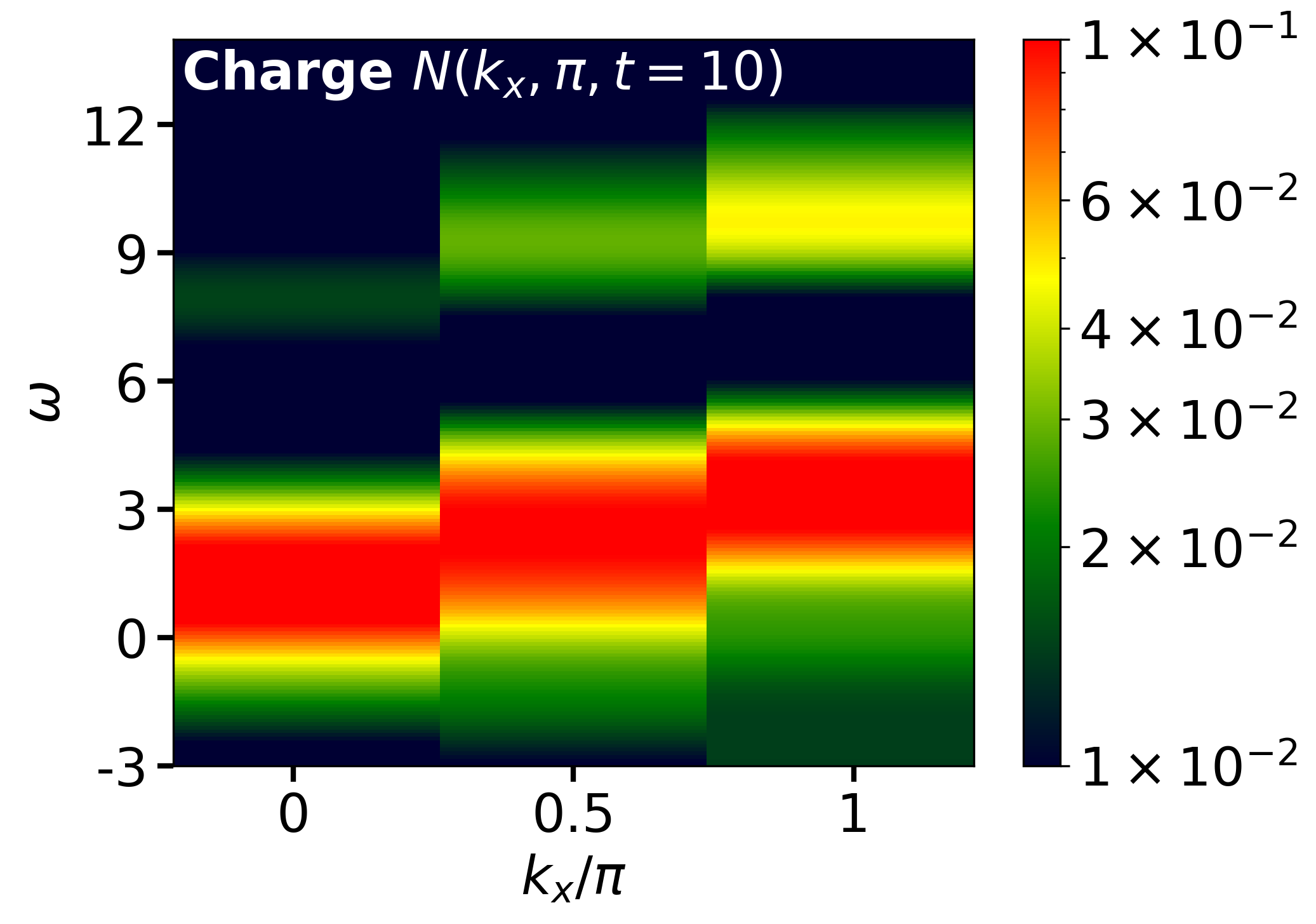}
  \end{minipage}
  \vspace{\baselineskip} 
  \begin{minipage}[b]{0.32\textwidth}
    \centering
    \includegraphics[width=\linewidth]{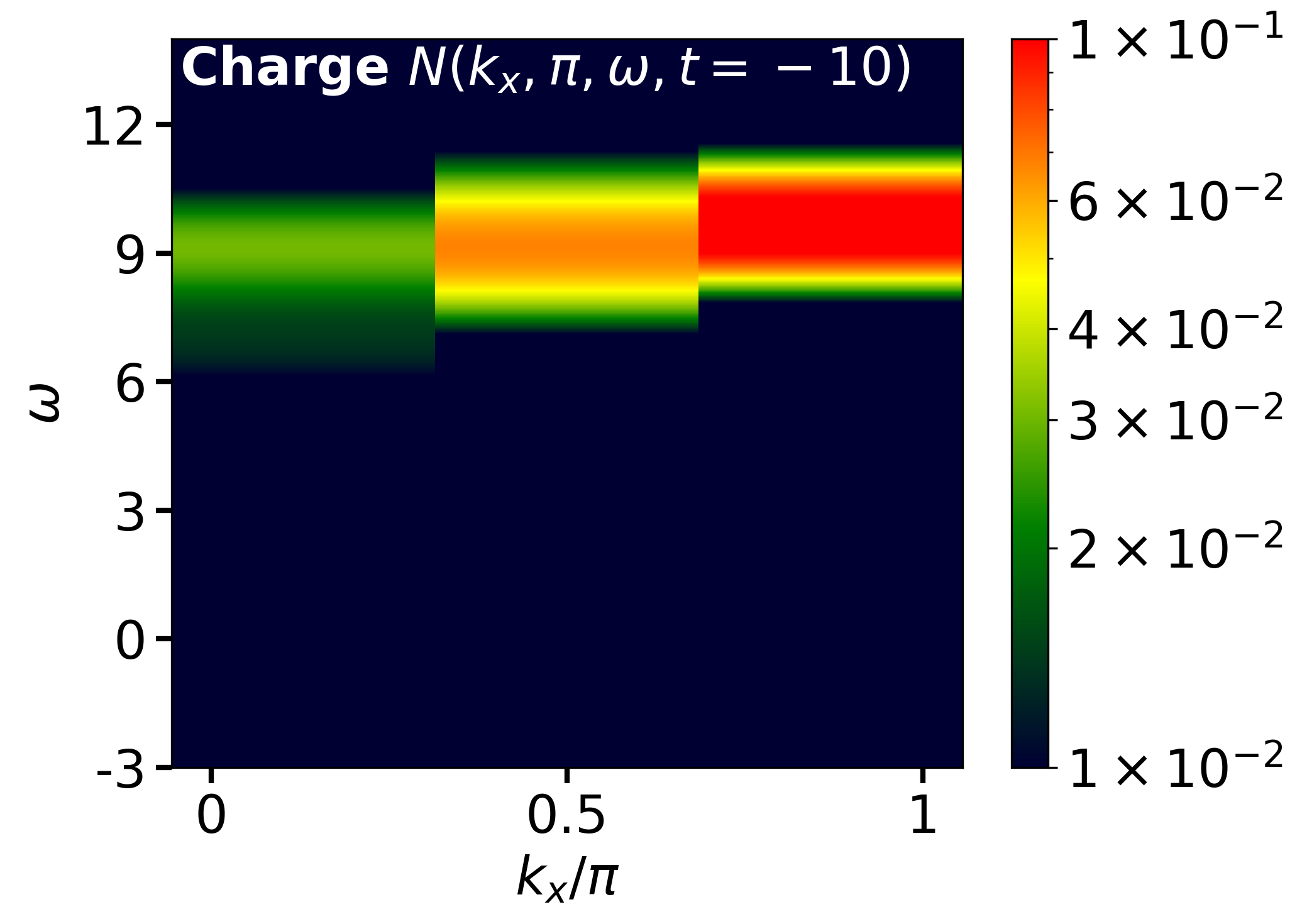}
  \end{minipage}
  \hfill
  \begin{minipage}[b]{0.32\textwidth}
    \centering
    \includegraphics[width=\linewidth]{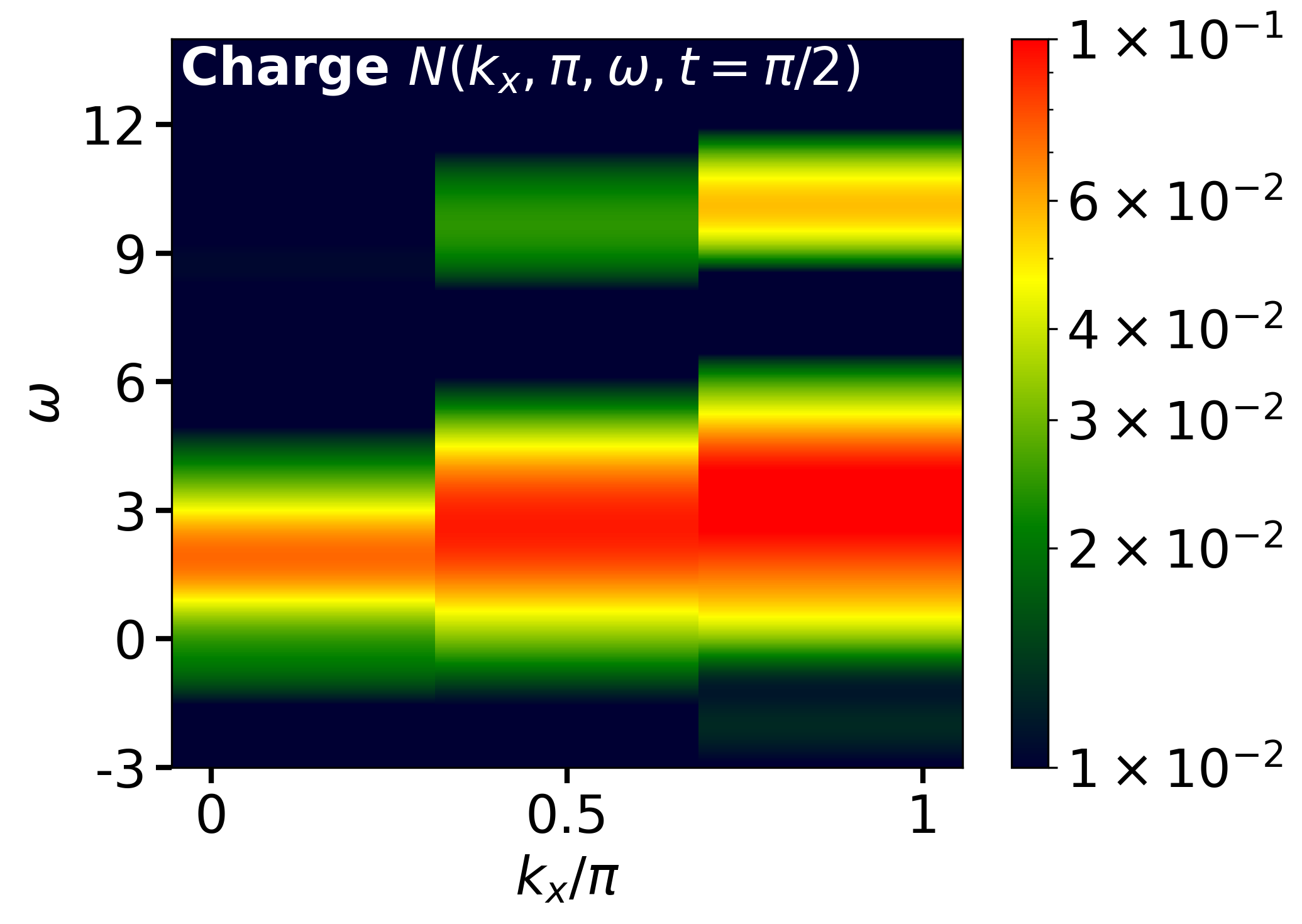}
  \end{minipage}
  \hfill
  \begin{minipage}[b]{0.32\textwidth}
    \centering
    \includegraphics[width=\linewidth]{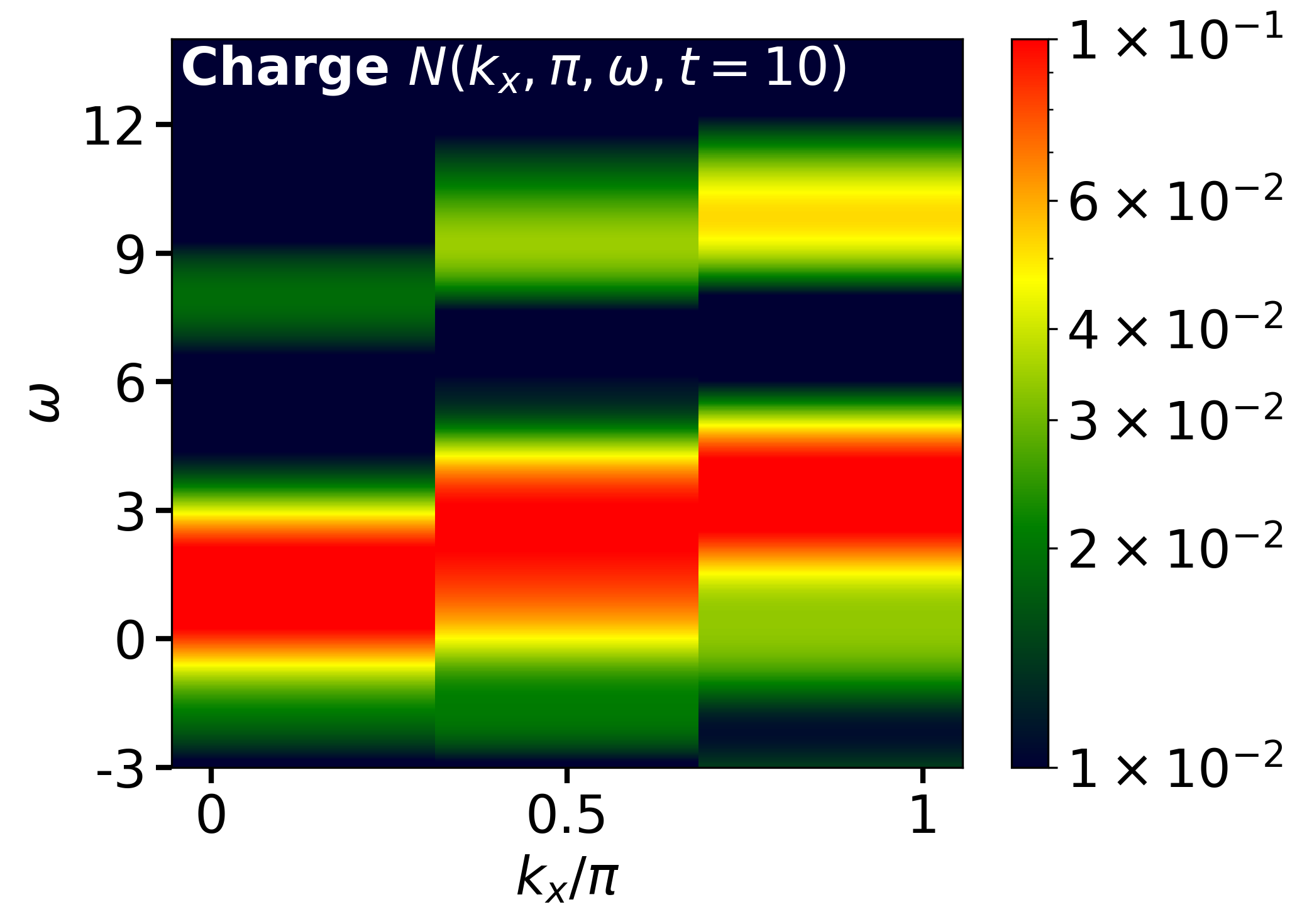}
  \end{minipage}
  \caption{Comparison between the full expression Eq.~\ref{eq:Idoubletime} and the approximation Eq.~\ref{eq:Isingletime} for the dynamical charge response of a 4 rung ladder using exact diagonalisation, for leg{-}direction pumping with $A_0=1$.}
  \label{fig:double_vs_single_approximation}
\end{figure*}

Fig.~\ref{fig:double_vs_single_approximation} shows a comparison between Eq.~\ref{eq:Idoubletime} and Eq.~\ref{eq:Isingletime} for the charge dynamics of a four rung ladder, using exact diagonalisation.
The approximation captures the intensity and spectral weight distribution (particularly the energy asymmetry of dispersing modes) of the response qualitatively, and in many cases quantitatively.
The worst agreement is during the pump ($t=\pi$) where the relatively rapid changes in the hopping parameter mean that the approximations made in arriving at Eq.~\ref{eq:Isingletime} are only obeyed on average.
Note that our pump protocols are idealised: a real pump would be composed of a (narrow) range of pump frequencies with a range of phase offsets, leading to phase averaging.
Therefore, we only hope to achieve a qualitative match during the pump between theory and a notional experiment.

\section{Effective temperature}
\label{app:effectiveT}
\begin{figure*}
    \centering
    \includegraphics[scale=0.6]{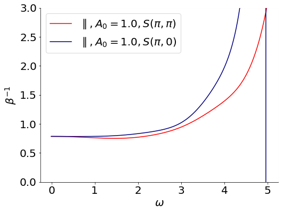}
    \includegraphics[scale=0.6]{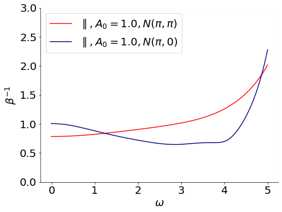}
    \caption{The temperature $\beta^{-1}$, extracted using Eq.~\ref{eq:invbeta} for the $A_0=1$, leg{-}direction pump.
    Left panel: results using the spin response at two different wave vectors, $(\pi,\pi)$ and $(\pi,0)$.
    Right panel: the corresponding results from the charge response.}
  \label{fig:thermalleg}
\end{figure*}

\begin{figure*}
    \centering
    \includegraphics[scale=0.6]{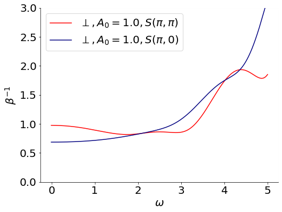}
    \includegraphics[scale=0.6]{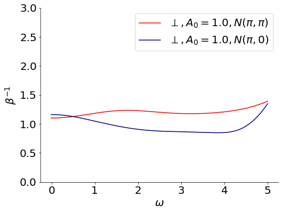}
    \caption{The temperature $\beta^{-1}$, extracted using Eq.~\ref{eq:invbeta} for the $A_0=1$, rung{-}direction pump.
    Left panel: results using the spin response at two different wave vectors, $(\pi,\pi)$ and $(\pi,0)$.
    Right panel: the corresponding results from the charge response.}
  \label{fig:thermalrung}
\end{figure*}

In an equilibrium system at finite temperature the dynamical response is time invariant, $I(\omega,\mathbf{k},t)=I(\omega,\mathbf{k})$ and obeys
\begin{align}
    I(-\omega,\mathbf{k})=e^{-\beta \omega} I(\omega,\mathbf{k})
    \label{}
\end{align}
as can be shown using the e.g. the Lehmann representation.
This is consistent with the requirement that $I(\omega,\mathbf{k})=0$ for negative frequencies at zero temperature, $\beta \to \infty$.
However, broadening due to the finite probe width $\sigma_X$ can yield a finite $\omega \le 0$ response even for a system at zero temperature (see e.g. the equilibrium cases in Fig.~\ref{fig:compare_rungs_vs_legs_quench_cutsPIPI} and Fig.~\ref{fig:compare_rungs_vs_legs_quench_cutsPI0} which appear as resolution limited Gaussians).
The above notwithstanding, a more significant negative frequency response is seen after pumping in both the spin and charge channels.
This broader response cannot be attributed to the probe width, and is a result of the system no longer being in its ground state.
The negative energy response can be used to try and attribute a temperature to the system by calculating
\begin{align}
\beta^{-1} = \omega \left(\ln \left[ \frac{I(\omega,\mathbf{k})}{I(-\omega,\mathbf{k})}\right]\right)^{-1}.
\label{eq:invbeta}
\end{align}
If the system is thermal, then it should display the same temperature at all momenta, and in both then spin and charge channels.
We show results for two of the pump protocols in Fig.~\ref{fig:thermalleg} and Fig.~\ref{fig:thermalrung}.
Numerical issues occur for larger $\vert\omega\vert$ when the response becomes very small, causing Eq.~\ref{eq:invbeta} to diverge.
At smaller $\omega$ the extracted temperature is finite, but even when $\beta^{-1}$ is slowly varying with $\omega$ there is not clear consistency in its value between different wave vectors and response channels, for either of the pump protocols shown.
We conclude that the response is not simply thermal.

\bibliographystyle{apsrev4-2}
\bibliography{Bibliography.bib}
\end{document}